\documentclass{JHEP3}
\usepackage{amsmath}
\usepackage{latexsym}
\usepackage{graphicx}
\usepackage{cite}
\usepackage{bbm}
\usepackage{mathrsfs}
\allowdisplaybreaks[1]

\numberwithin{equation}{section}

\def\AdSs5{$AdS_5$}
\def\AdSS5{$AdS_5$}
\def\AdS5s5{$AdS_5 \times S^5$}
\def\al{{\alpha^{\prime}}}
\def\gs{g_{st}}
\def\gy{g_{_{\rm YM}}}
\def\er{{\rm e}}
\def\dr{{\rm d}}
\def\Tr{{\rm Tr}}

\def\gs{g_{\rm s}}

\newcommand{\RR}{${\rm R}\!\otimes\!{\rm R}$\ }

\newcommand{\eg}{{\it e.g.~}}
\newcommand{\ie}{{\it i.e.~}}

\newcommand{\s}{\sigma}

\newcommand{\be}{\begin{equation}}
\newcommand{\ee}{\end{equation}}
\newcommand{\ba}{\begin{eqnarray}}
\newcommand{\ea}{\end{eqnarray}}
\newcommand{\bdm}{\begin{displaymath}}
\newcommand{\edm}{\end{displaymath}}
\newcommand{\ra}{\rangle}
\newcommand{\la}{\langle}
\newcommand{\pp}{\prime}

\newcommand\fr[1]{\frac{1}{#1}}
\newcommand{\barZ}{{\bar Z}}

\newcommand{\wtil}{\widetilde}
\newcommand{\what}{\widehat}
\newbox\SlashedBox
\def\fs#1{\setbox\SlashedBox=\hbox{#1}
\hbox to
0pt{\hbox to 1\wd\SlashedBox{\hfil/\hfil}\hss}{#1}}
\def\hboxtosizeof#1#2{\setbox\SlashedBox=\hbox{#1}
\hbox to
1\wd\SlashedBox{#2}}

\def\ms#1{\setbox\SlashedBox=\hbox{$#1$}
\hbox to 0pt{\hbox to
1\wd\SlashedBox{\hfil/\hfil}\hss}#1}

\newcommand{\Dsm}{\,{\raisebox{1pt}{$/$} \hspace{-8pt} D}}

\newcommand{\Z}{\mathbbm{Z}}

\def\t2{\tau_2}
\def\IZ{\relax\ifmmode\mathchoice {\hbox{\cmss Z\kern-.4em Z}}
{\hbox{\cmss Z\kern-.4em Z}}
{\lower.9pt\hbox{\cmsss Z\kern-.4em Z}}
{\lower1.2pt\hbox{\cmsss Z\kern-.4em Z}}
\else{\cmss Z\kern-.4em Z}\fi}

\def\S{\Sigma}

\def\xib{{\bar \xi}}

\def\b{\beta}
\def\a{{\alpha}}
\def\g{\gamma}
\def\veps{\varepsilon}
\def\vt{\vartheta}
\def\bvt{{\bar\vartheta}}

\def\adot{{\dot\alpha}}
\def\bdot{{\dot\beta}}

\def\d{\delta}

\def\D{\Delta}

\def\c1{{\chi^1}}

\def\v{\varphi}

\def\N4{{\cal N}=4}
\def\half{\frac{1}{2}}

\def\nn{\nonumber}

\def\nus{\begin{displaystyle}
    \bar\nu_{\rule{0pt}{2pt}}^{u[A}\nu^{B]}_u
    \end{displaystyle}}
\def\nut{\begin{displaystyle}
    \bar\nu_{\rule{0pt}{2pt}}^{u(A}\nu^{B)}_u
    \end{displaystyle}}
\def\nsix{(\bar\nu \nu)_{\bf 6}}
\def\nten{(\bar\nu \nu)_{\bf 10}}
\newcommand{\Git}{\mathit{\Gamma}}
\newcommand{\calD}{{\mathcal D}}

\newcommand{\calJ}{{\mathcal J}}

\newcommand{\calS}{{\mathcal S}}

\newcommand{\scrF}{{\mathscr F}}

\newcommand{\scrN}{{\mathscr N}}
\newcommand{\scrO}{{\mathscr O}}

\newcommand{\scrZ}{{\mathscr Z}}
\DeclareMathAlphabet{\mathpzc}{OT1}{pzc}{m}{it}

\newcommand{\scrm}{{\mathpzc m}}

\newcommand{\scrmf}{{\mathpzc m}_{\rm \, f}}
\newcommand\hsp[1]{\hspace*{#1 cm}}

\newcommand{\mb}[1]{\mathbf{#1}}
\newcommand{\mbb}[1]{\mathbf{\overline{#1}}}
\title{Non-perturbative effects in the BMN limit of $\scrN$=4
supersymmetric Yang--Mills}
\author{Michael B. Green and Aninda Sinha \\
Department of Applied Mathematics and
Theoretical Physics \\
Wilberforce Road, Cambridge CB3 0WA, UK \\
E-mail: \email{M.B.Green@damtp.cam.ac.uk, A.Sinha@damtp.cam.ac.uk}}
\author{Stefano Kovacs \\
Max-Planck-Institut f\"ur Gravitationsphysik \\
Albert-Einstein-Institut \\
Am M\"uhlenberg 1, D-14476 Potsdam, Germany  \\
E-mail: \email{stefano.kovacs@aei.mpg.de}}

\abstract{One-instanton contributions to the correlation functions
of two gauge-invariant single-trace operators in $\scrN=4$ SU($N$)
Yang--Mills theory are studied in semi-classical approximation in the
BMN limit. The most straightforward examples involve operators with
four bosonic impurities (whereas examples with two-impurity operators
pose technical problems). The explicit form for the correlation
functions, which determine the anomalous dimensions, follows after
integration over the large number of bosonic and fermionic moduli.
Our results demonstrate that the instanton contributions scale
appropriately in the BMN limit. We find impressive agreement with the
$D$-instanton contributions to mass matrix elements of the dual
plane-wave IIB superstring theory, obtained in a previous paper. Not
only does the dependence on the scaled coupling constants match, but
the dependence on the mode numbers of the states is also in striking
agreement.}

\preprint{DAMTP-2005-53 \\ AEI-2004-059}

\begin{document}

\section{Introduction}
\label{intro}

According to \cite{bmn} there is a  very interesting limit of the
AdS/CFT correspondence that relates a  special sector of the $\scrN$=4
supersymmetric Yang--Mills (SYM) theory to type IIB string theory in a
maximally supersymmetric plane-wave background. A notable feature of
this proposal is that it provides the first example of a gauge/gravity
duality which can be studied in a quantitative way beyond the
supergravity approximation. This is possible because string theory in
the relevant background, which is obtained as a Penrose limit of
AdS$_5\times S^5$ \cite{penrose}, can be quantised in the light cone
gauge \cite{met,mt}. Moreover there exists a regime in which both the
string and the gauge theory are weakly coupled, although there are
subtleties associated with the different way in which the limits
leading to this regime are taken on the two sides. This has allowed
very precise comparisons between perturbative corrections in the two
theories \cite{pert,recentgauge}.

The duality relates the string mass spectrum to the spectrum of
scaling dimensions of gauge theory operators in the so called BMN
sector of $\scrN$=4 SYM. This consists of gauge invariant operators of
large conformal dimension, $\D$, and large charge, $J$, with respect
to a U(1) subgroup of the SU(4) R-symmetry group. The duality involves
the double limit $\D\to\infty$, $J\to\infty$. The combination $\D-J$,
which is kept finite, is related to the string theory hamiltonian,
\be
\D-J = \fr{\mu} \, H^{(2)} \, ,
\label{dict}
\ee
where $\mu$ is the background value of the \RR five-form and is
related to the mass parameter, $m$, which appears in the light cone
string action by $m=\mu p_-\al$ \cite{met,mt}, where $p_-$ is the
light cone momentum.

The correspondence between the spectra of the two theories is thus the
statement that the eigenvalues of the operators on the two sides of
the equality (\ref{dict}) coincide.   A quantitative comparison is
possible if one considers the large $N$ limit in the gauge theory
focusing on operators in the BMN sector. As a result of combining the
large $N$ limit with the limit of large $\D$ and $J$, new effective
parameters arise \cite{kpss,mit1}, which are related to the ordinary
't Hooft parameters, $\lambda$ and $1/N$, by a rescaling,
\be
\lambda^\pp = \frac{\gy^2N}{J^2} \, , \qquad g_2 = \frac{J^2}{N} \, .
\label{g2lambpdef}
\ee
The correspondence relates these effective gauge theory couplings to
string theory parameters in the plane-wave background,
\be
m^2=(\mu p_-\al)^2 = \fr{\lambda^\pp} \, , \qquad
4\pi\gs m^2 = g_2 \, .
\label{paramid}
\ee
The double scaling limit, $N\to\infty$, $J\to\infty$ with $J^2/N$
fixed, connects the weak coupling regime of the gauge theory to string
theory at small $g_s$ and large $m$.

In this limit the leading perturbative corrections to the scaling
dimensions of BMN operators have been successfully compared to the
leading quantum corrections to the masses of the dual plane-wave
string states \cite{kpss,mit1,bkpss,mit2,pert,recentgauge,sz}, see
also the reviews \cite{reviews} for further references. In the present
paper we will study one-instanton effects in the $\scrN$=4 Yang--Mills
theory. These will be compared with $D$-instanton \cite{gg} induced
corrections to the plane-wave string mass spectrum that were computed
in \cite{gks} in order to check the validity of the BMN conjecture in
non-perturbative sectors.  In the original formulation of the AdS/CFT
correspondence very good agreement was found between the effects of
instantons in the $\scrN$=4 Yang--Mills theory and of $D$-instantons
in type IIB string theory in AdS$_5\times S^5$
\cite{bgkr,dhkmv,gk}. It is therefore of interest to see if similar
agreement can be established in the BMN limit and whether the results
of \cite{gks} can be reproduced from the study of instanton
contributions to the anomalous dimensions of BMN operators.

The possibility of testing the correspondence at the non-perturbative
level is especially relevant since several aspects of the
perturbative tests of the duality are only partially understood. A
precise holographic formulation of the duality connecting the dynamics
of the two theories beyond the identification of the spectra is still
lacking and even the explicit tests at the level of the spectrum are
not comprehensive. A limited class of states/operators has been
studied and agreement has been explicitly verified only at leading
order in $g_2$ (the planar limit). This is the limit in which the
string is free and the $\lambda'$ perturbation series on the gauge
side reproduces the free string spectrum. The first non-planar
contributions, of order $\lambda^\pp g_2^2$ or one-loop on the
string side, have also been compared successfully (although in this
case simplifying assumptions were made about the one-loop string
theory calculation, which have since been questioned \cite{gpa}). The
systematics of the perturbative expansion beyond these leading order
contributions has not been studied and the fact that the double
scaling parameters (\ref{g2lambpdef}) that arise at low orders indeed
represent the correct expansion parameters at all orders remains a
conjecture.

Results obtained in different but related limits of the AdS/CFT
duality, both in string theory \cite{nearbmnstring} and on the gauge
theory side \cite{nearbmngauge}, suggest the possibility that BMN
scaling, \ie the order by order reorganisation of the perturbative
expansion into a double series in $\lambda^\pp$ and $g_2$, might break
down at higher orders. In the strict BMN sector the scaling
(\ref{g2lambpdef}) has been verified to three loops in perturbation
theory \cite{bmn3loop}, but a deviation was observed in a related
matrix model calculation \cite{fkp} at four loops.

In this paper we will show that instanton contributions to the
conformal dimensions of BMN operators display BMN scaling. Two-point
functions of BMN operators computed in the semi-classical approximation
will be shown to be in striking agreement with the $D$-instanton induced
two-point amplitudes computed in \cite{gks}. The agreement includes
not only the dependence on the parameters $\lambda^\pp$ and $g_2$, but
also the dependence on the mode numbers characterising the states.  In
particular the agreement with \cite{gks} in the mode number dependence
is highly non-trivial and requires dramatic cancellations. These
results combined with the three loop perturbative result provide
substantial evidence indicating that BMN scaling should persist at all
orders.

Instanton contributions to the anomalous dimensions of BMN operators
are extracted from two-point correlation functions computed in the
semi-classical approximation. Conformal invariance determines the form
of two-point functions of primary operators, $\scrO$ and $\bar\scrO$,
to be
\be
\la\scrO(x_1)\bar\scrO(x_2)\ra = \frac{c}{(x_1-x_2)^{2\D}} \, ,
\label{conf2pt}
\ee
where $\D$ is the scaling dimension. In general in the quantum theory
$\D$ acquires an anomalous term, $\D(\gy)=\D_0+\g(\gy)$. At weak
coupling the anomalous dimension $\g(\gy)$ is small and substituting in
(\ref{conf2pt}) gives
\be
\la\scrO(x_1)\bar\scrO(x_2)\ra
=\frac{c\,\Lambda^{2\g(\gy)}}{(x_1-x_2)^{2\D_0}}
\left( 1-\g(\gy) \log\left[\Lambda^2(x_1-x_2)^2\right] + \cdots
\right) \, ,
\label{adimexp}
\ee
where $\Lambda$ is an arbitrary renormalisation scale. As a function of
the coupling constant the anomalous dimension admits an expansion
consisting of a perturbative series plus non-perturbative
corrections. The generic two-point function at weak coupling takes the
form
\ba
\la\scrO(x_1)\bar\scrO(x_2)\ra &=& \frac{c(\gy)}{(x_1-x_2)^{2\D_0}}
\left( 1-\gy^2\g^{(1)} \log\left[\Lambda^2(x_1-x_2)^2 \right] \right.
\nn \\
&& \left. + \cdots - \er^{2\pi i\tau}\g^{\rm (inst)}
\log\left[\Lambda^2(x_1-x_2)^2\right] + \cdots \right) \, .
\label{adimexp2}
\ea
Therefore perturbative and instanton contributions to the anomalous
dimension are extracted from the coefficients of the logarithmically
divergent terms in a two-point function. When there is more than one
operator with the same quantum numbers operator mixing occurs. In this
case the resulting set of two-point functions determines a matrix of
anomalous dimensions and the eigenvalues of this matrix are the
physical anomalous dimensions. The issue of operator mixing was first
discussed in the context of the BMN limit in \cite{bers}. 

The procedure for calculating the instanton-induced contribution to
the anomalous dimensions in semi-classical approximation is as
follows.  The gauge-invariant operators in the BMN sector are defined
by  colour traces involving a large number of elementary scalar fields
together with a finite number of bosonic or fermionic `impurities'. In
the semi-classical approximation correlation functions of such
operators are computed by replacing the fields by the solution to the
corresponding field equations in the presence of an instanton,
expressed in terms of the fermionic and bosonic moduli, and
integrating the resulting profiles over these moduli. These moduli
encode the broken superconformal symmetries together with the
(super)symmetries associated with the orientation of a SU($2$)
instanton within SU($N$). For large $N$ integration over these
moduli is carried out by a saddle point procedure (as in \cite{dhkm}).

The general  structure of  the anomalous dimensions of gauge invariant
operators in the $\scrN$=4 Yang--Mills theory with SU($N$) gauge group
is an expansion of the form
\be
\g(\gy,\theta, N) = \sum_{n=1}^\infty \g^{\rm pert}_n(N)\,\gy^{2n} +
\sum_{K>0}\sum_{m=0}^\infty \left[\g^{(K)}_m(N)\,\gy^{2m}\,
\er^{2\pi i\tau K} +\;\mathrm{c.c.} \, \right] \, ,
\label{gamma-exp}
\ee
where $\tau = \frac{\theta}{2\pi}+i\frac{4\pi}{\gy^2}$. The double
series in the second term in (\ref{gamma-exp}) contains the
contributions of multi-instanton sectors as well as the perturbative
fluctuations in each such sector.
One reason for studying instanton effects even though they are
 exponentially suppressed in the small coupling
limit, is that they determine the dependence on the
$\theta$-angle in $\scrN$=4 SYM.
They therefore  play an  essential r\^ole in implementing
 $S$-duality which is a symmetry of the theory, just as $D$-instantons
 are crucial for the $S$-duality in type IIB string theory.

If the BMN sector of the gauge theory scales appropriately
(\ref{gamma-exp}) becomes a series in the scaled couplings
$\lambda^\pp$ and $g_2$. In particular, we will show that the
leading one-instanton contribution to the two-point functions of a
class of four impurity BMN operators scales as it should in the BMN
limit and has the form
$(1/n_1n_2)^2\,g_2^{7/2}\exp\left(-8\pi^2/g_2\lambda^\pp+i\theta\right)$,
where the integers $n_1$ and $n_2$ correspond to the mode numbers of
the dual string state.  This result is in striking agreement with  the
corresponding $D$-instanton induced mass matrix on the string side
found in \cite{gks}.

For certain other classes of operators the leading one-instanton
contribution vanishes and the first non-zero correction is of higher
order in $\lambda^\pp$ (or a lower power of $m$ in the string
calculation). In such cases the calculation requires knowledge of a
non-leading term in the scalar solution -- a term involving six
fermionic moduli (whereas the leading term is quadratic in fermionic
moduli). We have not evaluated the precise form of this contribution
and so have not determined  the precise form of the matrix elements in
these cases. However, there is strong evidence that these also match
the string calculations. For example, for two impurity operators,
with some mild assumptions about the manner in which the fermion
moduli are distributed in the profile of the operators, we will find a
contribution to the two-point function of the form
${\lambda^\pp}^2g_2^{7/2}\exp\left(-8\pi^2/g_2\lambda^\pp+i\theta\right)$,
in accord with expectations from the string side. Later we will
comment on the systematics of the expansion in the one-instanton
sector and on how the higher order corrections can give rise to a
double series in $\lambda^\pp$ and $g_2$.

This paper is organised as follows. In section \ref{N4fieldops} we
review some general aspects of the $\scrN$=4 Yang--Mills theory and
the BMN limit. The general method for  evaluating instanton-induced
contributions to two-point functions of BMN operators in terms of zero
modes and the integration over super-moduli is described in section
\ref{inst2ptf}. The manner in which the profiles of the fields depend
on these moduli is presented in  section  \ref{N4instmult}. In
section \ref{2pt-BMN} we consider some specific examples of two-point
functions of BMN operators and derive expressions for the anomalous
dimensions that arise after integration over all the moduli. We first
consider the case of two-impurity operators (which presents the
technical difficulty alluded to above) and then four-impurity
operators. We conclude with a discussion in section \ref{concl},
which includes a comparison with the string results in \cite{gks}.
Some technical details of the calculations are presented in the
appendices.

\section{Fields and operators in the BMN limit}
\label{N4fieldops}

The purpose of this section is mainly to present the notation used in
the paper and to define the dictionary to be used for the comparison
with string theory in the plane-wave background. We will only consider
a small set of BMN operators with scalar impurities which are dual to
the string states studied in \cite{gks}. A more detailed discussion of
the various types of operators relevant for the comparison with string
theory in the plane-wave background can be found in the review papers
\cite{reviews}.

\subsection{Fields in $\scrN$=4 SYM}
\label{n4fields}

The $\scrN$=4 multiplet comprises six real scalars, $\hat\v^i$,
$i=1,\ldots,6$, four Weyl fermions, $\lambda^A_\a$, $A=1,\ldots,4$,
and a vector, $A_\mu$, with field strength $F_{\mu\nu}$, all
transforming in the adjoint representation of the gauge group. These
are the building blocks used to construct gauge invariant composite
operators which are classified according to the irreducible
representations of the superconformal group, SU(2,2$|$4). The latter
are identified by the quantum numbers $(\D,j_1,j_2;a,b,c)$ of the
maximal bosonic subgroup SO(2,4)$\times$SO(6), where $\D$ is the
scaling dimension, $j_1$ and $j_2$ the Lorentz spins and $[a,b,c]$ the
SU(4)$\sim$SO(6) Dynkin labels.

Under the SU(4) R-symmetry group the scalars transform in the $\mb6$,
the fermions in the $\mb4$ (and their conjugates in the $\mbb4$) and
the gauge field is a singlet. It is often convenient to label the
scalars by an antisymmetric pair of indices in the $\mb4$,
$\v^{[AB]}$, subject to the reality condition
\be
\bar\v_{AB} \equiv \left(\v^{AB}\right)^* = \fr2 \veps_{ABCD}\v^{CD}
\, . \label{realcond}
\ee
The two parametrisations of the $\scrN$=4 scalars, $\hat\v^i$ and
$\v^{AB}$, are related by
\be
\hat\v^i = \fr{\sqrt{2}}\,\S^i_{AB} \v^{AB} \, , \qquad
\v^{AB} = \fr{\sqrt{8}}\,\bar\S^{AB}_i \hat\v^i \, ,
\label{scalAB-scali}
\ee
where $\S^i_{AB}$ ($\bar\S^{AB}_i$) are Clebsch--Gordan coefficients
projecting the product of two $\mb4$'s ($\mbb4$'s) onto the
$\mb6$. They are defined in appendix \ref{useformulae}.  The
representation of the scalars  in terms of the $\v^{AB}$ fields is the
most convenient for instanton calculations since, as we shall see, it
makes manifest which fermion zero modes in a correlation function can
be soaked up by each scalar field.

In the limit relevant for the comparison with string theory in the
plane-wave background the symmetry group is a contraction of the
original group and the operators are classified according to
representations of the bosonic subgroup
SO(4)$\times$SO(4)$\times$U(1)$\times$U(1). We shall denote by $\calD$
the dilation operator and by $\calJ$ the U(1) generator selected by
the Penrose limit in the dual AdS background. The
SO(4)$\times$SO(4)$\times$U(1)$\times$U(1) quantum numbers are
$(s_1,s_2;s_1^\prime,s_2^\prime;\D,J)$, where $\D$ and $J$ refer to
$\calD$ and $\calJ$ and the spins $s_i$ and $s_i^\prime$ refer to the
two SO(4) factors. These can be considered to be respectively
subgroups of the original SO(6) and SO(2,4) groups.  This
identification is not completely correct.  The generators of the two
SO(4)'s corresponding to the isometries of the dual string background,
$\wtil G_i$, are related to the generators of the Euclidean Lorentz
group and to those of an SO(4) subgroup of the R-symmetry group,
$G_i$, $i=1,2$, by a similarity transformation, $\wtil
G_i=TG_iT^{-1}$. This distinction, however, will not be relevant for our
analysis.

Since a precise formulation of the gauge theory dual to the plane wave
string theory is not known, the rules for the decomposition of the
$\scrN$=4 fields according to representations of
SO(4)$\times$SO(4)$\times$U(1)$\times$U(1) are determined by the
quantum numbers of the dual string excitations. The gauge invariant
operators corresponding to states in the string spectrum will be
discussed in the next subsection. String excitations created by
bosonic and fermionic oscillators are associated respectively with the
insertion of bosonic and fermionic elementary fields (``impurities'')
in composite operators.

Bosonic excitations in the plane wave string theory originate from the
vector of SO(8) which decomposes under SO(4)$\times$SO(4) as
\be
\mb8_{\rm v} =
\left[\left(\half,\half\right);\left(0,0\right)\right] \oplus
\left[\left(0,0\right);\left(\half,\half\right)\right] \, ,
\label{so8vdecomp}
\ee
\ie they are vectors of one SO(4) and singlets of the second or vice
versa. Correspondingly in the $\scrN$=4 theory the six real scalars
are reorganised into a complex field, $Z$, and its conjugate, $\bar
Z$, which are singlets of SO(4)$\times$SO(4) and have $\D=1$ and
$J=\pm 1$ respectively, and four real fields which transform in the
$\mb4=(\half,\half)$ of the first SO(4) and are singlets with respect
to the second and have $J=0$ and $\D=1$. The insertion of the four
real scalars in a composite operator corresponds to the insertion of
bosonic creation operators with an index in one of the two SO(4)
factors in the dual string state.  States created by bosonic
oscillators which are vectors of the second SO(4) correspond to
operators involving insertions of $D_\mu Z$.  The fields $D_\mu Z$ are
in the $\mb4=(\half,\half)$ of the second SO(4) and have $J=1$ and
$\D=2$. Explicitly the scalar fields are
\be
\begin{array}{ll}
\displaystyle
Z = \phi^1 = 2\v^{14} \, , \quad & \displaystyle \bar Z =
\phi_1^\dagger = 2 \v^{23} \, , \rule{0pt}{20pt} \\
\displaystyle \v^1 = \hat\v^2 = \frac{1}{\sqrt{2}} \left( -\v^{13}
+ \v^{24} \right) \, , \quad & \displaystyle
\v^2 = \hat\v^3 = \frac{1}{\sqrt{2}} \left( \v^{12} + \v^{34} \right)
\, , \rule{0pt}{20pt} \\
\displaystyle \v^3 = \hat\v^5 = \frac{i}{\sqrt{2}} \left( -\v^{13} -
\v^{24} \right) \, , \quad & \displaystyle
\v^4 = \hat\v^6 = \frac{i}{\sqrt{2}} \left( \v^{12} - \v^{34} \right)
\, , \rule{0pt}{20pt}
\end{array}
\label{vi4}
\ee
Here, for convenience of notation, we have introduced the scalars
$\v^i$, $i=1,\ldots,4$, related to four of the $\hat\v^i$'s by a
relabelling. This should not cause any confusion since in the
following we shall only work with (\ref{vi4}) and we shall not need
the SO(6) fields (\ref{vphii-vphiAB}).

Unlike the scalar fields the fermions transform non trivially with
respect to both SO(4)'s. The four Weyl fermions of the $\scrN$=4 SYM
theory, $\lambda^A_\a$, transform in the $\mb4$ of SU(4)$\sim$SO(6)
and their conjugates, $\bar\lambda^\adot_A$, transform in the
$\mbb4$. Their decomposition with respect to
SO(4)$\times$SO(4)$\times$U(1) is dictated by the SO(4)$\times$SO(4)
decomposition of the SO(8) fermions of the light-cone string. The type
IIB fermions transform in the $\mb8_{\rm s}$ of SO(8), which under
SO(4)$\times$SO(4) decomposes as
\be
\mb8_{\rm s} =
\left[\left(0,\half\right);\left(0,\half\right)\right] \oplus
\left[\left(\half,0\right);\left(\half,0\right)\right] \, .
\label{so8sdecomp}
\ee
In terms of the $\mb8_{\rm s}$ fermions $S^a$ and $\wtil S^a$ the
decomposition is achieved via a projector \cite{met},
$\half(1\pm\Pi)$, 
\ba
&& S^a \to (S^-)^a_\a \oplus (S^+)^\adot_{\dot a} \nn \\
&& \wtil S^a \to (\wtil S^-)^a_\a \oplus (\wtil S^+)^\adot_{\dot a} \, .
\label{gsfermdecomp}
\ea
The Yang--Mills fermions, $\lambda^A_\a$, have $\D=\frac{3}{2}$ and
U(1) charge $\half$ for $A=1,4$ and $-\half$ for $A=2,3$. Similarly
their conjugates, $\bar\lambda_A^\adot$, have $\D=\frac{3}{2}$ and
U(1) charge $\half$ for $A=2,3$ and $-\half$ for $A=1,4$. To match
the quantum numbers of the string oscillators we choose the following
decomposition
\be
\lambda^A_\a \to \psi^{-\,a}_{+\half;\a} \oplus
\bar\psi^+_{-\half;\a a} \, , \qquad a=1,4 \, ,
\label{fermidec}
\ee
where the fermions $\bar\psi^+_{\a a}$ are defined as
\be
\bar\psi^+_{-\half;\a a} = \left(M^+ \lambda\right)_{-\half;\a a} \, ,
\label{fermiproj}
\ee
where the matrix $M^+$ is related to the matrix $\Pi$ used in the
plane-wave string theory to project the SO(8) fermions onto chiral
SO(4)$\times$SO(4) spinors. The spinors $\psi^{-\,a}_{+\half;\a}$
and $\bar\psi^+_{-\half;\a a}$ transform under SO(4)$\times$SO(4) in
the
$(\mb2_L;\mb2_L)=\left[\left(\half,0\right);\left(\half,0\right)\right]$
and have $(J=\half,\D=\frac{3}{2})$ and $(J=-\half,\D=\frac{3}{2})$
respectively.

Similarly
\be
\bar\lambda_A^\adot \to \psi_{+\half;\dot a}^{+\,\adot} \oplus
\bar\psi^{-\,\adot\dot a}_{-\half} \, ,  \qquad \dot a=2,3
\label{bfermidec}
\ee
where
\be
\bar\psi^{-\,\adot\dot a}_{-\half} = 
\left(M^- \bar\lambda\right)^{\adot\dot a}_{-\half} 
\label{bfermiproj}
\ee
and $M^-$ is also related to the projector used to define the
SO(4)$\times$SO(4) spinors in the dual plane-wave string theory.  The
fermions $\psi_{+\half;\dot a}^{+\,\adot}$ and
$\bar\psi^{-\,\adot\dot a}_{-\half}$ transform in the
$(\mb2_R;\mb2_R)=\left[\left(0,\half\right);\left(0,\half\right)\right]$
representation and have respectively $(J=\half,\D=\frac{3}{2})$
and $(J=-\half,\D=\frac{3}{2})$. Some aspects of instanton
contributions to BMN operators involving fermionic impurities will be
discussed in \cite{gks2}.

The field strength, $F_{\mu\nu}$, is a singlet with respect to the
first SO(4) and decomposes into $F^\pm_{\mu\nu}$ transforming in the
$\mb3^-=(1,0)$ and $\mb3^+=(0,1)$ with respect to the
second. $F^\pm_{\mu\nu}$ both have $J=0$ and $\D=2$.

In the string amplitudes in the plane-wave background $P_+$ and $P_-$
are conserved, so the operators in the gauge theory are conveniently
classified according to the dual quantities, \ie $\D-J$ and $\D+J$
respectively ($\D+J$ is actually infinite in the limit; it is
proportional to $P_-$, but the proportionality constant
diverges). Table \ref{quant-numb} summarises the $\D$, $J$ and
SO(4)$\times$SO(4) quantum numbers for the $\scrN$=4 elementary
fields. The notation SO(4)$_R$  and SO(4)$_C$ has been introduced to
denote the SO(4) groups descending from the original SO(6)
(R-symmetry) and SO(2,4) (conformal) groups respectively.

\TABLE[!htb]{
\begin{tabular}{|l|c|c|c|c|c|c|}
\hline
~~{\rm Field}~~~ & ~~~$\D$~~~ & ~~~$J$~~~
& ~~~ $\D-J$\rule[-8pt]{0pt}{23pt}~~~ & ~~~$\D+J$~~~
& ~~~SO(4)$_R$~~~ & ~~~SO(4)$_C$~~~ \\
\hline \hline
~~$Z$\rule[-8pt]{0pt}{23pt} & 1 & 1 & 0 & 2 & $(0,0)$ & $(0,0)$  \\
\hline
~~$\bar Z$ \rule[-8pt]{0pt}{23pt} & 1 & $-1$ & 2 & 0 & $(0,0)$
& $(0,0)$  \\
\hline
~~$\v^i$ \rule[-8pt]{0pt}{23pt} & 1 & 0 & 1 & 1 &
$\left(\half,\half\right)$ & $(0,0)$  \\
\hline
~~$D_\mu Z$ \rule[-8pt]{0pt}{23pt} & 2 & 1 & 1 & 3 & $(0,0)$ &
$\left(\half,\half\right)$ \\
\hline
~~$\psi^{-\,a}_\a$ \rule[-8pt]{0pt}{23pt} & $\frac{3}{2}$ & $\half$ & 1
& 2 & $\left(\half,0\right)$ & $\left(\half,0\right)$ \\
\hline
~~$\bar\psi^+_{\a a}$ \rule[-8pt]{0pt}{23pt} & $\frac{3}{2}$ &
$-\half$ & 2 & 1 & $\left(\half,0\right)$ & $\left(\half,0\right)$ \\
\hline
~~$\psi_{\dot a}^{+\,\adot}$ \rule[-8pt]{0pt}{23pt} & $\frac{3}{2}$ &
$\half$ & 1 & 2 & $\left(0,\half\right)$ & $\left(0,\half\right)$ \\
\hline
~~$\bar\psi^{-\,\adot\dot a}$ \rule[-8pt]{0pt}{23pt} &
$\frac{3}{2}$ & $-\half$ & 2 & 1 & $\left(0,\half\right)$ &
$\left(0,\half\right)$ \\
\hline
~~$F^-_{\mu\nu}$ \rule[-8pt]{0pt}{23pt} & 2 & 0 & 2 & 2 &
$(0,0)$ & $(1,0)$ \\
\hline
~~$F^+_{\mu\nu}$ \rule[-8pt]{0pt}{23pt} & 2 & 0 & 2 & 2 &
$(0,0)$ & $(0,1)$ \\
\hline
\end{tabular}
\caption{SO(4)$\times$SO(4)$\times$U(1)$\times$U(1) quantum numbers of the
$\scrN$=4 elementary fields.}
\label{quant-numb}
}

\subsection{BMN operators}
\label{bmnops}

The composite operators dual to states in the spectrum of string
theory in the plane wave background are also classified in terms of
the same quantum numbers. In particular, $\D-J$, which is dual to the
light-cone hamiltonian, measures the number of ``impurities'' and will
be used to classify the operators

At finite $J$ and $\D$ the selection rules of the $\scrN$=4 theory,
implied by the superconformal symmetry, apply. So only two-point
functions of (primary) operators of the same dimension can be
non-zero. More precisely the two-point functions that are relevant for
the calculation of anomalous dimensions are
\be
\la \bar\scrO^i_{{\rm \mathbf{\bar r}}_i,\D_i} (x)
\scrO^j_{{\rm \mathbf{r}}_j,\D_j} (y) \ra \, ,
\label{gen2pt}
\ee
where the subscripts denote the SU(4) representation and the scaling
dimension. The SU(2,2$|$4) symmetry imposes $\D_i=\D_j$ and  ${\rm
\mathbf{r}}_i={\rm \mathbf{r}}_j$ so that both $\D$ and $J$ are
conserved in two-point functions. In the case of the U(1) charge $J$
this means that the two operators in a non-zero two-point function
must have equal and opposite charges.

Gauge invariant composite operators which correspond to physical
string states are of the form
\ba
\scrO^{i_1\ldots i_k}_{J;n_1\ldots n_k} &=&
\fr{\sqrt{J^{\D-J+1}\left(\frac{\gy^2N}{8\pi^2}\right)^{J+k}}}
\hspace*{-0.1cm} \begin{array}[t]{c}
{\displaystyle \sum_{p_1,\ldots,p_k=0}^J} \\
{\scriptstyle p_1\le p_2\le\cdots\le p_k}
\end{array} \hspace*{-0.1cm}
\er^{2\pi i(p_1n_1+p_2n_2+\cdots+p_kn_k)/J} \nn \\
&& \hsp{3} \times \Tr\left(Z^{p_1}X_{\ell_1}Z^{p_2-p_1}X_{\ell_2}
\cdots Z^{p_k-p_{k-1}}X_{\ell_k}Z^{J-p_k}\right) \nn \\
&=& \fr{\sqrt{J^{\D-J-1}\left(\frac{\gy^2N}{8\pi^2}\right)^{J+k}}}
\hspace*{-0.1cm} \begin{array}[t]{c}
{\displaystyle \sum_{q_2,\ldots,q_k=0}^J} \\
{\scriptstyle q_2+\cdots+q_k \le J}
\end{array} \hspace*{-0.1cm} \er^{2\pi i[(n_2+\cdots+n_k)q_2+
(n_3+\cdots+n_k)q_3+\cdots+n_kq_k]/J} \nn \\
&& \hsp{3} \times \Tr\left(Z^{J-(q_2+\cdots+q_k)}X_{\ell_1}Z^{q_2}
X_{\ell_2}\cdots Z^{q_k}X_{\ell_k}\right) \, ,
\label{genBMNop}
\ea
where $\D-J$ denotes the total number of impurities.  Here the
cyclicity of the trace has been used and, after the change of
variables, $p_1\to q_1$, $p_i\to q_i-q_{i-1}$ ($i=2,\ldots,k$), the
sum over $q_1$ has been performed resulting in the condition
\be
n_1=-(n_2+\cdots+n_k) \, .
\label{levmatch}
\ee
In (\ref{genBMNop}) the $X_\ell$'s denote generic impurities, \ie any
of the elementary fields discussed in the previous subsection. String
states dual to these operators are created acting on the vacuum with
bosonic and fermionic oscillators. The integers $n_1,\ldots,n_k$ in
(\ref{genBMNop}) are identified with the mode numbers in the dual
string state and the relation (\ref{levmatch}) corresponds to the
level matching condition obeyed by the physical string states.

String creation operators are in one to one correspondence with
$\D-J=1$ impurities, see table \ref{quant-numb}. Bosonic oscillators
$\alpha^i_{-n}$ and $\alpha^\mu_{-n}$ in the $(\mb4;\mb1)$ and
$(\mb1;\mb4)$ of SO(4)$\times$SO(4) correspond to the insertion of
$\v^i$ and $D^\mu Z$ impurities respectively~\footnote{Here we are
using a different notation with respect to \cite{gks}, where the
oscillators $\a^\mu_{-n}$ were denoted by $\a^{i^\pp}_{-n}$.}.
Fermionic oscillators, $S^+_{-n}$ and $S^-_{-n}$, in the
$(\mb2_L;\mb2_L)$ and $(\mb2_R;\mb2_R)$ correspond to the insertion of
$\psi^{-\,a}_{+\half,\a}$ and $\psi^{+\,\adot}_{+\half,\dot a}$ impurities
respectively. In string theory for each type of excitation one must
consider left- and right-moving oscillators. These correspond to the
insertion of the same field, but with the associated $n_i$ in the
phase factor in (\ref{genBMNop}) being respectively positive or
negative.

The normalisation of operators of the form  (\ref{genBMNop}) involving
only $\D-J=1$ impurities is such that their tree level two-point
functions are of order 1 in the BMN limit, $N\to\infty$, $J\to\infty$
with $J^2/N$ finite. Operators involving $\D-J=2$ impurities have
vanishing two-point functions at tree level because for equal total
$\D-J$ they are normalised by the same prefactor but their definition
involves fewer sums. Therefore these operators do not correspond to
independent degrees of freedom in the BMN limit. In some cases,
however, the insertion of $\D-J=2$ impurities is necessary in order to
construct combinations which are well behaved in the double limit
$N\to\infty$, $J\to\infty$ at higher orders in perturbation
theory. For instance it is necessary to consider terms in which pairs
of $\v^i$ impurities are replaced by a $\bar Z$ insertion in order to
cancel divergences which arise at the level of the leading non planar
perturbative corrections \cite{bkpss}.

Operators with $\D-J=0,1$ are protected and so their
two-point functions do not receive instanton contributions. At the
level of two and more impurities the situation is more interesting.
The spectrum is significantly richer and more importantly there appear
unprotected operators. In the following we shall only discuss a small
selection of gauge invariant composite operators with scalar
impurities which are dual to the string states analysed in
\cite{gks}. A complete analysis would require computing the two-point
functions involving all the operators in each sector and then
diagonalising the resulting matrix of anomalous dimensions. We shall
not carry out this program in this paper, but we shall concentrate on
a few specific cases which illustrate the striking agreement with the
corrections to the string mass spectrum calculated in \cite{gks}. The
generic operator with $k$ scalar impurities is of the form
\ba
\scrO^{i_1\ldots i_k}_{J;n_1\ldots n_k}
&\!=\!& \fr{\sqrt{J^{k-1}\left(\frac{\gy^2N}{8\pi^2}\right)^{J+k}}}
\hspace*{-0.1cm} \begin{array}[t]{c}
{\displaystyle \sum_{p_1,\ldots,p_{k-1}=0}^J} \\
{\scriptstyle p_1+\cdots+p_{k-1} \le J}
\end{array} \hspace*{-0.2cm} \er^{2\pi i[(n_1+\cdots+n_{k-1})p_1+
(n_2+\cdots+n_{k-1})p_2+\cdots+n_{k-1}p_{k-1}]/J} \nn \\
&& \hsp{3} \times \Tr\left(Z^{J-(p_1+\cdots+p_{k-1})}\v^{i_1}Z^{p_1}
\v^{i_2}\cdots Z^{p_{k-1}}\v^{i_k}\right) \, ,
\label{k-scal-imp}
\ea
In our discussion we shall only consider operators with an even number
of impurities. This is because operators with odd $\D-J$, which
receive perturbative corrections \cite{gpl}, are expected not to
receive contributions in the one-instanton sector. This is a
prediction following from the calculation of string amplitudes in
\cite{gks}. In the plane-wave string theory the absence of instanton
contributions to two-point amplitudes of states with an odd number of
non-zero mode excitations is a straightforward consequence of the
structure of the $D$-instanton boundary state. In the $\scrN$=4 theory,
however, the corresponding statement is far from obvious.

\subsubsection{Two impurity operators}
\label{2impurity}

No field in the $\scrN$=4 multiplet has negative $\D-J$ hence the two
impurity operators are obtained with the insertion of either two
$\D-J=1$ fields or of a single field with $\D-J=2$. Because of the
normalisation (\ref{genBMNop}) only operators with two $\D-J=1$
insertions are relevant in the BMN limit.

Even restricting the attention to SO(4)$_C$ singlets, already at the
two impurity level there is a rather rich spectrum of operators, which
becomes even richer when multi-trace operators with the same quantum
numbers are taken into account. In the SO(4)$_R$ singlet sector one
can construct gauge invariant operators in the representations
$(0,0)\equiv\mb1$, $(1,0)\equiv\mb3^+$, $(0,1)\equiv\mb3^-$ and
$(1,1)\equiv\mb9$ of SO(4)$_R$. The singlet can be realised with the
insertion of two scalars, two gauge fields (through covariant
derivatives) or two fermions of the same chirality. Operators in the
$\mb3^\pm$ include combinations of two scalar or two fermionic
impurities. The $\mb9$ can only be obtained with the insertion of two
scalar impurities.

The operators with two $\v^i$ insertions are~\footnote{Here and in the
following we use square brackets to denote antisymmetrisation, curly
brackets to denote symmetrisation and subtraction of the trace and
parentheses to indicate symmetrisation without subtraction of the
trace part.}
\ba
\scrO_{\mb1;J;n} &=&
\fr{\sqrt{J\left(\frac{\gy^2N}{8\pi^2}\right)^{J+2}}} 
\left[ \sum_{p=0}^J
\er^{2\pi i pn/J} \, \Tr\left(Z^{J-p}\v^iZ^p\v^i\right) -
\Tr\left(Z^{J+1}\barZ\right) \right] \label{2imp-1} \\
\scrO^{[ij]}_{\mb3^\pm;J;n} &=& 
\fr{\sqrt{J\left(\frac{\gy^2N}{8\pi^2}\right)^{J+2}}}
\sum_{p=0}^J \er^{2\pi i pn/J} \, \Git^{ijkl}_\pm
\, \Tr\left(Z^{J-p}\v^{[k}Z^p\v^{l]}\right)
\label{2imp-3+-} \\
\scrO^{\{ij\}}_{\mb9;J;n} &=& 
\fr{\sqrt{J\left(\frac{\gy^2N}{8\pi^2}\right)^{J+2}}} \,
\sum_{p=0}^J \er^{2\pi ipn/J} \, \Tr\left(Z^{J-p}
\v^{\{i}Z^p\v^{j\}} \right) \label{2imp-9} \\
&\equiv & \fr{\sqrt{J\left(\frac{\gy^2N}{8\pi^2}\right)^{J+2}}} 
\,\sum_{p=0}^J \er^{2\pi ipn/J} \, \left[ \Tr\left(Z^{J-p}\v^{(i}
Z^p\v^{j)} \right) - \frac{\d^{ij}}{2} \, \Tr\left(Z^{J-p}
\v^{k}Z^p\v^{k} \right) \right] \, ,
\nn
\ea
where the projectors onto the $\mb3^+$ and $\mb3^-$ are defined as
$\Git^{ijkl}_\pm = \fr{4}\left(\d^{ik}\d^{jl}-\d^{il}\d^{jk}
\pm\veps^{ijkl}\right)$. The singlet operator (\ref{2imp-1}) provides
an example of what mentioned earlier about $\barZ$ insertions. In
order to define a well behaved BMN operator it is necessary to
consider the combination in (\ref{2imp-1}). The second term is needed
to cancel a divergent contribution to the two-point function of
$\scrO_{\mb1;J;n}$ arising at the leading non planar level
\cite{bkpss}.

The operators (\ref{2imp-1})-(\ref{2imp-9}) are dual to string states
in the plane-wave background of the form
\ba
&& \a^i_{-n}\wtil\a^i_{-n} |0\ra_h \label{str-2imp-1} \\
&& \Git_\pm^{ijkl} \a^k_{-n}\wtil\a^l_{-n} |0\ra_h
\label{str-2imp-3+-} \\
&& \a^{\{i}_{-n}\wtil\a^{j\}}_{-n} |0\ra_h \, ,
\label{str-2imp-9}
\ea
where $|0\ra_h$ denotes the BMN ground state and the indices
$i,j,\ldots$ are taken to be in one of the two SO(4) factors (to be
identified with SO(4)$_R$). The integer $n$ in
(\ref{2imp-1})-(\ref{2imp-9}) corresponds to the level of the dual
string excitation. The $n=0$ operators are protected and correspond to
supergravity states.

As already remarked, in order to compute the instanton induced
anomalous dimensions of the various operators in each sector it is in
principle necessary to diagonalise the appropriate matrix. In the
following we shall not consider the problem of mixing between single-
and multi-trace operators, since it is a subleading effect in the large
$N$ limit. In general, however, at the instanton level mixing occurs at
leading order among all the single trace operators in each sector
\cite{sk}. In the case of the two impurity operators it has been shown
\cite{bei} that all the operators in different sectors have the same
anomalous dimension as a consequence of superconformal
invariance. Therefore in the following we shall only analyse the
single operator in the $\mb9$ for which there is no mixing to
resolve. Superconformal symmetry guarantees that the results apply to
operators in others sectors as well.

In general the problem of resolving the operator mixing and computing
anomalous dimensions can be drastically simplified using the
constraints imposed by superconformal invariance, in particular, the
fact that all the operators in a multiplet have the same anomalous
dimension as the superconformal primary operator.

\subsubsection{Four impurity operators}
\label{4impurity}

The number of independent BMN operators grows very rapidly with the
number of impurities and at the four impurity level the spectrum is
already extremely rich. Bosonic operators in the SO(4)$_C$ singlet
sector exist in the following representations of SO(4)$_R$
\ba
&& (0,0)=\mb1 \, , \quad (0,1)=\mb3^+ \, , \quad (1,0)=\mb3^- \, ,
\quad (0,2)=\mb5^+ \, , \quad (2,0)=\mb5^- \, , \nn \\
&& (1,1)=\mb9 \, , \quad (1,2)=\mb{15^+} \, ,
\quad (2,1)=\mb{15^-} \, , \quad (2,2)=\mb{25} \, .
\label{4imp-irreps}
\ea
Operators relevant in the BMN limit involve four $\D-J=1$
impurities. The combinations which contribute to SO(4)$_C$ scalars are
listed (up to permutations of the four fields) in table
\ref{4impur}~\footnote{$t_{\mu_1\mu_2\mu_3\mu_4}$ is a projector onto
the singlet, \ie $\d_{\mu_1\mu_2}\d_{\mu_3\mu_4}$,
$\d_{\mu_1\mu_3}\d_{\mu_2\mu_4}$ or $\veps_{\mu_1\mu_2\mu_3\mu_4}$.}.

\TABLE[!h]{
{\small
\begin{tabular}{|l|l|l|}
\hline
$(i)$~~$\v^i\v^j\v^k\v^l$ \rule[-8pt]{0pt}{23pt} & $(ii)$
~~$t_{\mu_1\mu_2\mu_3\mu_4}D^{\mu_1}ZD^{\mu_2}ZD^{\mu_3}ZD^{\mu_4}Z$\,
& $(iii)$~~$\v^i\v^jD_\mu ZD^\mu Z$ \\
\hline
$(iv)$~~$\v^i\v^j\psi^{-\,\a a}\psi^{-\,b}_\a$ \rule[-8pt]{0pt}{23pt} &
$(v$)~~$\v^i\v^j\psi^+_{\adot\dot a}\psi^{+\,\adot}_{\dot b}$ &
$(vi)$~~$D_\mu ZD^\mu Z\psi^{-\,\a a}\psi^{-\,b}_\a$\, \\
\hline
$(vii)$~~$D_\mu ZD^\mu Z\psi^+_{\adot\dot a}\psi^{+\,\adot}_{\dot b}$
\rule[-8pt]{0pt}{23pt}&
$(viii)$~~$\v^iD_\mu Z\psi^{-\,a}_\a\psi^{+\,\adot}_{\dot a}$  &
$(ix)$~~$\psi^{-\,\a a}\psi^{-\,b}_\a\psi^{-\,\b c}\psi^{-\,d}_\b$ \\
\hline
$(x)$~~$\psi^+_{\adot\dot a}\psi^{+\,\adot}_{\dot b}
\psi^+_{\bdot\dot c}\psi^{+\,\bdot}_{\dot d}$
\rule[-8pt]{0pt}{23pt} & $(xi)$~~
$\psi^{-\,\a a}\psi^{-\,b}_\a\psi^+_{\adot\dot a}
\psi^{+\,\adot}_{\dot b}$ & ~ \\
\hline
\end{tabular}}
\caption{$\D-J=4$ combinations of impurities}
\label{4impur}
}

The SO(4)$_C$ singlet sector contains the largest number of operators,
involving all the combinations $(i)$-$(xi)$ in table \ref{4impur}.
Operators in the $\mb3^+$ can contain $(i),(ii),(iv)$-$(vi),(viii)$ and
$(ix)$ and those in the $\mb3^-$  $(i),(ii),(iv),(vii),(viii)$ and
$(x)$. The $\mb9$ involves $(i),(ii),(iv),(v)$ and $(viii)$. Operators
in the $\mb5^+$ and $\mb5^-$ can be obtained from $(i),(iv)$ and
$(ix)$ and from $(i),(v)$ and $(x)$ respectively, those in the
$\mb{15}^+$ and $\mb{15}^-$ from $(i)$ and $(iv)$ and from $(i)$ and
$(v)$ respectively. In the $\mb{25}$ there is only one operator
corresponding to the combination $(i)$ with indices fully symmetrised.

In the following we shall concentrate on a few specific two-point
functions corresponding to the amplitudes computed in \cite{gks}. This
will be sufficient to show how instanton contributions to gauge theory
correlation functions precisely reproduce various features observed in
string theory amplitudes. The operators we study in detail are those
containing four scalar impurities. These are of the form
\ba
\scrO_{\mb{r};J;n_1,n_2,n_3} &=&\frac{t^{\mb{r}}_{ijkl}}
{\sqrt{J^3\left(\frac{\gy^2N}{8\pi^2}\right)^{J+4}}}
\begin{array}[t]{c}
{\displaystyle \sum_{p_1,p_2,p_3=0}^J} \\
{\scriptstyle p_1+p_2+p_3 \le J}
\end{array} \hsp{-0.1} \er^{2\pi i[(n_1+n_2+n_3)p_1
+(n_2+n_3)p_2+n_3p_3]/J} \nn \\
&& \hsp{3} \times \Tr\left(Z^{J-(p_1+p_2+p_3)}\v^i
Z^{p_1}\v^jZ^{p_2}\v^kZ^{p_3}\v^l\right) \, ,
\label{4imp-4scal}
\ea
where $t^{\bf r}_{ijkl}$ is a projector onto the representation
${\bf r}$ of SO(4)$_R$. In particular in the singlet sector there are
three independent operators in this class, corresponding to the three
tensors
\be
t^{(1)}_{ijkl} = \veps_{ijkl} \, , \quad t^{(2)}_{ijkl} = \d_{ij}\d_{kl}
\, , \quad t^{(3)}_{ijkl} = \d_{ik}\d_{jl} \, .
\label{singlet-projs}
\ee
In section \ref{2pt-4impur} we study instanton contributions to
two-point functions of operators of the type (\ref{4imp-4scal}). We
discuss in detail the case of the singlet corresponding the choice of
the $t^{(1)}_{ijkl}$ projector. We will show that the dependence on
both the coupling constants, $\lambda^\pp$ and $g_2$, and the mode
numbers, $n_i$, is in exact agreement with the results of
\cite{gks}. We also find that for all the operators with four scalar
impurities in sectors other than the singlet instanton contributions
to the matrix of anomalous dimensions are suppressed by powers of
$\lambda^\pp$. This result is also in agreement with the string
prediction.

\section{Instanton contributions to two-point functions}
\label{inst2ptf}

In this section we recall some general aspects of the calculation of
instanton contributions to correlation functions, in particular
two-point functions, in $\scrN$=4 SYM.

In the semi-classical limit correlation functions are evaluated using
a saddle point approximation around the classical instanton
configuration. In this limit the computation of expectation values
reduces to an integration over the finite dimensional instanton moduli
space as parametrised in the ADHM construction \cite{adhm}. Before
focusing on operators of large dimension, $\D$, and R-charge, $J$, in
the following sections, we briefly discuss the general formalism for
extracting instanton contributions to the anomalous dimensions of
gauge invariant local operators. Comprehensive reviews of instanton
calculus in supersymmetric gauge theories can be found in
\cite{akmrv,dhkm} and instanton contributions to anomalous dimensions
of scalar operators in $\scrN$=4 SYM were studied in detail in
\cite{sk}.

Contributions to the (matrix of) anomalous dimensions are extracted
from the logarithmically divergent terms in two-point functions. In
the semi-classical approximation in the background of an instanton the
two-point function of a generic local operator, $\scrO(x)$, and its
conjugate takes the form
\be
\la\bar\scrO(x_1)\scrO(x_2)\ra_{\rm inst} = \int \dr \mu_{\rm inst}
(\scrm_{\rm \,b},\scrm_{\rm \,f}) \, \er^{-S_{\rm inst}} \;\,
\hat{\bar{\!\!\scrO}}(x_1;\scrm_{\rm \,b},\scrm_{\rm \,f})
\hat\scrO(x_2;\scrm_{\rm \,b},\scrm_{\rm \,f})\, ,
\label{semiclass}
\ee
where we have denoted the bosonic and fermionic collective coordinates
by $\scrm_{\rm \,b}$ and $\scrm_{\rm \,f}$ respectively. In
(\ref{semiclass}) $\dr\mu_{\rm inst}(\scrm_{\rm \,b}, \scrm_{\rm\,f})$
is the integration measure on the instanton moduli space, $S_{\rm
inst}$ is the classical action evaluated on the instanton solution and
$\hat\scrO$ and $\hat{\bar{\!\!\scrO}}$ denote the classical
expressions for the operators $\scrO$ and $\bar\scrO$ computed in the
instanton background.

A one-instanton configuration in SU($N$) Yang--Mills theory is
characterised by 4$N$ bosonic collective coordinates. With a
particular choice of parametrisation these bosonic moduli can be
identified with the size, $\rho$, and position, $x_0$, of the
instanton as well as its global gauge orientation. The latter can be
described by three angles identifying the iso-orientation of a SU(2)
instanton and 4$N$ additional constrained variables, $w_{u\adot}$ and
$\bar w^{\adot u}$ (where $u=1,\ldots,N$ is a colour index), in the
coset SU($N$)/(SU($N-2$)$\times$U(1)) describing the embedding of the
SU(2) configuration into SU($N$). In the one-instanton sector in the
$\scrN$=4 theory there are additionally 8$N$ fermionic collective
coordinates corresponding to zero modes of the Dirac operator in the
background of an instanton. They comprise the 16 moduli associated
with Poincar\'e and special supersymmetries broken by the instanton
and denoted respectively by $\eta^A_\a$ and $\bar\xi^{\adot A}$ (where
$A$ is an index in the fundamental of the SU(4) R-symmetry group) and
8$N$ additional parameters, $\nu^A_u$ and $\bar\nu^{Au}$, which can be
considered as the fermionic superpartners of the gauge orientation
parameters.  The sixteen superconformal moduli are exact, \ie they
enter the expectation values (\ref{semiclass}) only through the
classical profiles of the operators. The other fermion modes,
$\nu^A_u$ and $\bar\nu^{Au}$, appear explicitly in the integration
measure via the classical action, $S_{\rm inst}$. This distinction
plays a crucial r\^ole in the calculation of correlation functions.
The $\nu^A_u$ and $\bar\nu^{Au}$ modes satisfy the fermionic ADHM
constraints
\be \bar w^{\adot u}\nu^A_u = 0 \, , \quad
\bar\nu^{Au}w_{u\adot} = 0 \, ,
\label{nuconstraint}
\ee
which effectively reduce their number to $8(N-2)$.

In the one-instanton sector the gauge-invariant measure on the
instanton moduli space takes the form
\ba
&& \int \dr\mu_{\rm phys} \, \er^{-S_{\rm inst}}
\label{physmeasure} \\
&& = \frac{\pi^{-4N}\gy^{4N}\er^{2\pi i\tau}}{(N-1)!(N-2)!}
\int \dr\rho\,\dr^4x_0 \,\prod_{A=1}^4
\dr^2\eta^A\dr^2\bar\xi^A \, \dr^{N-2}\nu^A \dr^{N-2}\bar\nu^A
\,\rho^{4N-13} \er^{-S_{4F}} \, , \nn
\ea
where the instanton action is
\be
S_{\rm inst} = -2\pi i \tau + S_{4F}
= -2\pi i \tau + \frac{\pi^2}{2\gy^2\rho^2}
\veps_{ABCD} \scrF^{AB}\scrF^{CD}
\label{instaction1}
\ee
with
\be
\tau=\frac{4\pi i}{\gy^2}+\frac{\theta}{2\pi} \, , \quad
\scrF^{AB}=\frac{1}{2\sqrt{2}}(\bar\nu^{Au}\nu_u^B-\bar\nu^{Bu}\nu_u^A)
\,.
\label{instaction2}
\ee
In (\ref{physmeasure}) we have omitted an overall ($N$-independent)
numerical constant that will be reinstated in the final expression.

The two-point function (\ref{semiclass}) thus becomes
\ba
&& \hsp{-0.5} \la \bar\scrO(x_1)\scrO(x_2) \ra =
\frac{\pi^{-4N}\gy^{4N}\er^{2\pi i\tau}}{(N-1)!(N-2)!}
\int \dr\rho\,\dr^4x_0 \,\prod_{A=1}^4
\dr^2\eta^A\dr^2\bar\xi^A \, \dr^{N-2}\nu^A \dr^{N-2}\bar\nu^A
\,\rho^{4N-13} \nn \\
&& \hsp{1}\er^{\frac{\pi^2}{16\gy^2\rho^2}\veps_{ABCD}
(\bar\nu^{[A}\nu^{B]})(\bar\nu^{[C}\nu^{D]})}
\,\,\,\hat{\bar{\!\!\scrO}}\!\left(x_1;x_0,\rho,\eta,\bar\xi,\nu,
\bar\nu\right) \hat{\scrO}\!\left(x_2;x_0,\rho,\eta,\bar\xi,\nu,
\bar\nu\right) \, . \label{smclass2pt1}
\ea
Following \cite{dhkmv} the integration over the non-exact fermion
modes can be reduced to a gaussian form introducing auxiliary bosonic
coordinates, $\chi^i$, $i=1,\ldots,6$, to rewrite the gauge invariant
measure as
\ba
&& \hsp{-1} \frac{\pi^{-4N}\gy^{4N}\er^{2\pi i\tau}}{(N-1)!(N-2)!}
\int \dr\rho\,\dr^4x_0 \,\dr^6\chi \,\prod_{A=1}^4
\dr^2\eta^A\dr^2\bar\xi^A \, \dr^{N-2}\nu^A \dr^{N-2}\bar\nu^A \nn \\
&& \hsp{2.5} \rho^{4N-7} \exp\left[-2\rho^2\chi^i\chi^i
+\frac{4\pi i}{\gy}\chi_{AB}\scrF^{AB}\right] \, ,
\label{physmeaschi}
\ea
where $\chi_{AB}=\frac{1}{\sqrt{8}}\S^i_{AB}\chi^i$ and the symbols
$\S^i_{AB}$ are defined in (\ref{defsig6}).

The fermion modes $\nu^A_u$ and $\bar\nu^{B u}$ enter explicitly in
the classical profiles of the operators in the instanton background as
well as in the measure through the instanton action. It is thus
convenient to construct a generating function as in \cite{gk}, which
allows to deal easily with the otherwise complicated combinatorics
associated with the  integration over $\nu^A_u$ and $\bar\nu^{Au}$. We
introduce sources, $\bar\vt^u_A$ and $\vt_{Au}$, coupled to $\nu^A_u$
and $\bar\nu^{Au}$ and define
\ba
Z[\vt,\bar\vt] &\!\!=\!\!&\frac{\pi^{-4N}\gy^{4N}\er^{2\pi i\tau}}
{(N-1)!(N-2)!} \int \dr\rho \, \dr^4x_0 \, \dr^6\chi
\prod_{A=1}^4 \dr^2\eta^A \, \dr^2 {\bar\xi}^A\,
\dr^{N-2}{\bar\nu}^A\, \dr^{N-2}\nu^A \nn \\
&& \rho^{4N-7} \exp\left[-2\rho^2 \chi^i\chi^i+
\frac{\sqrt{8}\pi i}{\gy}{\bar\nu}^{Au}\chi_{AB}\nu^B_u +
{\bar\vt}^u_{A}\nu^A_u + \vt_{Au}{\bar\nu}^{Au} \right] \, .
\label{genfunct}
\ea
Performing the gaussian integrals over ${\bar\nu}$ and $\nu$ and
introducing polar coordinates,
\be
\chi^i \, \rightarrow \, (r,\Omega) \: , \quad
\sum_{i=1}^6 (\chi^i)^2 = r^2 \, ,
\label{polaromega}
\ee
we find
\ba
Z[\vt,\bar\vt] &\!\!=\!\!&
\frac{2^{-29}\pi^{-13}\,\gy^{8}\er^{2\pi i\tau}}{(N-1)!(N-2)!}
\int \dr\rho \, \dr^4x_0 \, \dr^5\Omega \prod_{A=1}^4 \dr^2\eta^A \,
\dr^2 {\bar\xi}^A \, \rho^{4N-7} \nn \\
&& \int_0^\infty \dr r \, r^{4N-3} \er^{-2\rho^2 r^2}
\scrZ(\vt,\bar\vt;\Omega,r) \, ,
\label{genfunctfin}
\ea
where all the numerical coefficients have been reinstated. In
(\ref{genfunctfin}) we have introduced the density
\be
\scrZ(\vt,\bar\vt;\Omega,r) = \exp\left[-\frac{i\gy}{\pi r}\,
{\bar\vt}_A^u \Omega^{AB} \vt_{Bu} \right] \, ,
\label{density}
\ee
where the symplectic form $\Omega^{AB}$ is given by
\be
\Omega^{AB}= \bar\Sigma^{AB}_i\Omega^i \, ,
\quad \sum_{i=1}^6 \left(\Omega^i\right)^2 = 1 \, .
\label{defOmega}
\ee
Gauge invariant operators depend on the $\nu^A_u$ and $\bar\nu^{Au}$
variables via colour singlet bilinears. These arise in symmetric or
anti-symmetric combinations transforming respectively in the $\mb{10}$
and $\mb6$ dimensional representations of the SU(4) R-symmetry
\ba
(\bar\nu^A \nu^B)_{\mb{10}} &\equiv&
\nut = ({\bar\nu}^{Au}\nu^B_u + {\bar\nu}^{Bu}\nu^A_u) \, ,
\label{10bilinear} \\
(\bar\nu^A \nu^B)_{\mb6} &\equiv&
\nus = ({\bar\nu}^{Au}\nu^B_u - {\bar\nu}^{Bu}\nu^A_u) \, .
\label{6bilinear}
\ea
Using the generating function defined in (\ref{genfunctfin}) the
$\bar\nu^A\nu^B$ bilinears in the operators $\scrO$ and $\bar\scrO$ in
(\ref{smclass2pt1}) can be rewritten in terms of derivatives of
$\scrZ(\vt,\bar\vt;\Omega,r)$ with respect to the sources, $\vt_A$ and
$\bar\vt_B$. The result for a two-point function in which the operator
insertions contain a total of $p$ $\nsix$ and $q$ $\nten$ bilinears is
of the form
\ba
&& \hsp{-0.5} \la\bar\scrO(x_1)\scrO(x_2)\ra =
\frac{\gy^{8}\er^{2\pi i \tau}}{(N-1)!(N-2)!} \int \dr\rho \,
\dr^{4}x_{0} \, \dr^{5}\Omega \prod_{A=1}^4 \dr^2\eta^A \,
\dr^2 {\bar\xi}^A \, \rho^{4N-7} \label{genernm} \\
&& \hsp{-0.5} \int \dr r \, r^{4N-3}\er^{-2\rho^{2}r^{2}}
\left.\frac{\d^{2p+2q}\scrZ[\vt,\bvt;\Omega,r]}{\d \vt_{u_{1}[A_{1}}
\d\bvt^{u_{1}}_{B_{1}]} \d \vt_{v_{1}(C_{1}} \d\bvt^{v_{1}}_{D_{1})}
\ldots}\right|_{\vt=\bvt=0}
\wtil{\bar{\!\!\scrO}}\!\left(x_{1};x_{0},\rho,\eta,\xib\right)
\wtil{\scrO}\!\left(x_{2};x_{0},\rho,\eta,\xib\right) \, , \nn
\ea
where $\wtil\scrO$ and $\,\,\wtil{\bar{\!\!\scrO}}$  contain the
dependence on the exact moduli, $\eta^A$ and $\xib^A$, and on the
bosonic collective coordinates after extracting the $\bar\nu^A\nu^B$
bilinears. Computing the $r$ integral gives
\ba
\la\bar\scrO(x_1)\scrO(x_2)\ra &\!\!\sim\!\!&
\a(p,q;N)\,\gy^{8+p+q}\,\er^{2\pi i\tau}
\int \dr\rho \, \dr^4x_0 \, \dr^5\Omega \prod_{A=1}^4 \dr^2\eta^A \,
\dr^2 {\bar\xi}^A \, \rho^{p+q-5} \nn \\
&& f(\Omega) \;\;\wtil{\bar{\!\!\scrO}}\!\left(x_1;\rho,x_0;
\eta,\bar\xi\right) \wtil\scrO\!\left(x_2;\rho,x_0;
\eta,\bar\xi\right) \, ,
\label{largeN-dep}
\ea
where $f(\Omega)$ contains the dependence on the $\Omega^{AB}$
variables obtained from the derivatives of
$\scrZ(\vt,\bar\vt;\Omega,r)$. The coefficient $\a(p,q;N)$ contains
the $N$ dependence and in the large $N$ limit we find
\ba
\a(p,q;N) &=&
\frac{2^{-2N+\half(p+q)}\,\pi^{-(p+q)}\,
\Gamma\left(2N-1-\half(p+q)\right)}
{(N-1)!(N-2)!} \left(N^{p+\frac{q}{2}}+
O(N^{p+\frac{q}{2}-1})\right) \nn \\
&=& \frac{N^{\half(p+1)}}{4\pi^{p+q+\half}}\left(1+O(1/N)\right) \, .
\label{largeN-coeff}
\ea
From (\ref{largeN-dep}) and (\ref{largeN-coeff}) it follows that the
insertion of a $\nten$ or $\nsix$ bilinear in a correlation function
produces a factor of $\gy$ or $\gy\sqrt{N}$ respectively.

In computing the moduli space integrations in expressions for
two-point functions of the type (\ref{largeN-dep}) it will prove
convenient to calculate first the fermionic integrals over $\eta^A$
and $\bar\xi^A$ and the angular integration over the five-sphere.
These give rise to selection rules that determine which operators
receive instanton contributions to their scaling dimensions. In
particular since the superconformal modes are exact a correlation
function can only receive instanton contribution if the operator
expressions contain exactly sixteen such modes in the combination
\be
\prod_{A=1}^4 \left(\eta^{\a A}\eta^A_\a\right)
\left(\bar\xi^A_\adot\bar\xi^{\adot A}\right) \, .
\label{saturmodes}
\ee
The integration over the five-sphere parametrised by the angular
variables $\Omega^{AB}$ factorises and gives rise to further selection
rules. It gives a non-vanishing result only if the SU(4) indices
carried by the $\Omega$'s in the two operators can be combined to form
a SU(4) singlet. The SU(4) indices are originally carried by the
fermion modes which are all in the $\mb4$, so the only possible
singlet combinations correspond to products of $\veps^{ABCD}$
tensors. The generic five-sphere integral is of the form
\be
\int \! \dr^5\Omega \, \Omega^{A_1B_1}\ldots
\Omega^{A_{2n}B_{2n}} = c(n) \left(\veps^{A_1B_1A_2B_2}
\ldots\veps^{A_{2n-1}B_{2n-1}A_{2n}B_{2n}} + {\rm permutations}
\right) \, , \label{spherintn}
\ee
where the normalisation constant $c(n)$ is
\be
c(n) = \frac{\pi^{5/2}\,\Gamma\!\left(n+\half\right)}
{2\Gamma\!\left(n+4\right)} \, .
\label{5spherecoeff}
\ee
Equations (\ref{saturmodes}) and (\ref{spherintn}) imply 
that a two-point function can receive a non-zero contribution only if
the combined profiles of the two operators contain fermion modes of
the four flavours with the same multiplicity.

The bosonic integrations over the position and size of the instanton
are left as a last step. In the case of two-point functions these
integrals are logarithmically divergent, signalling a contribution to
the matrix of anomalous dimensions.

\section{Fermion zero modes}
\label{N4instmult}

In order to evaluate instanton induced correlation functions we need
to integrate the classical profiles of the relevant composite
operators over the instanton moduli space. We are interested in the
dependence on the collective coordinates and of particular relevance
will be the way the fermionic modes enter into the expressions for the
various fields. The zero-mode dependence in the elementary fields of
the $\scrN$=4 SYM multiplet was reviewed in detail in \cite{sk}. Here
we briefly summarise the features which will be relevant for the
analysis of two-point functions of BMN operators.

The field equations of the $\scrN$=4 SYM theory admit a solution in
which the gauge potential corresponds to a standard instanton of
SU($N$) pure Yang--Mills theory and all the other fields vanish,
\be
A_\mu = A_\mu^I \, , \quad \v^{AB} = \lambda^A_\a =
\bar\lambda^\adot_A = 0 \, .
\label{bpstinst}
\ee
However, the Dirac operator has zero modes in the background of this
solution, \ie the equation $\bar{\Dsm}_{\adot\a}\lambda^{\a A}=0$ has
non-trivial solutions when the covariant derivative is evaluated in
the background of an instanton. The general solution to the Dirac
equation is linear in the instanton fermion zero modes. This
non-trivial solution gives rise to a non-zero solution for the scalar
fields when plugged into the corresponding equation,
$D^2\v^{AB}=\sqrt{2} [\lambda^A,\lambda^B]$. The latter admits a
solution for the scalar which is bilinear in the fermion
modes. Proceeding with this iterative solution of the field equations
one generates a complete supermultiplet and further iterations give rise
to additional terms with more fermion modes in each field. The general
solution obtained through this procedure is schematically of the form
\ba
&& A_\mu = \hspace*{-0.3cm} \begin{array}[t]{c}
{\displaystyle \sum_{n=0}} \\
{\scriptstyle 4n \le 8N}
\end{array} \hspace*{-0.3cm}
A_\mu^{(4n)} \, , \hsp{1.4}
\v^{AB} = \hspace*{-0.3cm} \begin{array}[t]{c}
{\displaystyle \sum_{n=0}} \\
{\scriptstyle 4n+2 \le 8N}
\end{array} \hspace*{-0.3cm}
\v^{(2+4n)AB} \nn \\
&& \lambda^A_\a = \hspace*{-0.3cm} \begin{array}[t]{c}
{\displaystyle \sum_{n=0}} \\
{\scriptstyle 4n+1 \le 8N}
\end{array} \hspace*{-0.3cm}
\lambda^{(1+4n)A}_\a \, , \hsp{0.5}
\bar\lambda_{\adot A} = \hspace*{-0.3cm} \begin{array}[t]{c}
{\displaystyle \sum_{n=0}} \\
{\scriptstyle 4n+3 \le 8N}
\end{array} \hspace*{-0.3cm}
\bar\lambda^{(3+4n)}_{\adot A} \, ,
\label{N4multizeromod}
\ea
where the notation $\Phi^{(n)}$ is used to denote a term in the
solution for the field $\Phi$ containing $n$ fermion zero modes.  It
is also understood that in (\ref{N4multizeromod}) the number of
superconformal modes in each field does not exceed 16 and the
remaining modes are of $\nu^A_u$ and $\bar\nu^{Au}$ type.

In computing the expressions for gauge invariant composite operators
we shall make use of the ADHM description in which the elementary
fields are written as $[N+2]\times[N+2]$ matrices. In particular, the
leading order term in the solution for the scalars $\v^{AB}$ is given
explicitly in appendix \ref{useformulae}. The solution of the iterative
equations becomes very involved after a few steps. However the flavour
structure of the combination of fermion zero modes in each term  can
be determined without actually solving the equations and is sufficient
to identify  which operators can get an instanton correction to their
scaling dimension.

All the fermion zero modes, both the superconformal ones, $\eta^A_\a$
and $\bar\xi^{\adot A}$, and the modes of type $\nu^A_u$ and
$\bar\nu^{Au}$, transform in the $\mb4$ of SU(4). We shall denote a
generic fermion mode by $\scrmf^A$.  The starting point for the
construction of the instanton supermultiplet is the classical
instanton, $A_\mu^{(0)}$, which has no fermions. The first term in
$\lambda^A_\a$ is linear in the fermion modes
\be
\lambda^{(1)A}_\a \sim \scrmf^A \, .
\label{lambda1su4}
\ee
For the term $\v^{(2)AB}$ in the scalar solution one finds
\be
\v^{(2)AB} \sim \scrmf^{[A}\scrmf^{B]} \, ,
\label{phi2su4}
\ee
\ie the two fermion modes are antisymmetrised in order to form a
combination in the $\mb6$.  The $\mbb4$ spinor
$\bar\lambda^{(3)\adot}_A$ contains three fermion modes in the
combination
\be
\bar\lambda^{(3)\adot}_A \sim \veps_{ABCD} \,\scrmf^B\scrmf^C\scrmf^D
\,, \label{blambda3su4}
\ee
so that the component $\lambda^{(3)}_A$ has one mode of each flavour
apart from $A$. Proceeding in the multiplet we find the quartic term
in the solution for the vector, $A_\mu^{(4)}$, which contains one
fermion mode of each flavour in a singlet combination
\be
A_\mu^{(4)} \sim \veps_{ABCD}\,\scrmf^A\scrmf^B\scrmf^C\scrmf^D \, .
\label{a4su4}
\ee
The following term is $\lambda^{(5)A}_\a$, which has flavour structure
\be
\lambda^{(5)A}_\a \sim \veps_{BCDE}
\, \scrmf^A \scrmf^B\scrmf^C\scrmf^D\scrmf^E \, ,
\label{lambda5su4}
\ee
\ie it involves a mode of flavour $A$ plus one of each
flavour. Then we find $\v^{(6)AB}$ that contains an antisymmetric
combination of a mode of flavour $A$ and one of flavour $B$ plus one
mode of each flavour antisymmetrised in a singlet,
\be
\v^{(6)AB} \sim \veps_{CDEF}
\, \scrmf^{[A}\scrmf^{B]} \scrmf^C\scrmf^D\scrmf^E\scrmf^F \, .
\label{phi6su4}
\ee
The previous expressions are symbolic and the products of $\scrmf$'s
in (\ref{phi2su4})-(\ref{phi6su4}) correspond to different
combinations of the modes $\eta^A_\a$, $\bar\xi^{\adot A}$, $\nu^A_u$
and $\bar\nu^{Au}$ in the various entries of the ADHM matrix for
each field. The structure of the terms with more fermion modes in the
multiplet can be determined analogously. The iterative solution of the
field equation to construct the first few terms in the multiplet was
carried out explicitly in \cite{bvv}.

From the above equations we can deduce the form of the component
fields in the decomposition relevant for the BMN limit. For the
scalars in (\ref{vi4}) we have
\ba
&& Z^{(2)} \sim \scrmf^{[1}\scrmf^{4]} \, , \qquad
\bar Z^{(2)} \sim \scrmf^{[2}\scrmf^{3]} \nn \\
&& \v^{(2)\,1,3} \sim \scrmf^{[1}\scrmf^{3]} +
\scrmf^{[2}\scrmf^{4]} \, , \qquad
\v^{(2)\,2,4} \sim \scrmf^{[1}\scrmf^{2]} +
\scrmf^{[3}\scrmf^{4]}
\, . \label{bmnscal2so4}
\ea
For the fermions in (\ref{fermidec}) and (\ref{bfermidec}) we have
respectively
\be
\psi^{-(1)\,a} \sim \scrmf^a \, ,  \qquad
\bar\psi^{+(1)}_a \sim \left(M^+\scrmf\right)_a
\, , \qquad a=1,4 \, \; \dot a =2,3
\label{bmnferm1so4}
\ee
and
\be
\psi^{+(3)}_{\dot a} \sim \veps_{\dot rBCD}
\,\scrmf^B\scrmf^C\scrmf^D \, ,  \, \qquad
\bar\psi^{-(3)\,\dot a} \sim \left(M^-\veps\right)^{\dot a}{}_{\!BCD}
\,\scrmf^B\scrmf^C\scrmf^D \, , \qquad a=1,4 \,,
\;\dot a =2,3  \, ,
\label{bmnbferm1so4}
\ee
so that
\ba
&& \hsp{-1}\psi^{-(1)\,1}\sim\scrmf^1 \,, \quad \psi^{-(1)\,4}
\sim\scrmf^4 \,,\quad \bar\psi^{+(1)\,1}\sim\scrmf^2 \,,
\quad \bar\psi^{+(1)\,4}\sim\scrmf^3 \,,
\label{bmnfer-so4c} \\
&& \hsp{-1} \psi^{+(3)\,2}\sim\scrmf^1\scrmf^3\scrmf^4 \,,
\quad \psi^{+(3)\,3}\sim\scrmf^1\scrmf^2\scrmf^4 \,,
\quad \bar\psi^{-(3)\,2}\sim\scrmf^2\scrmf^3\scrmf^4 \,,
\quad \bar\psi^{-(3)\,3}\sim\scrmf^1\scrmf^2\scrmf^3 \,.
\label{bmnbfer-so4c}
\ea
The terms of higher order are easily deduced from the previous general
discussion. Notice that we can assign U(1) charge $+\half$ to the
fermion modes $\scrmf^1$ and $\scrmf^4$ and charge $-\half$ to the
modes $\scrmf^2$ and $\scrmf^3$.

The dependence on the superconformal modes, $\eta^A_\a$ and
$\bar\xi^{\adot A}$, can be obtained using supersymmetry without
solving the field equations. These modes are associated with
superconformal symmetries broken in the instanton background. The
corresponding terms in the $\scrN$=4 supermultiplet can thus be
generated acting with the broken Poincar\'e and special
supersymmetries, $Q_{\a A}$ and $\bar S_{\adot A}$, on the classical
instanton solution for the gauge potential. In the case of SU(2) gauge
group there are no additional fermion modes and the complete solution
can be obtained in this way. In general, however, the dependence on
the $\nu^A_u$ and $\bar\nu^{Au}$ modes can be determined only by
solving the equations of motion.

It is useful to discuss the derivation of the dependence on the
superconformal modes using supersymmetry since it allows us to clarify
how different combinations of fermion modes appear in various
operators.

Substituting $A_\mu^{(0)}\equiv A_\mu^I$ in the supersymmetry
transformation of $\lambda^A_\a$ gives $\lambda^{(1)A}_\a$, which is
linear in $\eta^A_\a$ and $\bar\xi^{\adot A}$ and solves the
corresponding field equation.  Replacing $\lambda^A_\a$ by
 $\lambda^{(1)A}$ in the
variation of $\v^{AB}$ generates the solution $\v^{(2)AB}$ for the
scalar. The iteration of this procedure gives rise to
$\bar\lambda^{(3)}_{\adot A}$, then to the correction $A^{(4)}_\mu$ to
the gauge field and so on.

In the examples studied in \cite{bgkr,dhkmv,gk,sk} the superconformal
modes always appear in the expressions of gauge-invariant composite
operators in the combination
\be
\zeta^A_\a(x) = \frac{1}{\sqrt{\rho}}\left[ \rho\,\eta^A_\a -
(x-x_0)_\mu \s^\mu_{\a\adot} \,\bar\xi^{\adot A} \right] \, .
\label{zetadef}
\ee
In general, however, the dependence on the fermion superconformal modes,
even in gauge-invariant operators, is not only through this
combination and instead the moduli $\eta^A_\a$ and $\bar\xi^{\adot A}$
appear explicitly. This can be understood analysing the form of the
Poincar\'e and special supersymmetry variations of the fields. Under a
combination of the broken supersymmetries, $\eta^{\a A} Q_{\a A} +
\bar\xi^A_\adot \bar S^\adot_A$, we have
\ba
&& \hsp{-1.5} \d A_\mu = (\eta^A+\s\!\cdot\! x \,\bar\xi^A) \s_\mu
\bar\lambda_A  \label{QSAtransf} \rule{0pt}{14pt} \\
&& \hsp{-1.5} \d\lambda^A = F_{\mu\nu}\s^{\mu\nu}(\eta^A+\s\!\cdot\!
x\,\bar\xi^A) + [\v^{AB},\bar\v_{BC}](\eta^C+\s\!\cdot\! x\,\bar\xi^C)
\label{QSlamtransf} \rule{0pt}{14pt} \\
&& \hsp{-1.5} \d\v^{AB} = \lambda^A(\eta^B+\s\!\cdot\! x \,\bar\xi^B)
-(A\leftrightarrow B) \label{QSphitranf} \\
&& \hsp{-1.5} \d\bar\lambda_A = \Dsm\bar\v_{AB}(\eta^B+\s\!\cdot\!
x\,\bar\xi^B) + \bar\v_{AB}\bar\xi^B \, , \rule{0pt}{14pt}
\label{QSblamtransf}
\ea
which shows that, whereas the variations of $A_\mu$, $\lambda^A_\a$
and $\v^{AB}$ involve the combination $\zeta^A_\a$, the superconformal
variation of $\bar\lambda_A^\adot$ contains an extra term. Therefore
the profiles of operators involving $\bar\lambda_A^\adot$ in general
depend separately on $\eta^A_\a$ and $\bar\xi^A_\adot$. Since further
application of the broken supersymmetries generates new terms in the
solution for the elementary fields, it follows that not only operators
containing $\bar\lambda_{\adot A}$, but also those in which any
elementary field contain a non-minimal number of fermion modes (\eg
$A_\mu^{(4)}$, $\lambda^{(5)A}_\a$, $\v^{(6)AB}$) will depend on
$\eta^A_\a$ and $\bar\xi^A_\adot$ not only via $\zeta^A_\a$. This
observation will play an important r\^ole in the case of two impurity
BMN operators. As will be shown in the next section, a naive counting
of zero-modes including only terms with the minimal number of fermion
modes in each field would lead to conclude that these operators have
vanishing two-point functions in the instanton background. We will,
however, argue that the inclusion of the term $\v^{(6)AB}$ in the
solution is needed in order to compute the leading non-zero instanton
contributions to these two-point functions.

\section{Two-point functions of BMN operators}
\label{2pt-BMN}

In this section we analyse instanton contributions to two-point
functions of the BMN operators described in section
\ref{bmnops}. Using the results of the previous sections we shall
determine which operators have non-zero two-point functions in the
instanton background and the dependence of the instanton induced
anomalous dimensions on the parameters, $\gy$, $N$ and $J$ as well as
the integers corresponding to the mode numbers in the dual string
states. Zero and one impurity operators are protected, therefore their
two-point functions are not renormalised and receive no instanton
contribution. We shall therefore discuss two and four impurity
operators. Operators with an odd number of impurities are expected not
to receive instanton contributions. The analysis of two impurity
operators in the next subsection will be rather qualitative because
the leading non-zero contribution to their two-point functions
involves the six-fermion term in the scalar solution which is not
known explicitly. The four impurity case which is fully under control
will be discussed in the following subsection.

\subsection{Two impurity operators}
\label{2pt-2impur}

At the two impurity level we focus on the bosonic SO(4)$_C$ singlet
operator (\ref{2imp-9}) in the $\mb9$ of SO(4)$_R$. Since this sector
contains only one operator there is no problem of mixing and the
anomalous dimension of the operator $\scrO^{\{ij\}}_{J,\mb9;n}$ can be
read directly from the coefficient of the two-point function
$\la\scrO^{\{ij\}}_{J,\mb9;n}(x_1)\bar\scrO^{\{kl\}}_{-J,\mb9;m}(x_2)\ra$.

As usual it is convenient to compute this two-point function for a
particular choice of components, rather than work in a manifestly
SO(4)$_R$ covariant way. Therefore we consider~\footnote{The subscripts
indicating the SO(4)$_R$ representation and the U(1) charge will be
omitted except in situations where this may cause confusion.}
\be
G_\mb9(x_1,x_2) = \la\scrO^{\{13\}}_{n}(x_1)
\bar\scrO^{\{13\}}_{m}(x_2)\ra_{\rm inst} \, ,
\label{2pt-2imp-9-a}
\ee
so that there is no trace to subtract.

The component $\scrO^{\{13\}}_n$ is
\be
\scrO^{\{13\}}_n = \frac{i}{\sqrt{J\left(
\frac{\gy^2N}{8\pi^2}\right)^{J+2}}} \sum_{p=0}^J
\er^{2\pi ipn/J} \left[\Tr\left(Z^{J-p}\v^{13}Z^p\v^{13}\right)
-\Tr\left(Z^{J-p}\v^{24}Z^p\v^{24}\right)\right]
\label{2imp-9-13comp}
\ee
and the conjugate operator is
\be
\bar\scrO^{\{13\}}_n = \frac{-i}{\sqrt{J\left(
\frac{\gy^2N}{8\pi^2}\right)^{J+2}}} \sum_{p=0}^J
\er^{-2\pi ipn/J} \left[\Tr\left(\barZ^{J-p}\v^{13}\barZ^p\v^{13}\right)
-\Tr\left(\barZ^{J-p}\v^{24}\barZ^p\v^{24}\right)\right] \, .
\label{2imp-9b-13comp}
\ee
The semi-classical approximation in the one-instanton sector for
(\ref{2pt-2imp-9-a}) gives
\ba
&& \hsp{-0.2} G_\mb9(x_1,x_2) \!=\! \frac{\pi^{-4N}\gy^{4N}
\er^{2\pi i\tau}}{(N-1)!(N-2)!} \int \dr\rho\,\dr^4x_0 \,\prod_{A=1}^4
\dr^2\eta^A\dr^2\bar\xi^A \, \dr^{N-2}\nu^A \dr^{N-2}\bar\nu^A
\,\rho^{4N-13} \label{2pt-2imp-9-b} \\
&& \hsp{1.2}\er^{\frac{\pi^2}{16\gy^2\rho^2}\veps_{ABCD}
(\bar\nu^{[A}\nu^{B]})(\bar\nu^{[C}\nu^{D]})}\,
\hat{\scrO}^{\{13\}}_{J;n}\!\left(x_1;x_0,\rho,\eta,\bar\xi,\nu,
\bar\nu\right) \,\,\hat{\bar{\!\!\scrO}}^{\{13\}}_{-J;m}\!
\left(x_2;x_0,\rho,\eta,\bar\xi,\nu,\bar\nu\right)\, . \nn
\ea
In order to have a non-zero contribution to this two-point function
the classical profiles of the two operators must contain, when
combined, the sixteen fermion modes corresponding to the broken
supersymmetries.

It is easy to verify that substituting for each scalar field in
(\ref{2pt-2imp-9-a}) the leading order solution, $\v^{(2)}$, does not
allow to soak up all the superconformal modes. Using for each scalar
field the bilinear term (\ref{bmnscal2so4}) in the solution we find
that $\scrO^{\{13\}}$ contains the following combinations of fermion
modes
\be
\left(\scrmf^{[1}\scrmf^{4]}\right)^J \left(\scrmf^{[1}\scrmf^{3]}
\scrmf^{[1}\scrmf^{3]}+\scrmf^{[2}\scrmf^{4]}\scrmf^{[2}\scrmf^{4]}
\right) \, .
\label{2-9-fermodes}
\ee
Similarly, the conjugate operator $~\hat{\bar{\!\!\scrO}}^{\{13\}}$
contains
\be
\left(\scrmf^{[2}\scrmf^{3]}\right)^J \left(\scrmf^{[1}\scrmf^{3]}
\scrmf^{[1}\scrmf^{3]}+\scrmf^{[2}\scrmf^{4]}\scrmf^{[2}\scrmf^{4]}
\right) \, .
\label{2-b9-fermodes}
\ee
The argument given at the end of section \ref{N4instmult} shows that
in these traces, involving only $Z^{(2)}$ and $\v^{(2)AB}$, the
superconformal modes always appear in the combination $\zeta^A$ of
(\ref{zetadef}). This can be easily verified using the explicit
expression for the scalar ADHM matrices given in
(\ref{phi2-inst-sol}). Then, because of the condition
$\left(\zeta^A(x)\right)^3=0$ satisfied by the Weyl spinors $\zeta^A$,
in order to soak up the sixteen superconformal modes in
the two-point function (\ref{2pt-2imp-9-b}) each of the operators
should contain two factors of $\zeta^A$ for each flavour. In other
words the sixteen superconformal modes should appear in the two-point
function in the form
\be
\prod_{A=1}^4 \left[\zeta^A(x_1)\right]^2 \left[\zeta^A(x_2)\right]^2 \, .
\label{satscmodes}
\ee
Examining the combinations (\ref{2-9-fermodes}) and
(\ref{2-b9-fermodes}) it is clear that this is not
possible. $\hat\scrO^{\{13\}}$ cannot soak up the required
superconformal modes of flavour 2 and 3, since it does not contain two
factors of both $\zeta^2$ and $\zeta^3$, while
$~\hat{\bar{\!\!\scrO}}^{\{13\}}$ cannot soak up all the
superconformal modes of flavour 1 and 4, since it does not contain two
factors of both $\zeta^1$ and $\zeta^4$. This simple analysis of the
flavour structure of the superconformal modes in the classical
profiles of the operators shows that the two-point function
(\ref{2pt-2imp-9-a}) vanishes at leading order in the instanton
background. This argument does not rely on the way the remaining $J-2$
$(\bar\nu^A\nu^B)$ bilinears are distributed in the two operators.
According to the discussion in section \ref{inst2ptf} the leading
contribution in the large $N$ limit would come from terms in which all
the $(\bar\nu^A\nu^B)$ bilinears are antisymmetrised. However, since
the above argument is based only on the analysis of the superconformal
modes the conclusion that the leading $\gy$ contribution to the
two-point function (\ref{2pt-2imp-9-a}) vanishes is valid at all
orders in $1/N$.

In order to saturate the integrations over the superconformal modes in
(\ref{2pt-2imp-9-b}) we need to use for some of the scalar fields the
solution containing six fermionic modes, $\v^{(6)AB}$. Inspecting the
combinations (\ref{2-9-fermodes}) and (\ref{2-b9-fermodes}) found at
leading order and recalling (\ref{phi6su4}) it is easy to verify that
it is sufficient to consider one $\v^{(6)AB}$ (or $Z^{(6)}$ and
$\barZ^{(6)}$ respectively) insertion in each operator. These are the
leading order contributions, the insertion of more six-fermion scalars
leads to contributions of higher order in $\gy$ since in this case
more $\bar\nu^A\nu^B$ bilinears appear.

Recalling the form of $\v^{(6)AB}$ given in (\ref{phi6su4}) we find
that the combinations of fermionic modes in
$\Tr\left(Z^{J-p}\v^{AB}Z^p\v^{CD}\right)$ and
$\Tr\left(\barZ^{J-p}\v^{AB}\barZ^p\v^{CD}\right)$ are respectively
\be
\veps_{A^\pp B^\pp C^\pp D^\pp}\,\scrmf^{A^\pp}\scrmf^{B^\pp}
\scrmf^{C^\pp}\scrmf^{D^\pp}
\left(\scrmf^{[1}\scrmf^{4]}\right)^J \left(\scrmf^{[A}\scrmf^{B]}
\scrmf^{[C}\scrmf^{D]} \right)
\label{nonmin-O}
\ee
and
\be
\veps_{A^\pp B^\pp C^\pp D^\pp}\,\scrmf^{A^\pp}\scrmf^{B^\pp}
\scrmf^{C^\pp}\scrmf^{D^\pp} \left(\scrmf^{[2}\scrmf^{3]}\right)^J
\left(\scrmf^{[A}\scrmf^{B]} \scrmf^{[C}\scrmf^{D]} \right) \, .
\label{nonmin-Obar}
\ee
Notice that here the superconformal modes do not necessarily enter via
(\ref{zetadef}), since we are using the term with six fermions in the
solution for one of the fields in each operator. More precisely the
structure of the supersymmetry transformations
(\ref{QSAtransf})-(\ref{QSblamtransf}) shows that both traces contain
one single $\bar\xi$ mode which is not part of a $\zeta$. This is
crucial in order to get a non-zero result from the moduli space
integration, because it allows, for two flavours, to distribute three
fermionic superconformal modes in one operator and one in the other.

The resulting non-zero contribution to the two-point function is
\ba
&& G_{\mb9}(x_1,x_2) \sim 
\frac{\gy^4\er^{2\pi i\tau}}{JN^{3/2}}
\int \dr^4x_0\, \dr\rho \,\rho^{2J-5}  \, f(x_1,x_2;x_0,\rho)
\int\dr^5\Omega \, \left(\Omega^{14}\right)^{J-1}
\left(\Omega^{23}\right)^{J-1} \Omega^{13}\Omega^{24} \nn \\
&&\times \int \prod_{A=1}^4 \dr^2\eta^A\,\dr^2\bar\xi^A\,
\left\{\left[\left(\zeta^1\right)^2 \zeta^2 \left(\zeta^3\right)^2
\left(\zeta^4\right)^2\bar\xi^1\right]\!(x_1)
\left[\zeta^1\left(\zeta^2\right)^2 \left(\zeta^3\right)^2
\left(\zeta^4\right)^2\bar\xi^2\right]\!(x_2)\right. \nn \\
&& \left. \hsp{3.2} + \left[\left(\zeta^1\right)^2 \left(\zeta^2\right)^2
\zeta^3 \left(\zeta^4\right)^2\bar\xi^4\right]\!(x_1)
\left[\left(\zeta^1\right)^2\left(\zeta^2\right)^2
\left(\zeta^3\right)^2\zeta^4\bar\xi^3\right]\!(x_2)\right\} \, ,
\label{2pt-2imp-9-b1}
\ea
where the $(\bar\nu^A\nu^B)$ bilinears have been rewritten in terms of
$\Omega^{AB}$'s as described in section \ref{inst2ptf}. The overall
powers of $\gy$ and $N$ come from the normalisation of the operators,
the moduli space integration measure and the $\nsix$ bilinears, see
equations (\ref{largeN-dep}) and (\ref{largeN-coeff}). The fermion
superconformal modes are saturated and the corresponding integration
is non-zero. In (\ref{2pt-2imp-9-b1}) the dependence on the bosonic
moduli has been collected  in the function $f(x_1,x_2;x_0,\rho)$,
which can only be computed knowing the explicit form of the solution
$\v^{(6)AB}$ which we have not determined. The exact form of the
solution is also needed in order to compute the overall coefficient
and, in particular, the dependence on $J$. More details of the
derivation of (\ref{2pt-2imp-9-b1}) as well as of the evaluation of
the moduli space integrals are given in appendix \ref{calculations}.

The final result for the two-point function is of the form
\be
G_\mb9(x_1,x_2) \sim \frac{\gy^4J^3\er^{2\pi i\tau}}{N^{3/2}}
\fr{(x_1-x_2)^{2(J+2)}} \, I \, ,
\label{2pt-9-fin}
\ee
where $I$ is a logarithmically divergent integral, to be regulated \eg
by dimensional regularisation of the $x_0$ integral. The logarithmic
divergence is due to the bosonic integrations over $x_0$ and $\rho$,
as can be verified by dimensional analysis. The presence of this
divergence signals an instanton contribution to the anomalous
dimension of the operator $\scrO^{\{ij\}}_\mb9$.

As already observed there is only one operator in the representation
$\mb9$ of SO(4)$_R$ and thus there is no mixing to resolve and the
present analysis directly determines the instanton correction to the
scaling dimension. We thus find that the instanton induced anomalous
dimension of $\scrO_\mb9^{\{13\}}$ behaves as
\be
\g_\mb9^{\rm inst} \sim \frac{\gy^4 J^3}{N^{3/2}}
\,\er^{-\frac{8\pi^2}{\gy^2}+i\theta} \sim \left(g_2\right)^{7/2}
\left(\lambda^\pp\right)^2 \
\er^{-\frac{8\pi^2}{g_2\lambda^\pp}+i\theta} \, .
\label{9-adim}
\ee
This is in agreement with the non-perturbative correction to the mass
of the dual string state computed in \cite{gks}. In particular, the
anomalous dimension (\ref{9-adim}) is independent of the parameter $n$
corresponding to the mode number of the plane-wave string state.
Apart from the exponential factor characteristic of instanton effects,
(\ref{9-adim}) contains an additional factor of
$\left(\lambda^\pp\right)^2$. This is due to the inclusion of
six-fermion scalars which give rise to additional $\nsix$
bilinears, each of which brings one more power of $\gy$. As will be
shown in the next subsection in the case of four impurity SO(4)$_R$
singlets it is sufficient to consider the bilinear solution for all
the scalars and as a consequence we shall find a leading contribution
of order $(g_2)^{7/2}\er^{-8\pi^2/g_2\lambda^\pp}$.

Contributions in which some of the $\bar\nu\nu$ bilinears are in the
$\mb{10}$ of SU(4) give rise to subleading corrections which
are suppressed by powers of $1/N$.

Another class of contributions to (\ref{2pt-2imp-9-a}) which are
suppressed in the large $N$ limit  are those in which pairs of scalars
are contracted. In these terms the analysis of the superconformal
modes is unaltered and in order to soak them up it is again necessary
to use the solution $\v^{(6)AB}$ for two of the scalars. Two of the
scalars which were previously replaced by $\bar\nu\nu$ bilinears are
now contracted and do not contain any fermion modes. Hence the
integration over the moduli space produces one less power of
$\gy^2N$. However, with the normalisation we are using the propagator
is proportional to $\gy^2$, so that in conclusion the contribution of
these terms is down by $1/N$ with respect to (\ref{2pt-9-fin}) because
there is no power of $N$ associated with the contraction.

A careful analysis of both types of $1/N$ corrections shows that they
give a contribution to the anomalous dimension of the operator
$\scrO_\mb9$ of order $(g_2)^{9/2}(\lambda^\pp)^2$. These are the
leading terms in a power series in $g_2$. In general the corrections
to the semi-classical approximation in the BMN limit can be reorganised
into a double series in $g_2$ and $\lambda^\pp$.

Operators in different sectors can be studied along the same lines.
However, superconformal invariance implies that all the two impurity
operators have the same anomalous dimension \cite{bei} and thus the
above result can be extended to two impurity operators in all the
other sectors with no further calculations required.

Arguments similar to those discussed here, showing the vanishing of
the leading one-instanton contribution to the two-point function
$G_\mb9(x_1,x_2)$, have been used to prove various non-renormalisation
properties in \cite{bk,koni,sk}. In view of the results we found for
$\scrO_\mb9$, one can expect that some of the non-renormalisation
results of these papers may not be extended to higher orders in the
coupling.

\subsection{Four impurity operators}
\label{2pt-4impur}

The calculation of two-point functions of four impurity operators is
more involved than the corresponding calculation in the two impurity
case from the point of view of the combinatorial analysis. However, at
the four impurity level, in the case of SO(4)$_R$ singlets, the
leading instanton contributions do not involve the six fermion
solution for the scalar fields. A non-zero result is obtained using
only the bilinear solution, which is known explicitly and given in
(\ref{phi2-inst-sol}), in computing the classical profiles of the
operators. Therefore we can analyse in a quantitative way the
semi-classical contributions to the two-point functions. The fact that
non-zero correlation functions of singlet operators are obtained using
the minimal number of fermion modes for each field also implies that
in this case a contribution to the matrix of anomalous dimensions
arises at leading order in the instanton background. As we shall see
these operators have instanton induced anomalous dimension of order
$(g_2)^{7/2}\er^{2\pi i\tau}$. Another difference with respect to the
two impurity case studied in the previous section is that two-point
functions of four impurity operators depend explicitly on the integers
dual to the string mode numbers. We shall discuss in detail a
SO(4)$_R\times$SO(4)$_C$ singlet with four scalar impurities and show
that the behaviour of its two-point functions is in remarkable
agreement with the corresponding string calculation of \cite{gks}.
Other singlet operators can be analysed in a similar
fashion. Operators in other sectors will be shown to receive
contribution only at higher order in $\lambda^\pp$. This result
follows simply from the analysis of fermion zero modes and is also in
agreement with the string theory prediction.

\subsubsection{$\veps$-singlet operator}
\label{epsinglet}

In this subsection we present the calculation of the one-instanton
contribution to the two-point function of one particular SO(4)$_R$
singlet. More details are provided in appendix \ref{calculations}. We
focus on the four scalar impurity operator in which the SO(4)$_R$
indices are contracted via an $\veps$-tensor,
\ba
\scrO_{\mb1;J;n_1,n_2,n_3} &=&
\frac{\veps_{ijkl}}{\sqrt{J^3\left(\frac{\gy^2N}{8\pi^2}\right)^{J+4}}}
\begin{array}[t]{c}
{\displaystyle \sum_{q,r,s=0}^J} \\
{\scriptstyle q+r+s \le J}
\end{array} \hsp{-0.1} \er^{2\pi i[(n_1+n_2+n_3)q
+(n_2+n_3)r+n_3s]/J} \nn \\
&& \hsp{3} \times \Tr\left(Z^{J-(q+r+s)}\v^i
Z^{q}\v^jZ^{r}\v^kZ^{s}\v^l\right) \, .
\label{epsdef}
\ea
The string state in the plane-wave background which is naturally
identified as being dual to this operator is of the form
\be
\veps_{ijkl} \, \a^i_{-n_1}\a^j_{-n_2}\wtil\a^k_{-n_3}
\wtil\a^l_{-(n_1+n_2-n_3)} |0\ra_h \, ,
\label{dualeps}
\ee
where $|0\ra_h$ is the BMN ground state and the contraction runs over
values of the indices in one of the two SO(4) factors. $D$-instanton
contributions to the renormalisation of the mass of this state were
computed in \cite{gks}. We shall return to the comparison with the
string results at the end of this section. Notice, however, that the
state (\ref{dualeps}) is antisymmetric under the exchange of the two
left-moving or right-moving modes. The operator (\ref{epsdef}) on the
other hand has no definite symmetry under permutations of the
parameters $n_1$, $n_2$ and $n_3$. Therefore in order to construct a
gauge theory operator that can be identified with (\ref{dualeps}) it
will be necessary to explicitly antisymmetrise (\ref{epsdef}). This
point will prove crucial when comparing instanton corrections to the
scaling dimension of $\scrO_\mb1$ to $D$-instanton induced corrections
to the mass of the string state.

We are interested in the two-point function
\ba
&& \hsp{-0.3} G_\mb1(x_1,x_2;n_1,n_2,n_3;m_1,m_2,m_3) =
\la \scrO_{\mb1;n_1,n_2,n_3}(x_1) \,
\bar\scrO_{\mb1;m_1,m_2,m_3}(x_2) \ra_{\rm inst} \nn \\
&& \hsp{-0.3} = \frac{\pi^{-4N}\gy^{4N}
\er^{2\pi i\tau}}{(N-1)!(N-2)!} \int \dr\rho\,\dr^4x_0 \,\prod_{A=1}^4
\dr^2\eta^A\dr^2\bar\xi^A \, \dr^{N-2}\nu^A \dr^{N-2}\bar\nu^A
\,\rho^{4N-13} \label{eps1-2pt-def} \\
&& \hsp{-0.3} \times \er^{\frac{\pi^2}{16\gy^2\rho^2}\veps_{ABCD}
(\bar\nu^{[A}\nu^{B]})(\bar\nu^{[C}\nu^{D]})}
\hat{\scrO}_{\mb1;n_1,n_2,n_3}\!\left(x_1;x_0,\rho,\eta,\bar\xi,\nu,
\bar\nu\right) \,\hat{\bar{\!\!\scrO}}_{\mb1;m_1,m_2,m_3}\!
\left(x_2;x_0,\rho,\eta,\bar\xi,\nu,\bar\nu\right) \nn  .
\ea
As usual the semi-classical approximation requires the calculation of
the classical profiles of $\scrO_\mb1$ and $\bar\scrO_\mb1$ in the
instanton background.

Summing over the SO(4) indices in (\ref{epsdef}) and using the
relations (\ref{vi4}) we find that the operator $\scrO_\mb1$ contains
the independent traces
\ba
&& +\Tr\left(Z^p\v^{12}Z^q\v^{13}Z^r\v^{24}Z^s\v^{34}\right)
-\Tr\left(Z^p\v^{12}Z^q\v^{34}Z^r\v^{24}Z^s\v^{13}\right) \nn \\
&& + \Tr\left(Z^p\v^{12}Z^q\v^{24}Z^r\v^{34}Z^s\v^{13}\right)
+ \Tr\left(Z^p\v^{12}Z^q\v^{34}Z^r\v^{13}Z^s\v^{24}\right) \nn \\
&& - \Tr\left(Z^p\v^{12}Z^q\v^{13}Z^r\v^{34}Z^s\v^{24}\right)
- \Tr\left(Z^p\v^{12}Z^q\v^{24}Z^r\v^{13}Z^s\v^{34}\right) \, ,
\label{epssingexp}
\ea
where $p=J-(q+r+s)$, plus three other groups of six traces obtained by
cyclic permutations of the indices on the impurities in
(\ref{epssingexp}). The conjugate operator, $\bar\scrO_\mb1$, contains
the same terms, but with the $Z$'s replaced by $\barZ$'s.

It is straightforward to verify that these traces, when evaluated in
the instanton background, contain the correct combination of fermions
required to soak up the superconformal modes in a two-point function
and that this can be achieved using only the bilinear solution for all
the scalars. In this case all the $\eta^A$ and $\bar\xi^A$ modes in
the gauge invariant traces are combined into $\zeta^A$'s. In order to
give rise to a non-zero two-point function in the one instanton sector
both operators  should then contain the combination
$\prod_{A=1}^4\left(\zeta^A\right)^2$. To achieve this in each trace
in (\ref{epssingexp}) the four impurities must provide two
superconformal modes of flavours 2 and 3, whereas the modes of flavour
1 and 4 can be taken from the impurities or from the $Z$'s. Similarly
in the case of $\bar\scrO_\mb1$ the superconformal modes of flavour 1
and 4 come necessarily from the impurities and those of flavour 2 and
3 can be provided by the impurities or by the $\barZ$'s. As in the two
impurity case studied in the previous section the leading contribution
is obtained taking  all the remaining modes in $\nsix$ bilinears. In
all the traces appearing in both $\scrO_\mb1$ and $\bar\scrO_\mb1$ the
impurities contain two fermion modes of each flavour. The combination
of fermion modes entering into all the terms in $\scrO_\mb1$ is
\be
\left(\scrmf^1\right)^{J+2}\left(\scrmf^2\right)^2
\left(\scrmf^3\right)^2\left(\scrmf^4\right)^{J+2} \, ,
\label{fm-4imp-1}
\ee
whereas all the terms in the expansion of $\bar\scrO_\mb1$ contain
\be
\left(\scrmf^1\right)^2\left(\scrmf^2\right)^{J+2}
\left(\scrmf^3\right)^{J+2}\left(\scrmf^4\right)^2 \, .
\label{fm-4imp-1b}
\ee
The leading contribution to the two-point function $G_\mb1(x_1,x_2)$
in the semi-classical approximation arises from terms in the profiles
of the operators containing the  following combinations of fermion
modes
\ba
\scrO_\mb1 &\to& \left(\zeta^1\right)^2 \left(\zeta^2\right)^2
\left(\zeta^3\right)^2 \left(\zeta^4\right)^2
\left(\bar\nu^{[1}\nu^{4]}\right)^J \nn \\
\bar\scrO_\mb1 &\to& \left(\zeta^1\right)^2 \left(\zeta^2\right)^2
\left(\zeta^3\right)^2 \left(\zeta^4\right)^2
\left(\bar\nu^{[2}\nu^{3]}\right)^J \, .
\label{fmodcombs}
\ea
As previously observed these combinations can be obtained in many
different ways corresponding to the choice of which field, $Z$ or
$\v$, provides each of the $\zeta$'s of flavour 1 and 4 in $\scrO$.
An equal number of different contributions arises from the ways of
distributing the $\zeta$'s of flavour 2 and 3 among the impurities or
the $\barZ$'s in $\bar\scrO$.  In order to simplify the discussion of
the associated combinatorics it is convenient to introduce the
following notation. We denote by $\check\v^{AB}$ a scalar solution in
which only the $\nu$ and $\bar\nu$ modes are kept and all the
superconformal modes are set to zero; scalars containing only
bilinears in the superconformal modes are indicated by $\wtil\v^{AB}$;
the symbol $\what\v^{AB}$ is used for scalar profiles in which only
mixed terms, $\zeta\nu$ or $\zeta\bar\nu$, are included,
\ba
\check\v^{AB} &\equiv& \v^{AB}(x,x_0,\rho;\nu,\bar\nu;\eta=\bar\xi=0)
\label{checkscal} \\
\wtil\v^{AB} &\equiv& \v^{AB}(x,x_0,\rho;\eta,\bar\xi;\nu=\bar\nu=0)
\label{tilscal} \\
\what\v^{AB} &\equiv& \v^{AB}(x,x_0,\rho;\eta\nu,\eta\bar\nu,
\bar\xi\nu, \bar\xi\bar\nu;\zeta\zeta=\bar\nu\nu=0) \, .
\label{hatscal}
\ea
The same notation is also used for $Z\sim\v^{14}$ and $\barZ\sim\v^{23}$.

We are only interested in contributions to the two-point function
$G_\mb1(x_1,x_2)$ which survive in the BMN limit, $N\to\infty$,
$J\to\infty$, with $J^2/N$ fixed. The leading large $N$  contributions
are those in which the combinations (\ref{fmodcombs}) are selected,
\ie the superconformal modes are soaked up and all the remaining
fields are replaced by $\nsix$ bilinears. Within this class of terms
the dominant ones in the large $J$ limit are those in which as many
superconformal modes as possible are extracted from the $Z$'s and
$\barZ$'s, because there is roughly a multiplicity factor of $J$
associated with the choice of each $Z$ or $\barZ$ providing one such
mode. We first discuss these leading terms and we will then show that
these are the only non-vanishing contributions in the BMN limit.

As shown by the previous preliminary analysis, in the operator
$\scrO_\mb1$ the modes $\zeta^2$ and $\zeta^3$ necessarily come from
the impurities and thus the leading large $J$ terms arise from traces
in which we take the two $\zeta^1$ and the two $\zeta^4$ modes from
four distinct $Z$'s. Similar considerations apply to the
$\bar\scrO_\mb1$ operator with the r\^ole of the flavours $(1,4)$ and
$(2,3)$ exchanged. Using the notation introduced in
(\ref{checkscal})-(\ref{hatscal}) this means that we consider traces
in which all four impurities are $\what\v^{AB}$ matrices and we choose
four $Z$'s to be $\what Z$ matrices, with all the others being $\check
Z$'s. There is a total of 35 different traces of this type for each of
the $6\times4$ terms in the operator (\ref{epsdef}) and a similar
counting applies to its conjugate. The 35 traces correspond to the
inequivalent ways of choosing the four $\what Z$'s from the four
groups of $Z$'s in (\ref{epsdef}). For the generic term in the operator,
$\Tr\left(Z^p\v^{A_1B_1}Z^q\v^{A_2B_2}Z^r\v^{A_3B_3}Z^s\v^{A_4B_4}
\right)$, with $p=J-(q+r+s)$, we need to consider
\ba
&& \Tr\left(\check Z^{p_1}\what Z\check Z^{p_2}\what Z
\check Z^{p_3}\what Z\check Z^{p_4}\what Z\check Z^{p_5}
\what\v^{A_1B_1}\check Z^q\what\v^{A_2B_2}\check Z^r
\what\v^{A_3B_3}\check Z^s\what\v^{A_4B_4}\right) \nn \\
&& \Tr\left(\check Z^{p_1}\what Z\check Z^{p_2}\what Z
\check Z^{p_3}\what Z\check Z^{p_4}\what\v^{A_1B_1}
\check Z^{q_1}\what Z\check Z^{q_2}q\what\v^{A_2B_2}\check Z^r
\what\v^{A_3B_3}\check Z^s\what\v^{A_4B_4}\right) \nn \\
&& \ldots \nn \\
&& \Tr\left(\check Z^p\what\v^{A_1B_1}\check Z^q
\what\v^{A_2B_2}\check Z^r\what\v^{A_3B_3}\check Z^{s_1}
\what Z\check Z^{s_2}\what Z\check Z^{s_3}\what Z
\check Z^{s_4}\what Z\check Z^{s_5}\what\v^{A_4B_4}\right) \, ,
\label{35traces}
\ea
where in the first trace $\sum_i p_i=p-4=J-(q+r+s+4)$, in the second
$\sum_i p_i=p-3$ and $\sum_i q_i=q-1$ and so on until the last sum
where $\sum_i s_i=s-4$. The ellipsis in (\ref{35traces}) refers to
other combinations in which the four $\what Z$'s are gradually moved
to the right. All these traces can be evaluated using the ADHM
solution for the scalars given in (\ref{phi2-inst-sol}) and selecting
for each factor the matrix elements containing the appropriate fermion
bilinears. The calculation is rather involved. As explained in
appendix \ref{calculations} it can be carried out most efficiently
defining a more general trace from which the 35 distinct traces
(\ref{35traces}) can be obtained for different choices of indices.

In order to compute the relevant part of the profile of $\scrO_\mb1$
we need to sum the contributions of the traces (\ref{35traces})
corresponding to the 6$\times$4 choices of indices, $(A_i,B_i)$,
$i=1,\ldots,4$, on the impurities. A key feature of all these traces
is that they do not depend on the way the $Z$'s are grouped, \ie they
do not depend on the exponents, $(p_1,\ldots,p_5,q,r,s)$,
$(p_1,\ldots,p_4,q_1,q_2,r,s)$ etc. in (\ref{35traces}), but only on
the ordering of the four $\what Z$'s with respect to the four
impurities, $\what\v^{A_iB_i}$, $i=1,\dots,4$. This is a consequence
of the structure of the ADHM matrices and the restrictions imposed by
the ADHM constraints. Keeping only the terms with two $\zeta$'s of
each flavour all the traces in $\scrO_\mb1$ produce expressions which
after simple Fierz rearrangements can be brought to the form
\be
\frac{\rho^8}{[(x_1-x_0)^2+\rho^2]^{J+8}}
\left(\bar\nu^{[1}\nu^{4]}\right)^J\left[\left(\zeta^1\right)^2
\left(\zeta^2\right)^2\left(\zeta^3\right)^2\left(\zeta^4\right)^2
\right]\!(x_1) \, .
\label{restraces-O1}
\ee
Similarly all the contributions from the traces in $\bar\scrO_\mb1$
containing the required eight superconformal modes can be put in the
form
\be
\frac{\rho^8}{[(x_2-x_0)^2+\rho^2]^{J+8}}
\left(\bar\nu^{[2}\nu^{3]}\right)^J\left[\left(\zeta^1\right)^2
\left(\zeta^2\right)^2\left(\zeta^3\right)^2\left(\zeta^4\right)^2
\right]\!(x_2) \, .
\label{restraces-bO1}
\ee
Each set of indices $(A_i,B_i)$, $i=1,\dots,4$ on the impurities in
each of the 35 traces leads to a contribution of the form
(\ref{restraces-O1})-(\ref{restraces-bO1}) with a different numerical
coefficient.

The fact that the result of all the traces can be reduced to the above
expressions implies that when substituting into the definition of the
operator (\ref{epsdef}) and its conjugate a common factor
(\ref{restraces-O1}) or, respectively, (\ref{restraces-bO1}) can be
taken out of the traces. Associated with each of the 35 types of
traces there are, however, multiplicity factors which make the sums in
the definition of the operator non-trivial. For instance in the last
trace in (\ref{35traces}) there are $s$ choices for the first $\what
Z$ among the $Z$'s, $(s-1)$ choices for the second $\what Z$, $(s-2)$
for the third and $(s-3)$ for the fourth. After substituting into the
definition (\ref{epsdef}) and factoring out the moduli dependence in
the form (\ref{restraces-O1}), the contribution of the last trace in
(\ref{35traces}) involves the sums
\be
\begin{array}[t]{c}
{\displaystyle \sum_{q,r,s=0}^J} \\
{\scriptstyle q+r+s \le J}
\end{array} \hsp{-0.1} \er^{2\pi i[(n_1+n_2+n_3)q
+(n_2+n_3)r+n_3s]/J} s(s-1)(s-2)(s-3) \, ,
\label{sum35}
\ee
with a numerical coefficient resulting from the contributions of the
6$\times$4 permutations of indices of the impurities.  Repeating the
same analysis for all the traces means combining a huge number of
terms which makes the calculation extremely laborious. Completely
analogous steps go into the calculation of the profile of the
conjugate operator.

After lengthy algebraic manipulations and the use of the formalism
described in section \ref{inst2ptf}, the semi-classical result for the
two-point function (\ref{eps1-2pt-def}) takes the form
\ba
G_\mb1(x_1,x_2) &=& \frac{\er^{2\pi i\tau}}{J^3N^{7/2}}
\int \frac{\dr^4x_0\,\dr\rho}{\rho^5} \,
\frac{\rho^{J+8}}{[(x_1-x_0)^2+\rho^2]^{J+8}}
\frac{\rho^{J+8}}{[(x_2-x_0)^2+\rho^2]^{J+8}} \nn \\
&& \times \int \prod_{A=1}^4\dr^2\eta^A\dr^2\bar\xi^A \,
\prod_{B=1}^4 \left[\left(\zeta^B\right)^2(x_1)\right]
\left[\left(\zeta^B\right)^2(x_2)\right] \nn \\
&& \times \int \dr^5\Omega \, \left(\Omega^{14}\right)^J
\left(\Omega^{23}\right)^J \, K(n_1,n_2,n_3;J) K(m_1,m_2,m_3;J) \, ,
\label{2pt-4imp-1}
\ea
where following the discussion in section \ref{inst2ptf} the $\nsix$
bilinears have been expressed in terms of the angular variables
$\Omega^{AB}$. In (\ref{2pt-4imp-1}) overall numerical coefficients
have been omitted. The $J$ and $N$ dependence in the prefactor in
(\ref{2pt-4imp-1}) is obtained combining the normalisation of the
operators, the contribution of the measure on the instanton moduli
space and the factors of $\gy\sqrt{N}$ associated with $\nsix$
bilinears. The origin of the various factors which determine the
dependence on the parameters $\gy$, $N$ and $J$ will be summarised
shortly. The expression (\ref{2pt-4imp-1}) contains integrations over
the bosonic moduli, $x_0$ and $\rho$, the sixteen superconformal
fermion modes and the five-sphere coordinates $\Omega^{AB}$. The
dependence on the integers $n_i$, $m_i$, $i=1,2,3$, dual to the mode
numbers of the corresponding string state is contained in the functions
$K(n_1,n_2,n_3;J)$ and $K(m_1,m_2,m_3;J)$. These are given by the sum
of 35 terms,
\be
K(n_1,n_2,n_3;J) = \sum_{a=1}^{35} c_a \,\calS_a(n_1,n_2,n_3;J) \, ,
\label{modnumfunct}
\ee
where the symbols $\calS_a$ indicate 35 different sums over the indices
$q,r,s$ in which the summands are given by the phase factor $\exp
\{2\pi i[(n_1+n_2+n_3)q+(n_2+n_3)r+n_3s]/J\}$ times the multiplicity
factors associated with the different distributions of $\what Z$'s in
each case. The numerical coefficients $c_a$ are obtained combining
the contributions of the different permutations of indices on the
impurities for each of the 35 terms. See appendix
\ref{eps-singlet-detail} for more details.

In the large $J$ limit the leading order contribution to the sums
$\calS_a$ can be obtained using a continuum approximation by setting
$x=q/J$, $y=r/J$, $z=s/J$, so that $x,y,z\in[0,1]$. For instance the
sum (\ref{sum35}) is approximated as
\ba
&& \begin{array}[t]{c}
{\displaystyle \sum_{q,r,s=0}^J} \\
{\scriptstyle q+r+s \le J}
\end{array} \hsp{-0.1} \er^{2\pi i[(n_1+n_2+n_3)q
+(n_2+n_3)r+n_3s]/J} s(s-1)(s-2)(s-3) \nn \\
&& \to J^7 \int_0^1\dr x \int_0^{1-x}\dr y \int_0^{1-x-y} \dr z \,
z^4 \er^{2\pi i[(n_1+n_2+n_3)x+(n_2+n_3)y+n_4z]} \, ,
\label{continuum}
\ea
which shows that it behaves as $J^7$ for large $J$. We shall denote by
$\kappa(n_1,n_2,n_3)$ the function of the mode numbers arising from
these sums/integrals after extracting a factor of $J^7$,
\be
K(n_1,n_2,n_3;J) = J^7 \kappa(n_1,n_2,n_3) \, .
\label{defkappa}
\ee
The two-point function is then
\ba
G_\mb1(x_1,x_2) &=& \frac{J^{11}\,\er^{2\pi i\tau}}{N^{7/2}}
\int \frac{\dr^4x_0\,\dr\rho}{\rho^5} \,
\frac{\rho^{J+8}}{[(x_1-x_0)^2+\rho^2]^{J+8}}
\frac{\rho^{J+8}}{[(x_2-x_0)^2+\rho^2]^{J+8}} \nn \\
&& \times \int \prod_{A=1}^4\dr^2\eta^A\dr^2\bar\xi^A \,
\prod_{B=1}^4 \left[\left(\zeta^B\right)^2(x_1)\right]
\left[\left(\zeta^B\right)^2(x_2)\right] \nn \\
&& \times \int \dr^5\Omega \, \left(\Omega^{14}\right)^J
\left(\Omega^{23}\right)^J \, \kappa(n_1,n_2,n_3;J)
\kappa(m_1,m_2,m_3;J) \, .
\label{2pt-4imp-2}
\ea
Unlike the case of two impurity operators discussed in the previous
subsection, here the dependence on the instanton moduli is known
explicitly and we can compute the associated integrations. More
details are given in appendix \ref{eps-singlet-detail}. The
integration over the five-sphere in (\ref{2pt-4imp-2}) is a special
case of the general integral (\ref{spherintn}) and gives
\be
I_{S^5} = \int \dr^5\Omega \, \left(\Omega^{14}\right)^J
\left(\Omega^{23}\right)^J = \frac{\pi^3}{(J+1)(J+2)} \, .
\label{5sphere-4imp}
\ee
The integration over the superconformal modes is also
straightforward. It does not depend on $N$ or $J$. For each flavour
the result is
\be
I_{\zeta} = \int \dr^2\eta\dr^2\bar\xi \,
\left[\left(\zeta\right)^2(x_1)\right]
\left[\left(\zeta\right)^2(x_2)\right]
= -(x_1-x_2)^2 \, ,
\label{etabxiint}
\ee
so that the fermionic integrals contribute a factor of $(x_1-x_2)^8$.
The integration over the bosonic part of the moduli space must be
treated carefully since it is logarithmically divergent as expected in
the presence of a contribution to the matrix of anomalous
dimensions. The integrals need to be regulated for instance by
dimensional regularisation of the $x_0$ integral and can then be
computed using standard techniques, \eg introducing Feynman
parameters. The result is
\ba
I_{\rm b} &=& \int \frac{\dr^4x_0\,\dr\rho}{\rho^5} \,
\frac{\rho^{J+8}}{[(x_1-x_0)^2+\rho^2]^{J+8}}
\frac{\rho^{J+8}}{[(x_2-x_0)^2+\rho^2]^{J+8}} \nn \\
&=& \rule{0pt}{24pt}
\fr{\epsilon}\frac{\Gamma(J+6)\Gamma(J+8+\epsilon)}
{[\Gamma(J+8)]^2} \,\pi^{2-\epsilon} \fr{(x_{12}^2)^{J+8+\epsilon}}
\, , \qquad \epsilon\to 0 \, .
\label{regulint1}
\ea
The $1/\epsilon$ pole is the manifestation of a logarithmic divergence
in dimensional regularisation. The contribution (\ref{regulint1})
behaves as $1/J^2$ in the large $J$ limit.

Putting together all the contributions the dependence on the
parameters, $\gy$, $N$ and $J$, in the correlation function can be
summarised as follows
\ba
&& \underbrace{\left(\fr{\sqrt{J^3
(\gy^2N)^{J+4}}}\right)^2}_{\rm normalised ~ op.~ profile}
\;\underbrace{\left(\gy\sqrt{N}\right)^{2J}}_{\nu,\:\bar\nu
~ {\rm bilinears}}\;
\underbrace{\er^{2\pi i\tau}\gy^8\sqrt{N}}_{\rm measure}
\;\underbrace{\fr{J^2}}_{S^5 ~ {\rm integral}}
\;\underbrace{\fr{J^2}}_{x_0,\:\rho ~ {\rm integrals}}
\;\underbrace{(J^7)^2}_{\calS_a ~ {\rm sums}} \nn \\
&& \sim \frac{J^7}{N^{7/2}}\, \er^{2\pi i\tau}=
(g_2)^{7/2}\,\er^{-\frac{8\pi^2}{g_2\lambda^\pp}+i\theta} \, .
\label{gNJ-dep}
\ea
The final result for the two-point function is thus, up to a numerical
coefficient, 
\be
G_\mb1(x_1,x_2) =
(g_2)^{7/2}\,\er^{-\frac{8\pi^2}{g_2\lambda^\pp}+i\theta}
\kappa(n_1,n_2,n_3)\kappa(m_1,m_2,m_3) \, \fr{(x_{12}^2)^{J+4}}
\log\left(\Lambda^2x_{12}^2\right) \, .
\label{4imp-2pt-noas}
\ee
where the scale $\Lambda$ appears as a consequence of the $1/\epsilon$
divergence. It has no observable effect. The physical information
contained in the two-point function is in the contribution to the
matrix of anomalous dimensions which is read from the coefficient in
(\ref{4imp-2pt-noas}) and does not depend on $\Lambda$.

The result is expressed in terms of the double scaling parameters
$\lambda^\pp$ and $g_2$. Note that,  unlike the two-point functions
of two impurity operators (\ref{4imp-2pt-noas}) is independent of
$\lambda^\pp$ apart from the dependence in the exponential instanton
weight.

The non-perturbative mass correction computed in \cite{gks} for the
state (\ref{dualeps}), in terms of the same gauge theory parameters,
is of the form
\be
\d m\sim (g_2)^{7/2} \, \er^{-\frac{8\pi^2}{g_2\lambda^\pp}+i\theta}
\fr{(n_1n_2)^2} \, .
\label{4imp-string}
\ee
so that the $\lambda^\pp$ and $g_2$ dependence is in agreement with
the gauge theory calculation. The mode number dependence in
(\ref{4imp-string}) is remarkably simple and a special feature of the
string result, which is a direct consequence of the structure of the
$D$-instanton boundary state, is that it is non-vanishing only if the
mode numbers in both the incoming and the outgoing states are pairwise
equal. The only states which couple to the $D$-instanton are of the form
\be
\veps_{ijkl} \, \a^i_{-n_1}\a^j_{-n_2}\wtil\a^k_{-n_1}
\wtil\a^l_{-n_2} |0\ra_h \, .
\label{dualepspair}
\ee
On the other hand in the gauge theory result (\ref{4imp-2pt-noas})
obtained for the operator (\ref{epsdef}) the mode number dependence is
contained in  $\kappa(n_1,n_2,n_3)$ and $\kappa(m_1,m_2,m_3)$, which
are extremely complicated rational functions of their arguments. In
particular the condition that the integers $n_i$ be equal in pairs
does not seem to be required.

However, as observed after (\ref{dualeps}) in order to correctly match
the properties of the dual string state, the operator (\ref{epsdef})
must be explicitly antisymmetrised under the exchange of pairs of mode
numbers. This antisymmetrisation induces dramatic simplifications.
Working with the correctly antisymmetrised operators the result for
the two-point function is
\be
G_1(x_1,x_2) = \frac{3^2\,(g_2)^{7/2} \,
\er^{-\frac{8\pi^2}{g_2\lambda^\pp}+i\theta}}{2^{41}\,\pi^{9/2}}\,
\fr{(n_1n_2)(m_1m_2)} \fr{(x_{12}^2)^{J+4}} \,
\log\left(\Lambda^2x_{12}^2\right)
\label{4imp-2pt-as}
\ee
if the mode numbers in each of the two operators are equal in pairs
and vanishes otherwise. In (\ref{4imp-2pt-as}) we have reinstated all
the numerical coefficients coming from the profiles of the operators,
the combinatorics previous described and the moduli space measure.
This result is in perfect agreement with the string result
(\ref{4imp-string}) of \cite{gks}. It is worth stressing that the
simplification found after the  antisymmetrisation is extraordinary
given the complexity of the function $\kappa(n_1,n_2,n_3)$. Moreover
the condition of pairwise equal mode numbers which is also imposed in
this way is far from obvious and highly non-trivial from the point of
view of the gauge theory calculation.

As we have seen, in the two-point function computed in the
semi-classical approximation the mode number dependence factorises. A
consequence of this is that in $G_\mb1(x_1,x_2)$ the two independent
mode numbers in $\scrO_\mb1$, $n_1$ and $n_2$, do not have to equal
those in $\bar\scrO_\mb1$, $m_1$ and $m_2$. This appears to contradict
energy conservation in the dual string amplitude, which requires the
mode numbers of the outgoing state to match one to one those of the
incoming state. However, the fact that the condition $m_i=n_i$, $i=1,2$
does not arise is an effect of the semi-classical approximation. This
is valid in the $\lambda^\pp\to 0$ limit which corresponds to the
$m\to\infty$ limit in the plane-wave string theory (where $m$ is the
mass parameter entering the string action). In this strict limit
energy conservation in a two-point string amplitude only requires that
the number of oscillators in the incoming and outgoing states be
equal, with no constraint on the associated mode numbers. Therefore
(\ref{4imp-2pt-as}) is indeed in agreement with the string theory
result. On the other hand the instanton corrections discussed here
should be considered as subleading corrections on top of the
perturbative effects.  The condition $m_i=n_i$ on the operators in a
two-point function is already imposed at the perturbative level and
should therefore be assumed when computing instanton contributions in
the semi-classical approximation.

The calculation presented here is not sufficient to determine the
actual instanton induced anomalous dimension of the operator
$\scrO_\mb1$. This requires the diagonalisation of the matrix of
anomalous dimensions of which we have not computed all the entries.
Other entries are determined by the corresponding two-point functions
whose calculation follows the same steps described here and results in
expressions similar to (\ref{4imp-2pt-as}). From this we can conclude
that the behaviour of the leading instanton contribution to the
anomalous dimensions of singlet operators is
\be
\g_\mb1^{\rm inst} \sim  (g_2)^{7/2}\,
\er^{-\frac{8\pi^2}{g_2\lambda^\pp}+i\theta}
\fr{(n_1n_2)^2} \, .
\label{singletadim}
\ee
As a further test of the result we find that the two point function
vanishes in the limit of zero mode numbers. The function
$\kappa(n_1,n_2,n_3)$ is identically zero when $n_1=n_2=n_3=0$. This
is the expected behaviour because in this limit the operator is
expected to become protected and no corrections to its free theory
two-point functions should arise. The string theory counterpart of
this result is the decoupling of the supergravity modes (dual to the
protected operators with $\{n_i=0\}$), which was also verified in
\cite{gks}.

In the previous analysis we have considered only a class of
contributions in which in each operator as many superconformal modes
as possible were taken from the $Z$'s. It is easy to verify that these
are the only relevant terms at leading order in the BMN limit. All the
other types of traces are suppressed and vanish in the $J\to\infty$
limit. As an example consider a contribution to the profile of
$\scrO_\mb1$ in which the $\zeta^2$ and $\zeta^3$ modes as well as one
of either the $\zeta^1$ or $\zeta^4$ modes are taken from the
impurities. Instead of the last trace in (\ref{35traces}) we would
then consider traces of the type
\be
\Tr\left(\check Z^p\wtil\v^{A_1B_1}\check Z^q
\what\v^{A_2B_2}\check Z^r\what\v^{A_3B_3}\check Z^{s_1}
\what Z\check Z^{s_2}\what Z\check Z^{s_3}\what Z
\check Z^{s_4}\what\v^{A_4B_4}\right) \, ,
\label{subdomtrace}
\ee
where the first impurity contains two superconformal modes and thus
only three $\what Z$'s are needed. An analysis similar to that carried
out for the traces (\ref{35traces}) can be repeated in this case and one
finds that associated with such a trace there is sum of the form
\be
\begin{array}[t]{c}
{\displaystyle \sum_{q,r,s=0}^J} \\
{\scriptstyle q+r+s \le J}
\end{array} \hsp{-0.1} \er^{2\pi i[(n_1+n_2+n_3)q
+(n_2+n_3)r+n_3s]/J} \fr{3!} s(s-1)(s-2) \, ,
\label{subdomsum}
\ee
which behaves as $J^6$ in the large $J$ limit. The remaining $J$, $N$
and $\gy$ dependence in the correlation function is unmodified and
thus the combined behaviour of this contribution can be read from
(\ref{gNJ-dep}) replacing the last factor on the first line by
$(J^6)^2$ leading to
\bdm
\er^{2\pi i\tau}\frac{J^5}{N^{7/2}}
\sim \frac{\er^{2\pi i\tau}(g_2)^{7/2}}{J^2} \, ,
\edm
which vanishes in the BMN limit. Similar arguments can be repeated for
all the contributions other than those leading to (\ref{4imp-2pt-as}),
which is therefore the complete leading instanton contribution to this
singlet two-point function in the BMN limit. We shall briefly comment
on corrections to this result of higher order in $\lambda^\pp$ and
$g_2$ in the discussion section.

\subsubsection{Other four impurity singlets}
\label{othersingl}

There are two other independent four impurity singlet operators
involving four scalar impurities. They correspond to the two
inequivalent ways of contracting the SO(4)$_R$ indices with Kronecker
delta's,
\ba
\scrO^{(d_1)}_{\mb1;J;n_1,n_2,n_3} &=&
\frac{1}{\sqrt{J^3\left(\frac{\gy^2N}{8\pi^2}\right)^{J+4}}}
\begin{array}[t]{c}
{\displaystyle \sum_{q,r,s=0}^J} \\
{\scriptstyle q+r+s \le J}
\end{array} \hsp{-0.1} \er^{2\pi i[(n_1+n_2+n_3)q
+(n_2+n_3)r+n_3s]/J} \nn \\
&& \hsp{3} \times \Tr\left(Z^{J-(q+r+s)}\v^i
Z^{q}\v^iZ^{r}\v^jZ^{s}\v^j\right) \, ,
\label{deldel1def} \\
\scrO^{(d_2)}_{\mb1;J;n_1,n_2,n_3} &=&
\frac{1}{\sqrt{J^3\left(\frac{\gy^2N}{8\pi^2}\right)^{J+4}}}
\begin{array}[t]{c}
{\displaystyle \sum_{q,r,s=0}^J} \\
{\scriptstyle q+r+s \le J}
\end{array} \hsp{-0.1} \er^{2\pi i[(n_1+n_2+n_3)q
+(n_2+n_3)r+n_3s]/J} \nn \\
&& \hsp{3} \times \Tr\left(Z^{J-(q+r+s)}\v^i
Z^{q}\v^jZ^{r}\v^iZ^{s}\v^j\right) \, .
\label{deldel2def}
\ea
The calculation of instanton contributions to the two-point functions
of these operators proceeds in complete analogy with the discussion in
the previous subsection. These operators are expected to receive
contributions of the same type as the $\veps$-singlet (\ref{epsdef})
and to mix with the latter much in the same way as was found in the
string theory analysis of \cite{gks}.

Other singlet operators can be constructed using all the other
combinations of impurities in table \ref{4impur}. In all these cases
the analysis of fermion zero modes shows that a non-zero contribution
to the corresponding two-point functions can arise at the same leading
order as (\ref{singletadim}). Operators containing $D_\mu Z$
insertions correspond to string states involving bosonic oscillators
which are vectors of the second SO(4). Operators containing the
$\psi^{-\,a}_a$ and $\psi^{+\,\adot}_{\dot a}$ fermions are dual to
states created by the $S^\pm$ oscillators. The calculation of
two-point functions of all these operators is similar to that
described in the previous section with the additional technical
complication that in the presence of covariant derivatives the
solution $A_\mu^{(4)}$ for the gauge potential is needed and for
operators containing fermions the solution $\lambda^{(5)A}_\a$ is
needed.

As observed in the case of the operator (\ref{epsdef}), two-point
functions in the semi-classical approximation factorise, with the two
operators being related only by the five-sphere integration. Because
of this property mixing is expected among all the operators which
receive instanton contributions.

In the $\scrN$=4 theory it is in principle possible to construct a
large number of other operators which potentially mix with those
considered here, being SO(4)$_R\times$SO(4)$_C$ singlets with
$\D-J=4$. These involve $\D-J=2$ impurities and thus do not correspond
to new independent states having vanishing two-point functions in free
theory. However, it is known that the inclusion of such operators is
needed in perturbation theory to properly resolve the mixing beyond
the zeroth order approximation in the $g_2$ expansion. Since instanton
effects are exponentially suppressed in $g_2$ one should in principle
expect these operators to be relevant at leading order in the
instanton background. This is, however, not the case. The combinatorial
analysis involved in computing the classical profiles of the operators
shows that those containing $\D-J=2$ impurities are suppressed in the
large $J$ limit.

\subsubsection{Operators in other sectors}
\label{other4imp}

As observed in section \ref{4impurity} the spectrum of four impurity
BMN operators is rather rich. Instanton contributions to the anomalous
dimensions of operators in other sectors can be studied with the same
methods used for the singlets. $D$-instanton induced amplitudes for
string states in the plane wave background dual to non-singlet
operators are suppressed with respect to those in the singlet
sector. Hence string theory predicts that the leading instanton
contributions to the anomalous dimensions of non-singlet four impurity
operators should be suppressed with respect to (\ref{singletadim}).
More precisely the string prediction is that the leading non-zero
contributions should arise at order $\er^{2\pi i\tau}(g_2)^{7/2}
(\lambda^\pp)^2$. We shall not discuss in detail the calculation of
two-point functions needed to verify this prediction, but we present
here an argument indicating that the gauge theory result is indeed in
agreement with string theory. We focus on an operator with four scalar
impurities which is a singlet of SO(4)$_C$ and belongs to the
$\mb3^+\oplus\mb3^-$ of SO(4)$_R$,
\ba
\scrO^{[ij]}_{\mb3^+\oplus\mb3^-;J;n_1,n_2,n_3} &=&
\frac{1}{\sqrt{J^3\left(\frac{\gy^2N}{8\pi^2}\right)^{J+4}}}
\begin{array}[t]{c}
{\displaystyle \sum_{q,r,s=0}^J} \\
{\scriptstyle q+r+s \le J}
\end{array} \hsp{-0.1} \er^{2\pi i[(n_1+n_2+n_3)q
+(n_2+n_3)r+n_3s]/J} \nn \\
&& \hsp{3} \times \Tr\left(Z^{J-(q+r+s)}\v^k
Z^{q}\v^kZ^{r}\v^{[i}Z^{s}\v^{j]}\right) \, .
\label{4imp-3+3-def}
\ea
The study of other non-singlet operators is completely analogous.
Considering for concreteness the component $i=1$, $j=2$ in
(\ref{4imp-3+3-def}), we find that using for all the scalars the
bilinear solution the combinations of fermion modes contained in the
classical profiles of the operator and its conjugate are respectively
\ba
&& \left(\scrmf^1\right)^{J+3}\left(\scrmf^2\right)^{2}
\left(\scrmf^3\right)^{2}\left(\scrmf^4\right)^{J+1}
\label{O3+3-zm1} \\
&+&\left(\scrmf^1\right)^{J+2}\left(\scrmf^2\right)^{3}
\left(\scrmf^3\right)\left(\scrmf^4\right)^{J+2}
\label{O3+3-zm2} \\
&+& \left(\scrmf^1\right)^{J+2}\left(\scrmf^2\right)
\left(\scrmf^3\right)^{3}\left(\scrmf^4\right)^{J+2}
\label{O3+3-zm3} \\
&+& \left(\scrmf^1\right)^{J+1}\left(\scrmf^2\right)^{2}
\left(\scrmf^3\right)^{2}\left(\scrmf^4\right)^{J+3}
\label{O3+3-zm4}
\ea
and
\ba
&& \left(\scrmf^1\right)^{3}\left(\scrmf^2\right)^{J+2}
\left(\scrmf^3\right)^{J+2}\left(\scrmf^4\right)
\label{bO3+3-zm1} \\
&+& \left(\scrmf^1\right)^{2}\left(\scrmf^2\right)^{J+3}
\left(\scrmf^3\right)^{J+1}\left(\scrmf^4\right)^{2}
\label{bO3+3-zm2} \\
&+& \left(\scrmf^1\right)^{2}\left(\scrmf^2\right)^{J+1}
\left(\scrmf^3\right)^{J+3}\left(\scrmf^4\right)^{2}
\label{bO3+3-zm3} \\
&+& \left(\scrmf^1\right)\left(\scrmf^2\right)^{J+2}
\left(\scrmf^3\right)^{J+2}\left(\scrmf^4\right)^{3} \, .
\label{bO3+3-zm4}
\ea
This shows that the integrations over the superconformal modes in the
two-point function can be saturated, selecting terms containing
(\ref{O3+3-zm1}) or (\ref{O3+3-zm4}) in $\scrO^{[12]}$ and terms
containing (\ref{bO3+3-zm2}) and (\ref{bO3+3-zm3}) in
$\bar\scrO^{[12]}$. However, with these choices the resulting
five-sphere integrals vanish. For instance combining (\ref{O3+3-zm1})
and (\ref{bO3+3-zm2}) leads to the following moduli space integrals
\ba
&& \int \dr^8\eta \dr^8\bar\xi \left[\left(\zeta^1\right)^2
\left(\zeta^2\right)^2\left(\zeta^3\right)^2\left(\zeta^4\right)^2
\right]\!(x_1)\,\left[\left(\zeta^1\right)^2
\left(\zeta^2\right)^2\left(\zeta^3\right)^2\left(\zeta^4\right)^2
\right]\!(x_2) \nn \\
&\times& \int \dr^{4(N-2)}\nu\dr^{4(N-2)}\bar\nu \left(\bar\nu^{[1}
\nu^{4]}\right)^{J-1}\left(\bar\nu^{(1}\nu^{1)}\right)
\left(\bar\nu^{[2}\nu^{3]}\right)^{J-1}\left(\bar\nu^{(2}\nu^{2)}\right)
\, .  \label{zero-5sphere}
\ea
The integration over the five-sphere arising from the second line
of (\ref{zero-5sphere}) vanishes because the multiplicity of the
flavours 1 and 2 exceeds that of the flavours 3 and 4.

In order to soak up the superconformal modes while avoiding the
obstruction from the five sphere integral it is necessary to include a
six-fermion term in each operator. In this way the combinations of
modes in the two operators are the same as in
(\ref{O3+3-zm1})-(\ref{bO3+3-zm4}) with the addition of one mode of
each flavour. The same arguments given in section \ref{2pt-2impur} in
connection with two impurity operators can be repeated here and for
instance combining (\ref{O3+3-zm1}) and (\ref{bO3+3-zm4}) we get
moduli space integrations of the type
\ba
&& \int \dr^8\eta \dr^8\bar\xi \left[\left(\zeta^1\right)^2
\zeta^2\left(\zeta^3\right)^2\left(\zeta^4\right)^2\bar\xi^1
\right]\!(x_1)\,\left[\zeta^1\left(\zeta^2\right)^2
\left(\zeta^3\right)^2\left(\zeta^4\right)^2\bar\xi^2 \right]\!(x_2)
\nn \\ &\times& \int \dr^{4(N-2)}\nu\dr^{4(N-2)}\bar\nu
\left(\bar\nu^{[1}
\nu^{4]}\right)^{J+1}\left(\bar\nu^{[2}\nu^{3]}\right)^{J+1}
\left(\bar\nu^{[1}\nu^{2]}\right)\left(\bar\nu^{[3}\nu^{4]}\right) \, .
\label{nonzero-5sphere}
\ea
Just as in the two impurity case the resulting non-vanishing
contribution to the two-point function is suppressed by a factor of
$(\lambda^\pp)^2$ due to the additional $\nsix$ bilinears in
(\ref{nonzero-5sphere}). In conclusion the analysis of fermion zero
modes confirms that the leading non-zero instanton contribution to the
anomalous dimensions of four impurity operators in the $\mb3^+$ and
$\mb3^-$ representations behaves as
\be
\g_{\mb3^+\oplus\mb3^-}^{\rm inst} \sim
\er^{-\frac{8\pi^2}{g_2\lambda^\pp}+i\theta}\,(g_2)^{7/2}\,
(\lambda^\pp)^2 \, ,
\label{3+3-adim}
\ee
in agreement with the string prediction of \cite{gks}.

Other sectors can be analysed in a similar fashion and we find that
all non-singlet operators receive leading non-zero contributions of
the same order as (\ref{3+3-adim}).

\section{Discussion and conclusions}
\label{concl}

This paper has considered one-instanton contributions to two-point
correlation functions of gauge invariant operators in the BMN sector
of the $\scrN$=4 supersymmetric Yang--Mills theory. These determine
the leading instanton contributions to the anomalous dimensions of
operators dual to physical string states in the maximally
supersymmetric plane-wave background obtained as Penrose limit of
AdS$_5\times S^5$.  The basic message is that we find striking
agreement between these instanton effects in the gauge theory and
those of the plane-wave string theory calculated in \cite{gks}.

We focused on operators with two and four scalar impurities. The four
impurity case, although more involved, is fully under control, whereas
the two impurity case presents subtleties due to the fact the leading
semi-classical approximation vanishes and the first non-zero
contribution arises at higher order. We have explicitly computed a
two-point function of four impurity operators which are
SO(4)$_R\times$SO(4)$_C$ singlets.  Our analysis shows that instanton
induced contributions to the anomalous dimensions of operators in this
sector behave as
$1/(n_1n_2)^2\exp\left(-8\pi^2/g_2\lambda^\pp+i\theta\right)g_2^{7/2}$,
where $\lambda^\pp$ and $g_2$ are the effective coupling constant and
genus counting parameter in the BMN limit and $n_1$ and $n_2$
correspond to the mode numbers of the dual string state. The result is
in perfect agreement with the $D$-instanton correction to the mass
matrix elements of the corresponding states in the plane-wave string
theory which was computed in \cite{gks}. Even without directly
matching the numerical values of the anomalous dimensions and the
string mass renormalisation, the agreement with the string calculation
appears highly non-trivial. The correct dependence on the parameters
$\lambda^\pp$ and $g_2$ is obtained by combining contributions arising
from the integrations over the instanton moduli space and various
combinatorial factors. Even more impressively, the mode number
dependence found in \cite{gks} is reproduced after spectacular
cancellations. 

In the case of two impurity operators the leading
instanton correction vanishes.  The first non-zero contribution is
awkward to calculate completely, but with mild assumptions we showed
that it has the form
$\exp\left(-8\pi^2/g_2\lambda^\pp+i\theta\right)g_2^{7/2}{\lambda^\pp}^2$
and does not depend on the single mode number characterising the dual
string state.  This behaviour is also in agreement with the results of
\cite{gks}, although the subtleties presented by the  gauge theory
calculation did not arise on the string side.  Four impurity operators
in sectors other than the singlet have the same $\lambda^\pp$ and
$g_2$ dependence as two impurity operators, again in agreement with
the string prediction of \cite{gks}.

Our results provide a significant new test of the duality proposed in
\cite{bmn}. The fact that non-perturbative contributions obey BMN
scaling, \ie can be re-expressed in terms of the effective parameters
$\lambda^\pp$ and $g_2$, strongly supports the conjecture that this
property should hold at all orders. This can be further tested  by
analysing subleading effects in the instanton background. A class of
higher order contributions can easily be obtained from the
calculations presented in this paper, relaxing the requirement that
all the fermion modes of type $\nu$ and $\bar\nu$ be combined in
$\nsix$ bilinears. As already observed, replacing a $\nsix$ bilinear
with a $\nten$ leads to a suppression by a factor of $1/\sqrt{N}$, see
(\ref{largeN-coeff}). The profile of BMN operators of the type that we
considered contains $J$ $(\bar\nu\nu)$'s after the superconformal
modes have been soaked up. Hence each factor of $1/\sqrt{N}$ coming
from the replacement of a $\nsix$ by a $\nten$ is associated with a
factor of $J$ corresponding to the number of choices for the $\nten$
bilinear, resulting in a suppression by $g_2^{1/2}$. Moreover in order
to get a non-zero result from the five sphere integration an even
number of $\nten$ is required. Therefore contributions of this type
with an increasing number of $\nten$ insertions give rise to
subleading corrections which form a series in integer powers of
$g_2$. More complicated terms with the same behaviour correspond to
contributions in which pairs of fields are contracted between the
two-operators. We can also identify a class of subleading corrections
suppressed by powers of $\lambda^\pp$. These are generated by
including in the profile of the operators a number of fermion modes
greater than the minimal number required by the moduli space
integration. For instance in the case of the four impurity singlets
that we studied in section \ref{epsinglet} this is achieved by
including one six-fermion scalar in one of the two operators. In this
case the calculation is analogous to that of section \ref{2pt-2impur}
for the two impurity case and we can argue that the resulting
contribution should be suppressed by a factor of $\lambda^\pp$.
Including more six-fermion terms gives rise to higher powers of
$\lambda^\pp$. Although these arguments are rather qualitative they
indicate how the perturbative series of corrections to the
semi-classical one-instanton contributions can be reorganised into a
double series in $\lambda^\pp$ and $g_2$.

Some rather striking general properties of the $D$-instanton induced
corrections to the string mass spectrum of \cite{gks} can be
immediately deduced from the structure of the $D$-instanton boundary
state in the plane-wave background, whereas the corresponding effects
in the gauge theory are far from obvious.  In fact, the string theory
results suggest a number of extensions and generalisations of the
gauge theory results.  For example, one generic feature of the string
calculation is that only states with an even number of non-zero mode
insertions receive $D$-instanton corrections. Zero mode oscillators can
appear in odd numbers with the condition that they be contracted into
a SO(4)$\times$SO(4) scalar between the incoming and outgoing
states. The simplest example in which these properties can be verified
involves five impurity operators and the calculation of the necessary
two-point function in the gauge theory is extremely
complicated~\footnote{Three impurity operators present technical
difficulties similar to those encountered in the two impurity
case.}. In general, contributions to operators with a larger (even)
number of impurities are expected to be non-zero at leading order in
the instanton background. However, the complexity of the combinatorics
involved in such calculations grows rapidly with the number of
impurities.

Another peculiarity observed in the string theory calculation
\cite{gks} is that the $D$-instanton contribution to the masses of
certain states with a large number of fermionic non-zero mode
excitations involves large powers of the mass parameter $m$. When
expressed in terms of gauge theory parameters this corresponds to
large inverse powers of $\lambda^\pp$. As observed in \cite{gks} the
behaviour of these mass corrections is not pathological in the
$\lambda^\pp\to 0$ limit, because the inverse powers of $\lambda^\pp$
are accompanied by the instanton factor
$\exp\left(-8\pi^2/g_2\lambda^\pp\right)$. From the point of view of
the gauge theory this result is particularly intriguing not only
because of the unusual coupling constant dependence that the anomalous
dimensions of the dual operators should display, but also because
there are no other known examples of operators in $\scrN$=4 SYM whose
anomalous dimension receives instanton but not perturbative
corrections. We will study this particular class of BMN operators in a
future publication \cite{gks2}.

In the original formulation of the AdS/CFT duality, relating $\scrN$=4
SYM to type IIB string theory in AdS$_5\times S^5$, the effects of
multi-instantons in the large-$N$ limit of the gauge theory and of
multi $D$-instantons in string theory  were shown to be in remarkable
agreement \cite{dhkmv}. Clearly it would be of interest to generalise
the present work from the one-instanton sector to the multi-instanton
sector.  However, such a generalisation is technically very
challenging both on the string and on the gauge side.

 Instanton effects have been studied  in a number of different
supersymmetric field theories \cite{mg,hkm1,hk,gns,th,hkm2,gs} in the
context of the AdS/CFT correspondence, and agreement has been found
between string and gauge theory.  The example of the $\scrN$=2 Sp($N$)
superconformal field theory studied in \cite{mg,gns,th} is
particularly interesting in connection with our work because in this
case the analogue of the BMN limit has been studied in
\cite{bgmnn}. In this case the duality involves a theory of open and
closed strings in a plane-wave background and the dual gauge theory
has a rich spectrum of gauge-invariant operators and possesses a Higgs
branch \cite{gkk}. It would be interesting to study instanton effects
in the BMN sector of this theory.

In a conformal field theory, the problem of computing the scaling
dimensions of gauge invariant operators can be reformulated as an
eigenvalue problem for the dilation operator of the theory. At the
perturbative level this observation leads to a very efficient approach
to the calculation of anomalous dimensions in $\scrN$=4 SYM
\cite{dil}. Some comments about the possibility of extending this
approach to non-perturbative sectors were made in \cite{sk}, but there
has been no further progress in this direction. A remarkable
consequence of recasting the problem of computing anomalous dimensions
as an eigenvalue problem for the dilation operator is the emergence of
connections with integrable systems. In the planar limit the dilation
operator can be related to the hamiltonian of an integrable spin
chain, leading to the possibility of applying techniques such as the
Bethe ansatz to the computation of anomalous dimensions \cite{bs},
see \cite{brep} for a review and references. The integrability
structure observed in the $\scrN$=4 Yang--Mills theory appears,
however, to be spoiled by the inclusion of non-planar
contributions. Therefore instanton effects, which are exponentially
suppressed in the large-$N$ limit, are unlikely to be relevant in
connection with integrability.

Instantons play a special r\^ole in the $\scrN$=4 theory in connection
with the SL(2,$\Z$) $S$-duality symmetry, which transforms the complex
coupling so that $\tau\to\frac{a\tau+b}{c\tau+d}$ (with
$a,b,c,d\in\mathbbm{Z}$ satisfying $ad-bc=1$) thereby mixing
perturbative and non-perturbative effects. In the conformal phase
invariance of the theory requires that the full spectrum of scaling
dimensions be invariant. The anomalous dimensions are therefore
naturally expressed as functions $\g(\tau,\bar\tau)$. Similarly
$D$-instantons are instrumental in the implementation of $S$-duality
in type IIB string theory. Their r\^ole is well understood at the
level of the effective action for the supergravity states, but little
is known at the level of the massive string excitations. As in the SYM
case, invariance of the theory requires that the complete spectrum be
invariant. In general SL(2,$\Z$) transformations relate operators of
small and large dimension, just as in string theory they relate
fundamental strings to $D$-strings, which have large masses of order
$1/g_s$, in the limit of weak string coupling, $g_s\ll 1$. It would be
interesting to understand how $S$-duality is realised in type IIB
string theory in the plane-wave background. A corresponding symmetry
should exist in the BMN sector of $\scrN$=4 SYM and the instanton
effects which we studied in this paper should be important in its
implementation.

\acknowledgments{AS acknowledges financial support from PPARC and
Gonville and Caius college, Cambridge. We also wish to acknowledge
support from the European Union Marie Curie Superstrings Network
MRTN-CT-2004-512194.}

\appendix

\section{Useful formulae}
\label{useformulae}

This appendix contains some definitions and formulae used in the paper.
The Clebsch--Gordan coefficients $\S^i_{AB}$ ($\bar\S^{AB}_i$)
projecting the product of two $\mb4$'s ($\mbb4$'s) onto the $\mb6$ are
defined as
\ba
&& \S^i_{AB} = (\S^a_{AB},\S^{a+3}_{AB})
= (\eta^a_{AB},i\bar\eta^a_{AB}) \nn \\
&& \bar\S_i^{AB} = (\bar\S^a_{AB},\bar\S^{a+3}_{AB})
= (-\eta_a^{AB},i\bar\eta^{AB}_a) \, , \rule{0pt}{18pt}
\label{defsig6}
\ea
where $a=1,2,3$ and the 't Hooft symbols $\eta^a_{AB}$ and
$\bar\eta^a_{AB}$ are
\ba
&& \eta^a_{AB} = \bar\eta^a_{AB} = \veps_{aAB} \, ,
\qquad A,B=1,2,3 \, , \nn \\
&& \eta^a_{A4} = \bar\eta^a_{4A} = \d^a_A \, , \nn \\
&& \eta^a_{AB} = - \eta^a_{BA} \, , \qquad
\bar\eta^a_{AB} = - \bar\eta^a_{BA} \, .
\label{etadef}
\ea
In some situations the $\scrN$=1 formulation proves very useful. The
$\scrN$=1 decomposition of the $\scrN$=4 supermultiplet consists of
three chiral multiplets and one vector multiplet and under this
decomposition only a SU(3)$\times$U(1) subgroup of the SU(4)
R-symmetry group is manifest. The six scalars are combined into three
complex fields, $\phi^I$, $I=1,2,3$, according to
\ba
\phi^I &\!\!=\!\!& \fr{\sqrt{2}} \left(\hat\v^I+i\hat\v^{I+3}\right)
\nn \\
\phi^\dagger_I &\!\!=\!\!& \fr{\sqrt{2}} \left(\hat\v^I
-i\hat\v^{I+3}\right) \, . \label{complexscal}
\ea
The complex scalars $\phi^I$ and $\phi^\dagger_I$ transform
respectively in the $\mb3_{1}$ and $\mbb3_{-1}$ of SU(3)$\times$U(1)
(where the subscript indicates the U(1) charge). The fermions in the
chiral multiplets are
\be
\psi^I_\a = \lambda^I_\a \, , \qquad \bar\psi_I^\adot =
\bar\lambda_I^\adot \, , \qquad I=1,2,3 \, ,
\label{n1chifermions}
\ee
transforming in the $\mb3_{3/2}$ and $\mbb3_{-3/2}$. The fourth
fermion and the vector form the $\scrN$=1  vector multiplet,
$\{\lambda_\a = \lambda^4_\a \, , A_\mu\}$, and are SU(3)$\times$U(1)
singlets.

Using these definitions we find the following relations among the
scalars in the different formulations
\be
\begin{array}{lll}
\displaystyle \hat\v^1 = \frac{\sqrt{2}}{2}\left(\v^{14}
+\v^{23}\right) \, , \; & \displaystyle \hat\v^2 =
\frac{\sqrt{2}}{2} \left( -\v^{13}+\v^{24}\right) \, ,
\; & \displaystyle \hat\v^3 = \frac{\sqrt{2}}{2} \left(
\v^{12}+\v^{34}\right) \, , \\
\displaystyle \hat\v^4 = \frac{i\sqrt{2}}{2} \left( -\v^{14}
+\v^{23}\right) \, , \; & \displaystyle \hat\v^5 =
\frac{i\sqrt{2}}{2} \left(-\v^{13}-\v^{24}\right) \, , \;
& \displaystyle \hat\v^6 = \frac{i\sqrt{2}}{2} \left(
\v^{12}-\v^{34}\right) \rule{0pt}{23pt}
\end{array}
\label{vphii-vphiAB}
\ee
and
\be
\begin{array}{lll}
\displaystyle \phi^1 = 2 \v^{14} \, , \; & \displaystyle
\phi^2 = 2 \v^{24} \, , \; & \displaystyle \phi^3 = 2 \v^{34} \, , \\
\displaystyle \phi^\dagger_1 = 2 \v^{23} \, , \; & \displaystyle
\phi^\dagger_2 = -2 \v^{13} \, , \; & \displaystyle \phi^\dagger_3
= 2 \v^{12} \, . \rule{0pt}{20pt}
\end{array}
\label{phiI-vphiAB}
\ee

In the ADHM formalism the expressions for the $\scrN$=4 elementary
fields in the background of an instanton are conveniently given as
$[N+2]\times[N+2]$ matrices. In particular, the two-fermion solution
for the scalar field $\v^{AB}$ in the one instanton sector can be
written in the block-form
\be
({\hat\v}^{AB})_{u\b;}{}^{v\g} = \left( \begin{array}{cc} \left(
{\hat\v}^{(1)AB} \right)_{u;}{}^{v} & \quad
\left({\hat\v}^{(2)AB}\right)_{u;}{}^{\g} \\
\left({\hat\v}^{(3)AB}\right)_{\b;}{}^{v} & \quad
\left({\hat\v}^{(4)AB}\right)_{\b;}{}^{\g} \end{array} \right) \, ,
\label{hatscalar}
\ee
where $u,v=1,2,\ldots,N$, $\a,\b=1,2$ and  
\ba
\left({\hat\v}^{(1)AB}\right)_{u;}{}^{v} &\!\!=\!\!&
\frac{1}{4(y^2+\rho^2)^2} \left\{ y^2 \left[ -16
({\bar\xi}^{\adot B}{\bar\xi}^A_\bdot - {\bar\xi}^{\adot A}
{\bar\xi}^B_\bdot)w_{u;\adot}{\bar w}^{\bdot;v}
\right. \right. \nn \\
&\!\!+\!\!& \left. 4 w_{u;\adot}
({\bar\xi}^{\adot B} {\bar\nu}^{A\,v}-{\bar\xi}^{\adot A}
{\bar\nu}^{B\,v}) \right] \nn \\
&\!\!+\!\!&(y^2+\rho^2) \left[ -4({\bar\xi}^B_\adot
\nu^A_u-{\bar\xi}^A_\adot \nu^B_u){\bar w}^{\adot;v} +
(\nu^B_u{\bar\nu}^{A\,v} - \nu^A_u{\bar\nu}^{B\,v})\right]  \nn \\
&\!\!+\!\!& \left. y^{\adot\d}\left[ 16(\eta^B_\d{\bar\xi}^A_\bdot -
\eta^A_\d{\bar\xi}^B_\bdot)w_{u;\adot}{\bar w}^{\bdot;v} -
4w_{u;\adot} (\eta^B_\d {\bar\nu}^{Av} - \eta^A_\d {\bar\nu}^{Bv})
\right]\right\} \nn \\
\left({\hat\v}^{(2)AB}\right)_{u;}{}^{\g} &\!\!=\!\!&
\frac{1}{4(y^2+\rho^2)^2} \left\{ 16y^2
w_{u;\adot}({\bar\xi}^{\adot B}\eta^{\g A} - {\bar\xi}^{\adot A}
\eta^{\g B}) + 4 (y^2+\rho^2)(\nu^B_u \eta^{\g A} - \nu^A_u
\eta^{\g B})  \rule{0pt}{18pt} \right. \nn \\
&\!\!-\!\!& \left. w_{u;\adot} \left[
16y^{\adot\d} (\eta^B_\d \eta^{\g A} - \eta^A_\d \eta^{\g B}) +
\frac{1}{2} \frac{y^2+\rho^2}{\rho^2} y^{\adot\g} ({\bar\nu}^{Au}
\nu^B_u - {\bar\nu}^{Br} \nu^A_r) \right] \right\} \nn \\
\left({\hat\v}^{(3)AB}\right)_{\b;}{}^{v} &\!\!=\!\!&
\frac{1}{4(y^2+\rho^2)^2}\left\{ \rho^2
\left[ 16y_{\b\adot}({\bar\xi}^{\adot B}{\bar\xi}^A_\bdot
- {\bar\xi}^{\adot A}{\bar\xi}^B_\bdot) {\bar w}^{\bdot;v}
- 4y_{\b\adot} ({\bar\xi}^{\adot B} {\bar\nu}^{Av} -
{\bar\xi}^{\adot A} {\bar\nu}^{Bv} )  \right. \right. \nn \\
&\!\!-\!\!& \left.\left. 16 (\eta^B_\b{\bar\xi}^A_\adot -
\eta^A_\b{\bar\xi}^B_\adot) {\bar w}^{\adot;v} + 4(\eta^B_\b
{\bar\nu}^{Av} - \eta^A_\b {\bar\nu}^{Bv}) \right]\right\} \nn \\
\left({\hat\v}^{(4)AB}\right)_{\b;}{}^{\g} &\!\!=\!\!&
\frac{\rho^2}{4(y^2+\rho^2)^2} \left[
-16y_{\b\adot}({\bar\xi}^{\adot B}\eta^{\g A} - {\bar\xi}^{\adot A}
\eta^{\g B}) +16 (\eta^B_\b \eta^{\g A} - \eta^B_\b \eta^{\g A})
\right.  \nn \\
&\!\!+\!\!& \left. \frac{1}{2}\frac{y^2+\rho^2}{\rho^2}
\delta_\b^\g ({\bar\nu}^{Ar} \nu^B_r - {\bar\nu}^{Br} \nu^A_r)
\right] \, .
\label{phi2-inst-sol}
\ea

\section{Instanton induced two-point functions of BMN operators}
\label{calculations}

In this appendix we present some details of the calculations of
one-instanton contributions to the two-point functions of BMN
operators discussed in section \ref{2pt-BMN}.

\subsection{Two-impurity operator in the $\mb9$ of SO(4)$_R$}
\label{2imp9-detail}

As shown in section \ref{2pt-2impur} the leading semi-classical
contribution to the two-point functions of two impurity operators in
the $\mb9$ of SO(4)$_R$ vanishes because the superconformal modes
cannot be soaked up. A non-zero result is obtained including for one
scalar field in each operator the six fermion solution.

In the case of the component considered in section \ref{2pt-2impur}
the terms in the two-point function which contain the correct
combination of fermion modes to give a non vanishing contribution are
\ba
G_{\mb9}(x_1,x_2) &\!\!=\!\!& 
\fr{J\left(\frac{\gy^2N}{8\pi^2}\right)^{J+2}} \sum_{p,q=0}^J
\cos\!\left(\frac{2\pi ipn}{J}\right)\cos\!\left(\frac{2\pi iqm}{J}
\right) \nn \\
&& \times \left\{ \rule{0pt}{15pt}
\la \Tr\left[\left(Z^{J-p}\v^{13}Z^p\v^{13}\right)
(x_1)\right] \Tr\left[\left(\barZ^{J-q}\v^{24}\barZ^q\v^{24}\right)
(x_2)\right]\ra \right. \nn \\
&& \left. \rule{0pt}{15pt}\hsp{0.2}
+ \la \Tr\left[\left(Z^{J-p}\v^{24}Z^p\v^{24}\right)
(x_1)\right] \Tr\left[\left(\barZ^{J-q}\v^{13}\barZ^q\v^{13}\right)
(x_2)\right]\ra \right\} \, .
\label{2pt-2imp-9-c}
\ea
The other terms vanish in the instanton background either because they
do not contain all the required superconformal modes or because of the
integration over the five-sphere. For instance if one considers
$\la\Tr\left[\left(Z^{J-p}\v^{13}Z^p\v^{13}\right) (x_1)\right]
\Tr\left[\left(\barZ^{J-q}\v^{13}\barZ^q\v^{13}\right)
(x_2)\right]\ra$ it is easy to verify that the  superconformal modes
can be soaked up, but the resulting five-sphere integral vanishes
because among the remaining fermion modes different flavours appear
with different multiplicities.

Let us consider the terms in (\ref{2pt-2imp-9-c}) where in each trace
one scalar is understood to be replaced with the six-fermion
solution. In the first expectation value the two traces contain
respectively the following combinations of fermion modes
\ba
&& \left(\scrmf^1\right)^{J+3}\left(\scrmf^2\right)^1
\left(\scrmf^3\right)^3\left(\scrmf^4\right)^{J+1} \nn \\
&& \left(\scrmf^1\right)^1\left(\scrmf^2\right)^{J+3}
\left(\scrmf^3\right)^{J+1}\left(\scrmf^4\right)^3 \, .
\label{fm-2imp-9-1}
\ea
Using the fact that each trace contains one $\bar\xi$ mode not part of
a $\zeta$ we can soak up the superconformal modes selecting
the following combinations of fermion modes in the two traces
\ba
\Tr\left( Z^{J-p}\v^{13}Z^p\v^{13} \right) &\to&
\left(\zeta^1\right)^2 \zeta^2 \left(\zeta^3\right)^2
\left(\zeta^4\right)^2 \bar\xi^1
\left(\bar\nu^{[1}\nu^{4]}\right)^{J-1}
\left(\bar\nu^{[1}\nu^{3]}\right) \nn \\
\Tr\left( \barZ^{J-p}\v^{24}\barZ^p\v^{24} \right) &\to&
\zeta^1 \left(\zeta^2\right)^2 \left(\zeta^3\right)^2
\left(\zeta^4\right)^2 \bar\xi^2
\left(\bar\nu^{[2}\nu^{3]}\right)^{J-1}
\left(\bar\nu^{[2}\nu^{4]}\right) \, .
\label{fm-2imp-9-2}
\ea
Similarly the two traces in the second term in (\ref{2pt-2imp-9-c})
contain respectively
\be
\left(\scrmf^1\right)^{J+1}\left(\scrmf^2\right)^3
\left(\scrmf^3\right)^1\left(\scrmf^4\right)^{J+3}
\label{fm-2imp-9-3a}
\ee
and
\be
\left(\scrmf^1\right)^3\left(\scrmf^2\right)^{J+1}
\left(\scrmf^3\right)^{J+3}\left(\scrmf^4\right)^1 \,
\label{fm-2imp-9-3b}
\ee
and we need to consider
\ba
\Tr\left( Z^{J-p}\v^{24}Z^p\v^{24} \right) &\to&
\left(\zeta^1\right)^2 \left(\zeta^2\right)^2 \zeta^3
\left(\zeta^4\right)^2 \bar\xi^4
\left(\bar\nu^{[1}\nu^{4]}\right)^{J-1}
\left(\bar\nu^{[2}\nu^{4]}\right) \nn \\
\Tr\left( \barZ^{J-p}\v^{13}\barZ^p\v^{13} \right) &\to&
\left(\zeta^1\right)^2 \left(\zeta^2\right)^2
\left(\zeta^3\right)^2\zeta^4\bar\xi^3
\left(\bar\nu^{[2}\nu^{3]}\right)^{J-1}
\left(\bar\nu^{[1}\nu^{3]}\right) \, .
\label{fm-2imp-9-4}
\ea
These expressions contain the correct combinations of fermion
superconformal modes such that the corresponding integration is
non-zero. The two terms in (\ref{2pt-2imp-9-c}) give rise to
\ba
&& \int \prod_{A=1}^4 \dr^2\eta^A\,\dr^2\bar\xi^A\,
\left\{\left[\left(\zeta^1\right)^2 \zeta^2 \left(\zeta^3\right)^2
\left(\zeta^4\right)^2\bar\xi^1\right]\!(x_1)
\left[\zeta^1\left(\zeta^2\right)^2 \left(\zeta^3\right)^2
\left(\zeta^4\right)^2\bar\xi^2\right]\!(x_2)\right. \nn \\
&& \left. \hsp{2.8} + \left[\left(\zeta^1\right)^2 \left(\zeta^2\right)^2
\zeta^3 \left(\zeta^4\right)^2\bar\xi^4\right]\!(x_1)
\left[\left(\zeta^1\right)^2\left(\zeta^2\right)^2
\left(\zeta^3\right)^2\zeta^4\bar\xi^3\right]\!(x_2)\right\} \nn \\
&& \hsp{2.9} \sim (x_1-x_0)\cdot(x_2-x_0)\,(x_1-x_2)^4 \, .
\label{suconfint2}
\ea
After re-expressing the $(\bar\nu^A\nu^B)$ bilinears in terms of
$\Omega^{AB}$'s as described in section \ref{inst2ptf} both
(\ref{fm-2imp-9-2}) and (\ref{fm-2imp-9-4}) lead to the same
five-sphere integral,
\be
I_{S^5}=\int\dr^5\Omega \, \left(\Omega^{14}\right)^{J-1}
\left(\Omega^{23}\right)^{J-1} \Omega^{13}\Omega^{24} \, .
\label{2imp-5sphere-1}
\ee
This can be evaluated rewriting it as
\be
I_{S^5}=\int_{\sum_{i=1}^6\Omega_i^2=1} \dr^6\Omega \,\left(\S^{14}_i
\Omega^i\right)^{J-1} \left(\S^{23}_j\Omega^j\right)^{J-1}
\left(\S^{13}_k\Omega^k\right)\left(\S^{24}_l\Omega^l\right) \,
\label{2imp-5sphere-2}
\ee
where the symbols $\S^{AB}_i$ are defined in (\ref{defsig6}).
Defining $\Omega\equiv\S^{14}_i\Omega^i=(\Omega^1+i\Omega^4)$,
$\bar\Omega\equiv\S^{23}_i\Omega^i=(\Omega^1-i\Omega^4)$, $\wtil\Omega
\equiv\S^{13}_i\Omega^i=(\Omega^2+i\Omega^5)$ and $\bar{\wtil\Omega}
\equiv\S^{24}_i\Omega^i=(\Omega^2-i\Omega^5)$ the integral reduces to
\ba
I_{S^5} &=& \int \dr^6\Omega \: \d\!\left(\sum_{i=1}^6\Omega_i^2-1\right)
\left(\Omega\bar\Omega\right)^{J-1}\left(\wtil\Omega\bar{\wtil\Omega}
\right) \nn \\
&=& \int \dr\Omega\,\dr\bar\Omega\,\dr\wtil\Omega\,\dr\bar{\wtil\Omega}
\,\dr^2\Omega^I \: \d(\Omega^I\Omega^I+\Omega\bar\Omega+\wtil\Omega
\bar{\wtil\Omega}-1) \left(\Omega\bar\Omega\right)^{J-1}
\left(\wtil\Omega\bar{\wtil\Omega} \right) \, ,
\label{2imp-5sphere-3}
\ea
where $\Omega^I=(\Omega^3,\Omega^6)$. Introducing polar coordinates
for the $\Omega^I$ directions
\ba
I_{S^5} &=& 2\pi \int \dr r \, r \, \dr\Omega\,\dr\bar\Omega\,
\dr\wtil\Omega\,\dr\bar{\wtil\Omega} \: \d(r^2+\Omega\bar\Omega+
\wtil\Omega\bar{\wtil\Omega}-1) \left(\Omega\bar\Omega\right)^{J-1}
\left(\wtil\Omega\bar{\wtil\Omega} \right) \nn \\
&=& \pi \int_{\Omega\bar\Omega+\wtil\Omega\bar{\wtil\Omega}\le 1}
\dr\Omega\,\dr\bar\Omega\,\dr\wtil\Omega\,\dr\bar{\wtil\Omega} \:
\left(\Omega\bar\Omega\right)^{J-1}\left(\wtil\Omega\bar{\wtil\Omega}
\right) \, .
\label{2imp-5sphere-4}
\ea
The remaining integrals are straightforward
\ba
I_{S^5} &=& 2\pi^2 \int_{\Omega\bar\Omega\le 1} \dr\Omega\,
\dr\bar\Omega \: \left(\Omega\bar\Omega\right)^{J-1}
\int_0^{\sqrt{1-\Omega\bar\Omega}}\dr z z^3 \nn \\
&=& \frac{\pi^2}{2} \int_{\Omega\bar\Omega\le 1} \dr\Omega\,
\dr\bar\Omega \: \left(1-\Omega\bar\Omega\right)^2
\left(\Omega\bar\Omega\right)^{J-1} = \frac{\pi^3}{2}
\int_0^1\dr z \, z \left(1-z^2\right)^2 z^{2J} \nn \\
&=& \frac{\pi^3}{J(J+1)(J+2)} \, .
\label{2imp-5sphere}
\ea
As observed in section \ref{2pt-2impur} the exact dependence on the
bosonic moduli in the two-point function $G_\mb9(x_1,x_2)$ cannot be
determined without knowing the six-fermion solution. Dimensional
analysis indicates that the final bosonic integrations over position
and size of the instanton are logarithmically divergent as expected in
the presence of an instanton contribution to the anomalous dimension
of the operator $\scrO^{\{ij\}}$.

\subsection{$\veps$-singlet four impurity operator}
\label{eps-singlet-detail}

In order to compute the profiles of the operator $\scrO_\mb1$ in
(\ref{epsdef}) and its conjugate, which are needed in the calculation
of the two-point function (\ref{eps1-2pt-def}), we have to evaluate the
traces (\ref{35traces}). In the instanton background these are
rewritten as traces of $[N+2]$-dimensional ADHM matrices. To compute
these traces more efficiently it is convenient to define the
$[N+2]\times[N+2]$ matrix
\be
\left[U_{k_1,k_2}^{C_1,D_1;C_2,D_2}(\zeta,\nu,\bar\nu)
\right]_{u,\a}{}^{v,\b} = \left[\left(\check\v^{14}\right)^{k_1}
\what\v^{C_1D_1}\left(\check\v^{14}\right)^{k_2}
\what\v^{C_2D_2}\right]_{u,\a}{}^{v,\b} \, .
\label{Udef}
\ee
where the notation used is that introduced in
(\ref{checkscal})-(\ref{hatscal}). This has the standard
block-form of ADHM matrices and the range of the indices here is the
same as in (\ref{phi2-inst-sol}) for the elementary  scalar fields.

In terms of the matrix
$U_{k_i,k_j}^{C_i,D_i;C_j,D_j}(\zeta,\nu,\bar\nu)$
all the 35 traces we are interested in can be written as
\be
\Tr\left[U_{k_1,k_2}^{C_1,D_1;C_2,D_2}(\zeta,\nu,\bar\nu)
U_{k_3,k_4}^{C_3,D_3;C_4,D_4}(\zeta,\nu,\bar\nu)
U_{k_5,k_6}^{C_5,D_5;C_6,D_6}(\zeta,\nu,\bar\nu)
U_{k_7,k_8}^{C_7,D_7;C_8,D_8}(\zeta,\nu,\bar\nu)
\right] \, ,
\label{gen35trace}
\ee
for appropriate choices of the indices $C_i$, $D_i$ and the exponents
$k_i$, $i=1,\ldots,8$. For example the three traces written explicitly
in (\ref{35traces}) become
\bdm
\Tr\left[U_{p_1,p_2}^{1,4;1,4}(\zeta,\nu,\bar\nu)
U_{p_3,p_4}^{1,4;1,4}(\zeta,\nu,\bar\nu)
U_{p_5,q}^{A_1,B_1;A_2,B_2}(\zeta,\nu,\bar\nu)
U_{r,s}^{A_3,B_3;A_4,B_4}(\zeta,\nu,\bar\nu)\right] \, ,
\edm
\bdm
\Tr\left[U_{p_1,p_2}^{1,4;1,4}(\zeta,\nu,\bar\nu)
U_{p_3,p_4}^{1,4;A_1,B_1}(\zeta,\nu,\bar\nu)
U_{q_1,q_2}^{1,4;A_2,B_2}(\zeta,\nu,\bar\nu)
U_{r,s}^{A_3,B_3;A_4,B_4}(\zeta,\nu,\bar\nu)\right] \nn \\
\label{U35traces}
\edm
and
\bdm
\Tr\left[U_{p,q}^{A_1,B_1;A_2,B_2}(\zeta,\nu,\bar\nu)
U_{r,s_1}^{A_3,B_3;1,4}(\zeta,\nu,\bar\nu)
U_{s_2,s_3}^{1,4;1,4}(\zeta,\nu,\bar\nu)
U_{s_4,s_5}^{1,4;A_4,B_4}(\zeta,\nu,\bar\nu)\right] \, .
\edm
The generic trace (\ref{gen35trace}) is thus the only one that needs
to be evaluated. It can be computed using the building blocks
(\ref{phi2-inst-sol}) and the result is
\ba
&& \hsp{-0.2} \Tr\left[U_{k_1,k_2}^{C_1,D_1;C_2,D_2}(\zeta,\nu,\bar\nu)
U_{k_3,k_4}^{C_3,D_3;C_4,D_4}(\zeta,\nu,\bar\nu)
U_{k_5,k_6}^{C_5,D_5;C_6,D_6}(\zeta,\nu,\bar\nu)
U_{k_7,k_8}^{C_7,D_7;C_8,D_8}(\zeta,\nu,\bar\nu)\right] \nn \\
&& \hsp{-0.2} = \fr{2^{3J-8}}
\frac{\rho^8}{[(x-x_0)^2+\rho^2]^{J+8}}
\left(\bar\nu^{[1}\nu^{4]}\right)^{J-8} \left\{ \left[
\left(\zeta^{D_1}\zeta^{D_2}\right)\left(\zeta^{D_3}\zeta^{D_4}\right)
\left(\zeta^{D_5}\zeta^{D_6}\right)\left(\zeta^{D_7}\zeta^{D_8}\right)
\right] \rule{0pt}{14pt} \right. \nn \\
&& \hsp{-0.2} \left[\left(\bar\nu^{[C_8}\nu^{4]}\right)
\left(\bar\nu^{[1}\nu^{C_1]}\right) + \left(\bar\nu^{[C_8}\nu^{1]}\right)
\left(\bar\nu^{[4}\nu^{C_1]}\right)\right]
\left[\left(\bar\nu^{[C_2}\nu^{4]}\right)
\left(\bar\nu^{[1}\nu^{C_3]}\right) + \left(\bar\nu^{[C_2}\nu^{1]}\right)
\left(\bar\nu^{[4}\nu^{C_3]}\right)\right] \nn \\
&& \hsp{-0.2} \left[\left(\bar\nu^{[C_4}\nu^{4]}\right)
\left(\bar\nu^{[1}\nu^{C_5]}\right) + \left(\bar\nu^{[C_4}\nu^{1]}\right)
\left(\bar\nu^{[4}\nu^{C_5]}\right)\right]
\left[\left(\bar\nu^{[C_6}\nu^{4]}\right)
\left(\bar\nu^{[1}\nu^{C_7]}\right) + \left(\bar\nu^{[C_6}\nu^{1]}\right)
\left(\bar\nu^{[4}\nu^{C_7]}\right)\right] \nn \\
&& \hsp{-0.2} + \left[
\left(\zeta^{D_2}\zeta^{D_3}\right)\left(\zeta^{D_4}\zeta^{D_5}\right)
\left(\zeta^{D_6}\zeta^{D_7}\right)\left(\zeta^{D_8}\zeta^{D_1}\right)
\right] \left[\left(\bar\nu^{[C_1}\nu^{4]}\right)
\left(\bar\nu^{[1}\nu^{C_2]}\right) + \left(\bar\nu^{[C_1}\nu^{1]}\right)
\left(\bar\nu^{[4}\nu^{C_2]}\right)\right] \nn \\
&& \hsp{-0.2} \left[\left(\bar\nu^{[C_3}\nu^{4]}\right)
\left(\bar\nu^{[1}\nu^{C_4]}\right) + \left(\bar\nu^{[C_3}\nu^{1]}\right)
\left(\bar\nu^{[4}\nu^{C_4]}\right)\right]
\left[\left(\bar\nu^{[C_5}\nu^{4]}\right)
\left(\bar\nu^{[1}\nu^{C_6]}\right) + \left(\bar\nu^{[C_5}\nu^{1]}\right)
\left(\bar\nu^{[4}\nu^{C_6]}\right)\right] \nn \\
&& \hsp{-0.2} \left. \left[\left(\bar\nu^{[C_7}\nu^{4]}\right)
\left(\bar\nu^{[1}\nu^{C_8]}\right) + \left(\bar\nu^{[C_7}\nu^{1]}\right)
\left(\bar\nu^{[4}\nu^{C_8]}\right)\right] + {\rm permutations}
\rule{0pt}{14pt} \right\} \, ,
\label{explUtrace}
\ea
where the permutations not indicated explicitly correspond to
antisymmetrisation in all the $(C_i,D_i)$ pairs.

The key feature of (\ref{explUtrace}) is that it depends on the set of
indices $(C_i,D_i)$, but not on the exponents $k_i$. The traces
(\ref{35traces}) require four pairs of $(C_i,D_i)$ indices to be
$(1,4)$ while the remaining four pairs correspond to the $(A_k,B_k)$
indices carried by the impurities. The fact that (\ref{explUtrace})
does not depend on the $k_i$'s means that the traces (\ref{35traces})
do not depend on the exponents on the $Z$'s, but only on the relative
positions of the $\what Z$'s with respect to the impurities. Therefore
when substituting into the definition (\ref{epsdef}) of the operator
the traces (\ref{explUtrace}) can be taken out of the sums over the
indices $q$, $r$, $s$. After substituting the values of the indices
corresponding to the various terms in the expansion (\ref{epssingexp})
and some simple Fierz rearrangements the profile of the operator
$\scrO_\mb1$ takes the form of a common factor containing the
dependence on the bosonic and fermionic moduli, multiplying the
combination $K(n_1,n_2,n_3;J)$ of 35 sums which contain the dependence
on the mode numbers $n_1$, $n_2$ and $n_3$, see (\ref{modnumfunct}).
To illustrate more concretely how this works let us describe
explicitly one particular term. We consider the first trace in
(\ref{epssingexp}) and compute the contribution of the last type in
(\ref{35traces}) for this trace. We have to evaluate
\ba
&& \frac{1}{\sqrt{J^3\left(\frac{\gy^2N}{8\pi^2}\right)^{J+4}}}
\begin{array}[t]{c}
{\displaystyle \sum_{q,r,s_1,\ldots,s_5=0}^J} \\
{\scriptstyle q+r+s_1+\cdots+s_5 \le J-4} \\
{\scriptstyle s_1+\cdots+s_5=s-4}
\end{array} \hsp{-0.1} \er^{2\pi i[(n_1+n_2+n_3)q
+(n_2+n_3)r+n_3s]/J} \nn \\
&& \hsp{3} \times \Tr\left(\check Z^{J-(q+r+s)}\what\v^{12}
\check Z^{q}\what\v^{13}\check Z^{r}\what\v^{24}\check Z^{s_1}
\what Z\check Z^{s_2}\what Z\check Z^{s_3}\what Z\check Z^{s_4}
\what Z\check Z^{s_5}\what \v^{34}\right) \nn \\
&& = \frac{1}{\sqrt{J^3\left(\frac{\gy^2N}{8\pi^2}\right)^{J+4}}}
\begin{array}[t]{c}
{\displaystyle \sum_{q,r,s_1,\ldots,s_5=0}^J} \\
{\scriptstyle q+r+s_1+\cdots+s_5 \le J-4} \\
{\scriptstyle s_1+\cdots+s_5=s-4}
\end{array} \hsp{-0.1} \er^{2\pi i[(n_1+n_2+n_3)q
+(n_2+n_3)r+n_3s]/J} \nn \\
&& \hsp{6} \times \Tr\left(U_{p,q}^{1,2;1,3}
U_{r,s_1}^{2,4;1,4}U_{s_2,s_3}^{1,4;1,4}
U_{s_4,s_5}^{1,4;3,4}\right) \, .
\label{1term1}
\ea
Using (\ref{explUtrace}) with the particular choice of indices in the
trace in (\ref{1term1}) we get (up to a numerical constant)
\ba
&& \hsp{-0.5} \Tr\left(U_{p,q}^{1,2;1,3}
U_{r,s_1}^{2,4;1,4}U_{s_2,s_3}^{1,4;1,4}
U_{s_4,s_5}^{1,4;3,4}\right) 
= \fr{2^{3J+8}}\,\frac{\rho^8}{[(x_1-x_0)^2+\rho^2]^{J+8}}
\left(\bar\nu^{[1}\nu^{4]}\right)^J \nn \\
&& \hsp{0.5} \times \left[\left(\zeta^1\zeta^1
\right)\!\left(\zeta^2\zeta^3\right)\!\left(\zeta^2\zeta^3\right)
\!\left(\zeta^4\zeta^4\right) - \left(\zeta^1\zeta^4\right)\!
\left(\zeta^4\zeta^4\right)\!\left(\zeta^2\zeta^3\right)\!
\left(\zeta^2\zeta^3\right) \right] \rule{0pt}{18pt}\nn \\
&& \hsp{0.5} = -\frac{3}{2}\,\fr{2^{3J+8}}\, 
\frac{\rho^8}{[(x_1-x_0)^2+\rho^2]^{J+8}}
\left(\bar\nu^{[1}\nu^{4]}\right)^J \left[\left(
\zeta^1\right)^2 \left(\zeta^2\right)^2 \left(\zeta^3\right)^2
\left(\zeta^4\right)^2\right] \, , \rule{0pt}{18pt}
\label{1termtrace}
\ea
where the last line has been obtained using simple Fierz
rearrangements on the $\zeta$'s. As anticipated the trace is
independent of the exponents, $q$, $r$, $s_i$. Equation
(\ref{1term1}) then becomes
\ba
&& -\frac{3}{2} \, \fr{2^{3J+8}}\,
\frac{1}{\sqrt{J^3\left(\frac{\gy^2N}{8\pi^2}\right)^{J+4}}}
\frac{\rho^8}{[(x_1-x_0)^2+\rho^2]^{J+8}} \,
\left(\bar\nu^{[1}\nu^{4]}\right)^J \left[\left(
\zeta^1\right)^2 \left(\zeta^2\right)^2 \left(\zeta^3\right)^2
\left(\zeta^4\right)^2\right] \nn \\
&&  \times \begin{array}[t]{c}
{\displaystyle \sum_{q,r,s=0}^J} \\
{\scriptstyle q+r+s \le J}
\end{array} \hsp{-0.1} \er^{2\pi i[(n_1+n_2+n_3)q
+(n_2+n_3)r+n_3s]/J} \fr{4!} s(s-1)(s-2)(s-3) \, ,
\label{1term2}
\ea
where we have used the fact that there is no dependence on the single
exponents, $s_1$,\ldots,$s_5$, so that the result is independent of
the way the four $\what Z$ are distributed among the last $s$
$Z$'s. This leads to the factor $\fr{4!} s(s-1)(s-2)(s-3)$ which  is a
multiplicity coefficient associated with the number of ways of picking
four identical $\what Z$'s out of $s$ $Z$'s. Equation (\ref{1term2})
illustrates the factorisation of the result into two terms, the first
line containing the dependence on the instanton moduli and the second
line containing the dependence on the mode numbers.

The function $K(n_1,n_2,n_3;J)$ takes the form
\be
K(n_1,n_2,n_3;J) = \sum_{a=1}^{35} c_a\,\calS_a(n_1,n_2,n_3;J) \, ,
\label{modnumfunct2}
\ee
where each of the $\calS_a(n_1,n_2,n_3;J)$ is a sum similar to the
second line of  (\ref{1term2}) with different summand corresponding to
the different multiplicity factors associated with the distributions
of $\what Z$'s in the traces (\ref{35traces}). Table \ref{coefftab}
summarises the contributions to (\ref{modnumfunct2}).

\TABLE[!h]{
{\small
\begin{tabular}{|c|c||c|c|}
\hline
~~{\rm Summand}~~ & Coefficient ($c_a$) & ~~{\rm Summand}~~
&  Coefficient ($c_a$) \rule[-5pt]{0pt}{18pt} \\
\hline \hline
~ $\fr{4!}p(p-1)(p-2)(p-3)$\rule[-5pt]{0pt}{18pt} & $-18$ &
~$\fr{2!}prs(s-1)$\rule[-5pt]{0pt}{18pt} & $0$  \\
\hline
~$\fr{3!}p(p-1)(p-2)q$\rule[-5pt]{0pt}{18pt} & $-9$ &
~$\fr{3!}ps(s-1)(s-2)$\rule[-5pt]{0pt}{18pt} & $9$  \\
\hline
~$\fr{3!}p(p-1)(p-2)r$\rule[-5pt]{0pt}{18pt} & $-18$ &
~$\fr{4!}q(q-1)(q-2)(q-3)$\rule[-5pt]{0pt}{18pt} & $18$  \\
\hline
~$\fr{3!}p(p-1)(p-2)s$\rule[-5pt]{0pt}{18pt} & $-9$ &
~$\fr{3!}q(q-1)(q-2)r$\rule[-5pt]{0pt}{18pt} & $9$  \\
\hline
~$\fr{(2!)^2}p(p-1)q(q-1)$\rule[-5pt]{0pt}{18pt} & $0$ &
~$\fr{3!}q(q-1)(q-2)s$\rule[-5pt]{0pt}{18pt} & $18$  \\
\hline
~$\fr{2!}p(p-1)qr$\rule[-5pt]{0pt}{18pt} & $-9$ &
~$\fr{(2!)^2}q(q-1)r(r-1)$\rule[-5pt]{0pt}{18pt} & $0$  \\
\hline
~$\fr{2!}p(p-1)qs$\rule[-5pt]{0pt}{18pt} & $0$ &
~$\fr{2!}q(q-1)rs$\rule[-5pt]{0pt}{18pt} & $9$  \\
\hline
~$\fr{(2!)^2}p(p-1)r(r-1)$\rule[-5pt]{0pt}{18pt} & $-18$ &
~$\fr{(2!)^2}q(q-1)s(s-1)$\rule[-5pt]{0pt}{18pt} & $18$  \\
\hline
~$\fr{2!}p(p-1)rs$\rule[-5pt]{0pt}{18pt} & $-9$ &
~$\fr{3!}qr(r-1)(r-2)$\rule[-5pt]{0pt}{18pt} & $-9$  \\
\hline
~$\fr{(2!)^2}p(p-1)s(s-1)$\rule[-5pt]{0pt}{18pt} & $0$ &
~$\fr{2!}qr(r-1)s$\rule[-5pt]{0pt}{18pt} & $0$  \\
\hline
~$\fr{3!}pq(q-1)(q-2)$\rule[-5pt]{0pt}{18pt} & $9$ &
~$\fr{2!}qrs(s-1)$\rule[-5pt]{0pt}{18pt} & $9$  \\
\hline
~$\fr{2!}pq(q-1)r$\rule[-5pt]{0pt}{18pt} & $0$ &
~$\fr{3!}qs(s-1)(s-2)$\rule[-5pt]{0pt}{18pt} & $18$  \\
\hline
~$\fr{2!}pq(q-1)s$\rule[-5pt]{0pt}{18pt} & $9$ &
~$\fr{4!}r(r-1)(r-2)(r-3)$\rule[-5pt]{0pt}{18pt} & $-18$  \\
\hline
~$\fr{2!}pqr(r-1)$\rule[-5pt]{0pt}{18pt} & $-9$ &
~$\fr{3!}r(r-1)(r-2)s$\rule[-5pt]{0pt}{18pt} & $-9$  \\
\hline
~$pqrs$\rule[-5pt]{0pt}{18pt} & $0$ &
~$\fr{(2!)^2}r(r-1)s(s-1)$\rule[-5pt]{0pt}{18pt} & $0$  \\
\hline
~$\fr{2!}pqs(s-1)$\rule[-5pt]{0pt}{18pt} & $9$ &
~$\fr{3!}rs(s-1)(s-2)$\rule[-5pt]{0pt}{18pt} & $9$  \\
\hline
~$\fr{3!}pr(r-1)(r-2)$\rule[-5pt]{0pt}{18pt} & $-18$ &
~$\fr{4!}s(s-1)(s-2)(s-3)$\rule[-5pt]{0pt}{18pt} & $18$  \\
\hline
~$\fr{2!}pr(r-1)s$\rule[-5pt]{0pt}{18pt} & $-9$  &
~ & ~ \\
\hline
\end{tabular}}
\caption{Contributions to the function $K(n_1,n_2,n_3;J)$. $\calS_a$
  are sums over the indices $q,r,s\in[0,J]$ (with the constraint
  $q+r+s\le J$) in which the summands are those indicated in the table
  multiplied by the phase factor $\exp\left(2\pi
  i[(n_1+n_2+n_3)q+(n_2+n_3)r+n_3s]/J\right)$. Here $p=J-(q+r+s)$. The
  $c_a$'s are the combined coefficients taking into account all the
  terms in the operator.}
\label{coefftab}
}

Using the coefficients given in table \ref{coefftab}, and noting that
the phase factor factorises, (\ref{modnumfunct2}) can be written as
\ba
K(n_1,n_2,n_3)&=&\begin{array}[t]{c}
{\displaystyle -\frac{3}{4}\sum_{p,q,r,s=0}^J} \\
{\scriptstyle q+r+s+p=J}
\end{array} \hsp{-0.1} \er^{2\pi i[(n_1+n_2+n_3)q
+(n_2+n_3)r+n_3s]/J}(p-q+r-s)(p+q+r+s)^3 \nonumber \\
&=&\begin{array}[t]{c}
{\displaystyle -\frac{3J^3}{4}\sum_{p,q,r=0}^J} \\
{\scriptstyle p+q+r \le J}
\end{array} \hsp{-0.1} \er^{2\pi i[(n_1+n_2+n_3)p
+(n_2+n_3)q+n_3r]/J}(2p+2r-J) \, .
\label{combsums}
\ea
The sums in (\ref{combsums}) can be approximated with integrals in the
$J\to\infty$ limit, which can then be evaluated differentiating a
generating function. The relevant generating function is given by
\ba
g(a_1,a_2,a_3,a_4)&=&\int_0^1\dr x \int_0^{1-x}\dr y \int_0^{1-x-y}
\dr z \,\er^{2\pi i[a_1 x+a_2 y+a_3 z+a_4 (1-x-y-z)]} \nn \\
&=& \frac{i}{8\pi^3} \sum_{i=1}^4
\frac{\er^{2\pi i a_i}}{\prod_{j=1,j\neq i}^4(a_i-a_j)} \, .
\label{gfunction}
\ea
Thus the above sum requires evaluating
\be
f(a_1,a_2,a_3,a_4)=\frac{3i}{8\pi} \left(2
\frac{\partial}{\partial a_1}+2\frac{\partial}{\partial a_3}-2\pi i
\right) g(a_1,a_2,a_3,a_4) \, .
\label{combsum-gf}
\ee
As discussed in section \ref{epsinglet} in order to get the correct
mode number dependence we need to antisymmetrise (\ref{combsum-gf})
with respect to the exchange of pairs of mode numbers. Therefore we
need to compute
\ba
&&\lim_{n_3\rightarrow -n_1}\left[ f(n_1+n_2+n_3,n_2+n_3,n_3,0) -
f(n_1+n_2+n_3,n_1+n_3,n_3,0) \right. \nn \\
&& \hsp{1} \left. - f(-n_3,-n_1-n_3,-n_1-n_2-n_3,0) + 
f(-n_3,-n_2-n_3,-n_1-n_2-n_3,0) \right] \nn \\
&& \hsp{1} = \frac{3}{8\pi^2}\frac{1}{(n_1 n_2)} \, ,
\label{antisym-gf}
\ea
where only in the case where we impose pairwise equality do we get a
non-zero result. In conclusion the mode number dependence in the
profile of the operator $\scrO_\mb1$ is
\be
K(n_1,n_2,n_3;J) = \frac{3}{8\pi^2}\frac{J^7}{(n_1 n_2)} \, ,
\label{ni-dep}
\ee
where the factor of $J^7$ is the combination of the $J^3$ in
(\ref{combsums}) and a $J^4$ arising from the conversion of the sums
into integrals in the continuum limit.

The two-point function $G_\mb1(x_1,x_2)$ thus becomes
\ba
&&\hsp{-0.2} G_\mb1(x_1,x_2) = \frac{J^{11}\,\er^{2\pi i\tau}}{N^{7/2}}
\fr{(n_1n_2)(m_1m_2)} \int \frac{\dr^4x_0\,\dr\rho}{\rho^5} \,
\frac{\rho^{J+8}}{[(x_1-x_0)^2+\rho^2]^{J+8}}
\frac{\rho^{J+8}}{[(x_2-x_0)^2+\rho^2]^{J+8}} \nn \\
&& \hsp{-0.2} \times \int \prod_{A=1}^4\dr^2\eta^A\dr^2\bar\xi^A \,
\prod_{B=1}^4 \left[\left(\zeta^B\right)^2(x_1)\right]
\left[\left(\zeta^B\right)^2(x_2)\right] \int \dr^5\Omega \,
\left(\Omega^{14}\right)^J \left(\Omega^{23}\right)^J \, .
\label{2pt-4imp-2a}
\ea
The integrations in the second line of (\ref{2pt-4imp-2a}) are
straightforward. The integrals over the fermion superconformal modes
give $(x_1-x_2)^8$, see (\ref{etabxiint}). The five sphere integral is
similar to that encountered in the two impurity case and can be
calculated in a similar fashion. Proceeding as in
(\ref{2imp-5sphere-1})-(\ref{2imp-5sphere}) we get
\be
I_{S^5} = \int \dr^5\Omega \,
\left(\Omega^{14}\right)^J \left(\Omega^{23}\right)^J
= \int \dr\Omega\,\dr\bar\Omega\,\dr^4\Omega^I \:
\d(\Omega^I\Omega^I+\Omega\bar\Omega-1)
\left(\Omega\bar\Omega\right)^J \, ,
\label{4imp-5sphere-1}
\ee
where $\Omega=(\Omega^1+i\Omega^4)$, $\bar\Omega=(\Omega^1-i\Omega^4)$
and $\Omega^I=(\Omega^2,\Omega^3,\Omega^5,\Omega^6)$, so that introducing
spherical coordinates
\ba
I_{S^5}&=& 2\pi^2 \int \dr r \, r \, \dr\Omega\,\dr\bar\Omega\,
\: \d(r^2+\Omega\bar\Omega-1) \left(\Omega\bar\Omega\right)^J \nn \\
&=& 2\pi^2 \int_{\Omega\bar\Omega\le 1} \dr\Omega\,\dr\bar\Omega \:
(\left(1-\Omega\bar\Omega\right)\left(\Omega\bar\Omega\right)^J
= \frac{\pi^3}{(J+1)(J+2)} \, .
\label{4imp-5sphere}
\ea
The integration over the bosonic part of the moduli space must be
treated carefully since it is logarithmically divergent as expected in
the presence of a contribution to the matrix of anomalous
dimensions. The integrals need to be regulated for instance by
dimensional regularisation of the $x_0$ integral. Introducing Feynman
parameters we get
\ba
I_{\rm b} &=& \int \frac{\dr^4x_0\,\dr\rho}{\rho^5} \,
\frac{\rho^{J+8}}{[(x_1-x_0)^2+\rho^2]^{J+8}}
\frac{\rho^{J+8}}{[(x_2-x_0)^2+\rho^2]^{J+8}} \nn \\
&=& \frac{\Gamma(2J+16)}{[\Gamma(J+8)]^2} \int_0^1\dr\a_1\dr\a_2 \,
\d(1-\a_1-\a_2)\a_1^{J+7}\a_2^{J+7} \nn \\
&\times& \int\dr^4x_0\,\dr\rho \,
\frac{\rho^{2J+11}}{[(x_0-\a_1x_1-\a_2x_2)^2+\rho^2
+\a_1\a_2x_{12}^2]^{2J+16}} \, .
\label{bosonint1}
\ea
After dimensional regularisation,
\ba
I_{\rm b} \to I^{(\epsilon)}_{\rm b} &=&
\frac{\Gamma(J+6)\Gamma(J+8+\epsilon)}{[\Gamma(J+8)]^2}
\,\pi^{2-\epsilon} \fr{(x_{12}^2)^{J+8+\epsilon}}
\int_0^1\dr\a \, \fr{[\a(1-\a)]^{1+\epsilon}} \nn \\
 &=& \fr{\epsilon}\frac{\Gamma(J+6)\Gamma(J+8+\epsilon)}
{[\Gamma(J+8)]^2} \,\pi^{2-\epsilon} \fr{(x_{12}^2)^{J+8+\epsilon}}
\, . \label{regulint}
\ea
The $1/\epsilon$ pole corresponds to a logarithmic divergence in
dimensional regularisation.

Substituting into (\ref{2pt-4imp-2a}) we finally get
\be
G_1(x_1,x_2) \sim (g_2)^{7/2}\,\er^{-\frac{8\pi^2}{g_2\lambda^\pp}+i\theta}
\fr{(n_1n_2)(m_1m_2)} \fr{(x_{12}^2)^{J+4}} \,
\log\left(\Lambda^2x_{12}^2\right) \, ,
\label{4imp-2pt-as2}
\ee
where the exact numerical coefficient was given in (\ref{4imp-2pt-as}).

\end{document}